\documentclass[acmsmall,authorsversion]{acmart}
\makeatletter
\def\@ACM@checkaffil{%
  \if@ACM@instpresent\else\ClassWarningNoLine{\@classname}{No institution present for an affiliation}\fi
  \if@ACM@citypresent\else\ClassWarningNoLine{\@classname}{No city present for an affiliation}\fi
  \if@ACM@countrypresent\else\ClassWarningNoLine{\@classname}{No country present for an affiliation}\fi}
\def\@folioblob{}
\makeatother
\usepackage{graphicx} % Required for inserting images

\newcommand{\RM}[1]{}
\newcommand{\GDchange}[1]{#1}


\title{GRay: Ray Tracing 3D Gaussians Near the Speed of Splats}

\author{Yohan Poirier-Ginter}
\orcid{0009-0001-5806-5136}
\affiliation{%
  \institution{Université Laval}
  \city{Quebec}
  \country{Canada}
}
\affiliation{%
  \institution{Inria, Université Côte d'Azur}
  \city{Nice}
  \country{France}
}
\email{yohan.poirier-ginter.1@ulaval.ca}

\author{Jean-François Lalonde}
\orcid{0000-0002-6583-2364}
\affiliation{
  \institution{Université Laval}
  \city{Quebec}
  \country{Canada}
}
\email{jean-francois.lalonde@gel.ulaval.ca}

\author{George Drettakis}
\orcid{0000-0002-9254-4819}
\affiliation{%
  \institution{Inria, Université Côte d'Azur}
  \city{Nice}
  \country{France}
}
\email{george.drettakis@inria.fr}

\setcopyright{rightsretained}
\acmJournal{PACMCGIT}
\acmYear{2026} \acmVolume{9} \acmNumber{1} \acmArticle{14}
\acmMonth{5} \acmDOI{10.1145/3804496}

\usepackage{microtype}
\usepackage{pifont}    % \ding marks used in the appendix tables
\usepackage{adjustbox} % wide appendix tables
\usepackage{cleveref}
\crefname{section}{sec.}{secs.}
\Crefname{section}{Sec.}{Secs.}
\crefname{paragraph}{sec.}{secs.}
\Crefname{paragraph}{Sec.}{Secs.}
\crefname{table}{tab.}{tabs.}
\Crefname{table}{Tab.}{Tabs.}
\crefname{figure}{fig.}{figs.}
\Crefname{figure}{Fig.}{Figs.}
\crefname{equation}{eq.}{eqs.}
\Crefname{equation}{Eq.}{Eqs.}

\newcommand{\changed}[1]{#1}   % added text
\newcommand{\deleted}[1]{}

\begin{abstract}
3D Gaussian Splatting (3DGS) is a popular representation for radiance field reconstruction, distinguished by the rendering speed of its rasterization-based renderer. While 3D Gaussians can also be ray traced, this approach has so far been slower, with 3D Gaussian Ray Tracing (3DGRT) \cite{3dgrt} taking nearly one order of magnitude longer to optimize. To address this, we present GRay, a fast ray tracer for 3D Gaussians designed to close this performance gap and match 3DGS's speed. 
Our method leverages the algorithmic difference between both approaches: unlike rasterization, ray tracing evaluates only Gaussians that are actually intersected by a ray, 
leading to potentially logarithmic---rather than linear---scaling in the number of primitives. 
This property allows ray tracing to better exploit dense scenes composed of numerous tiny Gaussians, a configuration which has largely been overlooked. Notably, we show that dense initialization---which creates many small Gaussians---slows down rasterization, but instead \emph{speeds up} ray tracing. Designed to leverage this effect, GRay renders nearly $4{\times}$ faster and optimizes nearly $10{\times}$ faster than 3DGRT while maintaining similar quality, and has competitive speed with 3DGS albeit at somewhat lower quality. Code is available at \url{https://repo-sam.inria.fr/nerphys/gray}.

\end{abstract}

\begin{document}

\newcommand{\todo}[1]{\textcolor{red}{\bfseries TODO: #1}}
\newcommand{\notsure}[1]{\textcolor{orange}{\bfseries NOTSURE: #1}}
\newcommand{\iftime}[1]{\textcolor{purple}{\bfseries IFTIME: #1}}
\acmSubmissionID{20}

\begin{teaserfigure}
\vspace{-.5em}
\centering
\includegraphics[width=\textwidth]{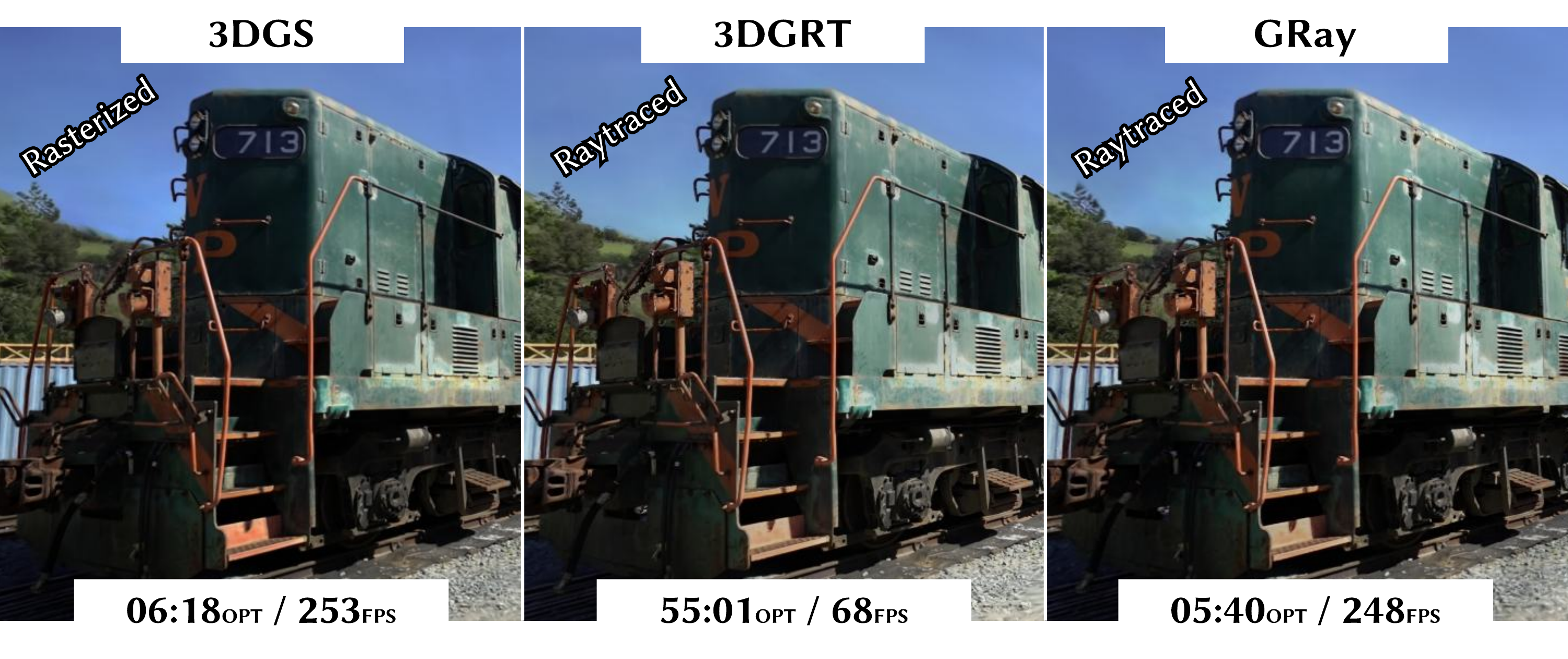}
\centering
\resizebox{\textwidth}{!}{%
\begin{tabular}{@{}lccccccccc@{}}
\toprule
Method & PSNR$_\uparrow$ & SSIM$_\uparrow$ & LPIPS$_\downarrow$ & Init Time$_\downarrow$ & Opt Time$_\downarrow$ & FPS$_\uparrow$ & Init \#G$_\downarrow$ & Final \#G$_\downarrow$ & \#Iterations \\
\midrule
3DGS          & \textbf{27.10} & \textbf{0.831} & 0.262 & \textbf{00{:}00} & 06{:}18 & \textbf{253} & \textbf{0.11M}  & 2.25M  & 30000 \\
3DGRT         & 26.77 & 0.828 & 0.258 & \textbf{00{:}00} & 55{:}01 & 68         & \textbf{0.11M}  & 3.24M  & 30000 \\
GRay          & 26.47 & 0.819 & \textbf{0.236} & 01{:}58 & \textbf{05{:}40 } & 248 & 3.27M       & \textbf{1.52M} & 15000 \\
\bottomrule
\end{tabular}
}
\caption{Reconstruction quality and performance of 3DGS~\cite{3dgs} vs. 3DGRT~\cite{3dgrt} vs. our method GRay averaged over the 13 standard 3DGS evaluation scenes; an example test view is also provided as reference. By leveraging dense initialization to its full extent, GRay runs nearly $4{\times}$ faster and optimizes nearly $10\times$ faster than 3DGRT while improving LPIPS.}\label{fig:teaser}
\Description{Table comparing PSNR, SSIM, LPIPS, training speed, and Gaussian counts for 3DGS, 3DGRT, and GRay alongside a test image from the Train scene for each method.}
\end{teaserfigure}  
\maketitle

\section{Introduction}

3D Gaussian Splatting (3DGS)~\cite{3dgs} is a well-established representation for radiance field reconstruction and rendering, known for its speed. Its efficiency arises from its highly optimized CUDA tile-based rasterizer which accumulates and composites Gaussians in screen space: by sharing memory among neighboring pixels within a tile, it minimizes memory traffic, and by sorting with a consistent global ordering, it avoids redundant work. However, rasterization faces fundamental limitations: it is restricted to pinhole cameras and unable to compute arbitrary light paths. While extensions such as 3DGUT~\cite{3dgut} can adapt rasterization to more general camera models, full ray tracing remains indispensable for future research, especially for inverse rendering and relighting~\cite{r3dg,envgs,interreflective-gs,reflective-gs}. Moreover, it provides a conceptually simpler formulation, free of popping artifacts and of the affine approximation. Unfortunately, ray tracing also forfeits many of the efficiencies that make splatting fast: sorting must be done per pixel, incurring more work, and shared memory can no longer be exploited, incurring higher memory traffic. Consequently, it is often regarded as inherently slower than splatting: the most popular ray tracing method, 3DGRT~\cite{3dgrt}, optimizes about $9\times$ slower than 3DGS in our experiments.

In this work, we show that the speed gap between splatting and ray tracing can be closed by leveraging their difference in algorithmic scaling. At its core, ray tracing uses a BVH to select and evaluate exactly the set of Gaussians intersecting each pixel. As such, its cost is governed by the number of true intersections, which can yield logarithmic scaling in the number of primitives if they are sufficiently small. In contrast, 3DGS's rasterizer operates on tiles, evaluating \emph{all} Gaussians overlapping a tile for every pixel within it; this results in substantial wasted computation for small Gaussians, and linear scaling regardless of their size.\footnote{Prior work has identified this inefficiency and proposed partial remedies~\cite{adr-gaussian,speedy-splat,stop-the-pop,aaa-gaussians}, but ray tracing’s pixel-wise selection provides a precise and principled solution.}
This key algorithmic difference allows ray tracing to better exploit dense scenes composed of numerous small Gaussians; in principle, ray tracing should scale better than splatting at high Gaussian densities. This has largely been overlooked in existing works and has never been verified empirically.

First, we show that algorithmic scaling matters in practice by switching ray tracing to dense initialization (DI), which was recently proposed in EDGS~\cite{edgs} as a superior alternative to Surface from Motion initialization with iterative densification. With DI, instead of progressively refining a sparse set of large Gaussians, matching networks are used to predict millions of small Gaussians densely covering all visible surfaces at the outset. While it was proposed to eliminate the densification stage and accelerates convergence, we show that DI further benefits ray tracing. Notably, while DI slows down rasterization, it instead \emph{speeds up} ray tracing as predicted by complexity analysis. In fact, it reduces ray tracing workloads even at the start of optimization with drastically inflated Gaussian counts. Hence, we argue that algorithmic scaling matters and that ray tracing is inherently well-suited to the DI regime.

We further propose a novel ray tracing method designed to leverage ray tracing's algorithmic scaling and exploit tiny Gaussians. Our method curates and extends 3DGRT~\cite{3dgrt} and EPBRR~\cite{epbrr} for dense initialization, and beyond the ray tracer itself, we also revisit the training regimen through improved initialization, pruning, learning rates and regularization. Our ray tracing solution, dubbed GRay, exploits DI by design.

In summary, our main contributions are:
\begin{itemize}
\item Showing that algorithmic complexity matters in Gaussian ray tracing, by scaling ray tracing to high counts of tiny Gaussians via dense initialization (DI). We are the first to clearly demonstrate DI’s synergistic effect with ray tracing and suggest that Gaussian ray tracing could outscale splatting at very high Gaussian densities;
\item A novel fast ray tracer for 3D Gaussians, GRay, with optimized hyperparameters and implementation, which nearly matches 3DGS's frame rate while maintaining the highest possible quality;
\item An extensive investigation of ray tracing techniques in the context of DI, validated through comprehensive ablation studies and experimental analysis.
\end{itemize}

GRay achieves nearly $4{\times}$ faster frame rates than 3DGRT and nearly $10{\times}$ faster optimization with similar (or slightly lower) quality---that is, at speeds matching 3DGS with and without dense initialization. Code is available at \url{https://repo-sam.inria.fr/nerphys/gray}.

\section{Related Work}
In this section we first \changed{briefly describe 3DGS and provide an overview of its initialization and densification strategy, while comparing the latter} to the dense initialization proposed by Kotovenko~et~al.~\shortcite{edgs} (EDGS). We then survey prior work on ray tracing 3D Gaussians.

\paragraph{3D Gaussian Splatting.}
3DGS~\cite{3dgs} introduced a representation for radiance fields based on rasterized anisotropic 3D Gaussians. Given target images, 3DGS first estimates their camera poses using Structure From Motion~\cite{sfm,colmap-sfm} (SfM), then optimizes its Gaussians to match the targets. In 3DGS's tile-based rasterization, pixels are grouped into tiles of $16 \times 16$ pixels, for which the relevant Gaussians are depth-sorted once and then composited efficiently with shared memory. Its high quality and speed, as well as the explicit nature of its representation, have led to widespread adoption~\cite{a-survey-3dgs,impact-and-outlook-of-3dgs}.

3DGS proposed initializing Gaussians from the sparse point cloud obtained from SfM’s triangulated matches, which greatly improved on random initialization. In this approach, initial sizes are determined based on the average distance between the approximate 3 nearest neighbors. However, because this point cloud is sparse, areas where the sparse initialization is lacking must be ``filled in'' with additional Gaussians. As such, 3DGS employs adaptive density control throughout optimization: to increase detail, Gaussians with high error indicated by high positional gradients are split and cloned; to reduce waste, Gaussians with low opacity are pruned. While effective, this method requires a large number of steps (30000) to converge, and while it does detect many “dead’’ Gaussians, it does not catch them all. Hence, numerous followup works have since proposed alternative pruning criteria to be used during or after training~\cite{spotlesssplats,pup3dgs,epbrr,speedy-splat,radsplat,eagles,mini-splatting,reducing-the-memory-footprint}.

Noting the inherent limitations of sparse initialization, Kotovenko~et~al.~\shortcite{edgs} (EDGS) proposed dense initialization (DI) as an alternative. In their approach, dense matching networks (notably RoMa~\cite{roma}) are used to find pixel correspondences between views. These correspondences are then triangulated into 3D with the SfM camera poses as usual. To keep computational expenses reasonable, they first \RM{subset}\GDchange{subsample} the input images to a set of reference images, and for each reference image, only compute matches against a few nearest neighbors determined by comparing camera poses. EDGS's dense initialization only takes a few minutes to run.

Because DI’s correspondences already cover most surfaces reasonably well, densification is no longer required: quality levels similar to sparse initialization become possible without it. Because densification is omitted and surfaces are better approximated from the start, convergence by iteration count is much faster, even though individual iterations can be costlier due to the high primitive count, especially before any pruning takes place. 

\deleted{Recent works on feedforward 3D reconstruction have proven capable of producing metrically consistent 3D point maps and camera poses without running classical SfM pipelines~\cite{mast3r,vggt}, and have also been used for initializing 3DGS~\cite{instant-splat}. Approaches for predicting Gaussians directly from input images while omitting optimization are also showing promise~\cite{anysplat,lvt}. As such, network-based initialization will likely become increasingly common in the near future, offering more control over Gaussian initialization.}

\paragraph{3D Gaussian Raytracing.} 3D Gaussian Raytracing (3DGRT)~\cite{3dgrt} first studied hardware-accelerated ray tracing with OptiX~\cite{optix} as an alternative for rendering 3D Gaussians. Their method bounds each Gaussian with a proxy triangle mesh and batches their evaluation, accumulating contributions in fixed-size groups of $K$ Gaussians \changed{with repeated BVH traversals}, as required to avoid register spilling. They also investigated different approaches for accelerating ray tracing, including techniques which traded some quality for speed; we explain the ones we retain directly in \cref{sec:method}. On our hardware, 3DGRT still optimizes about $9\times$ slower than a recent version of 3DGS (\cref{fig:teaser}) despite quality reduction.

More recently, Poirier-Ginter et al.~\shortcite{epbrr} (EPBRR) proposed various improvements to 3DGRT’s ray tracer, while also noting the apparent value of DI. However, these techniques were not tested against proper dense initialization and were not included in a full ray-tracing solution. \changed{They only offered limited comparisons to 3DGRT which excluded DI and trained at reduced iteration counts without Spherical Harmonics (SH)}, and were not validated individually. \changed{Their method} used a crude custom implementation of DI which can be summarized as ``binning the depth maps''---experiments in EDGS, on the other hand, argued against this approach. We detail all inclusions from EPBRR in \cref{sec:method}. Furthermore, our work thoroughly investigates the performance impact of their contributions under proper DI. 

\paragraph{Other Gaussian Ray Tracers} Several other works have proposed ray-traced variants of 3D Gaussians, but none achieve speeds comparable to 3DGS. 

RayGauss~\cite{raygauss} attains very high quality at the cost of performance by treating 3D Gaussians as volumetric primitives and ray marching them in a NeRF-like manner~\cite{nerf}. To this end, it introduces a slab-by-slab integration scheme and augments spherical harmonics with a spherical Gaussian encoding to better capture view-dependent color. While this yields higher PSNR than 3DGS, it comes with a $6\times$ training slowdown and approximately $25\times$ lower frame rates. RayGaussX~\cite{raygaussx} later extends this approach with several optimizations aimed at accelerating ray marching, including empty-space skipping, adaptive sampling, and improved ray coherence through Morton-order sorting. \changed{We briefly experiment with their Morton-order sorting but did not find it to bring a significant improvement when applied to our method.} It also introduces a regularizer to limit anisotropic Gaussians as these are highly inefficient since they bound Gaussians by axis-aligned bounding boxes. These additions reduce training time to be comparable with 3DGRT, but frame rates remain approximately $2.5\times$ slower. Most of these techniques are either orthogonal to our approach or inapplicable since they are designed for ray marching; for instance, we do not require anisotropy regularization due to our use of oriented bounding boxes (\cref{sec:method}). More crucially, our goals differ: while RayGauss and RayGaussX prioritize improving quality over 3DGS while accepting a performance loss, our goal is instead to match the speed of 3DGS while tolerating a modest reduction in quality.

RaySplats~\cite{raysplats} notably accelerates ray tracing by capping the maximum number of Gaussians accumulated along each ray, making it possible to store all Gaussians in a fixed-size local memory array. This cap may trigger early termination even when the transmittance threshold has not yet been reached. To stabilize training under this approximation, RaySplats continues accumulating additional Gaussians beyond the cap and uses their contributions to correct the gradients of the front-most Gaussians. The technique we refer to as \emph{detached hybrid transparency} follows the same underlying principle, but is based on EPBRR's implementation which excludes the fixed cap on the number of Gaussians and the secondary threshold; also note that the use of the per-pixel-linked-lists (\cref{sec:method}) relieves the need for fixed size arrays. While they report good reconstruction quality, the authors of RaySplats make no claims regarding rendering performance.

Sun et al.~\shortcite{stochastic-ray-tracing} proposes stochastic subsampling to accelerate the ray tracing of pretrained 3D Gaussians, but doesn't offer a backward pass and only reaches frame rates about $3 \times$ below 3DGS.

Other works include custom ray tracing solutions for their specific needs: Condor et al.~\shortcite{dont-splat-your-gaussians} uses ray tracing for physically accurate volumetric integration of 3D Gaussians (potentially even including scattering), while using icosahedrons to bound Gaussians much like 3DGRT. They also proposed an alternative kernel function to increase compactness (we instead use the modified kernel proposed by 3DGRT). EVER~\cite{ever} uses ray tracing to obtain the exact per-pixel sorting required for this improved blending algorithm, but uses axis-aligned bounding boxes and does not reach high performance ($4-20 \times$ slower FPS than 3DGS). REdiSplats~\cite{redisplats} ray traces flat 3D Gaussians aligned to mesh-based disks, which allows for controlling Gaussians by editing mesh vertices; however, they do not compare FPS or training times to existing approaches. 

\changed{Finally, the concurrent work GRTX~\cite{grtx} proposed a hardware extension to accelerate 3DGRT-style repeated BVH traversals, and proposed using icosahedron meshes with instancing --- akin to icosahedrons \emph{inside} oriented bounding boxes.}

\paragraph{Gaussian Ray Tracing's Applications.} Finally, ray tracing Gaussians finds use in many downstream applications like inverse rendering and relighting~\cite{r3dg,envgs,interreflective-gs,reflective-gs}, which for the moment tend to use custom-built solutions and often ray trace 2D Gaussian primitives. While ray tracing 2D Gaussians seems promising, we stick to 3D Gaussians in this work to compare more clearly against 3DGS/3DGRT. Besides, alternative approaches for ray tracing radiance fields have also been proposed \cite{radiant-foam,salf}, but these are out of scope for this paper, since they introduce different representations.

\section{Can Ray Tracing Surpass Splatting?}
\label{sec:outscale}

The current literature on Gaussian rendering portrays ray tracing as inherently slower than splatting, a price that must be paid for its numerous advantages. However, the two approaches are algorithmically distinct and exhibit fundamentally different scaling behavior. At its core, 3DGS's tile-based rasterizer scales linearly with the number of Gaussian primitives in a tile, since sorting involves all primitives and since each primitive is evaluated for every pixel in that tile. In contrast, ray tracing selects only the Gaussians relevant to each pixel through BVH traversal, and therefore scales with the number of true ray-Gaussian intersections rather than the number of primitives. This distinction has two important consequences. First, shrinking Gaussians consistently accelerates ray tracing by reducing intersection counts, whereas rasterization only benefits when Gaussians are large enough to cover multiple tiles. Second, when the Gaussian count grows without increasing the likelihood of intersection—namely when surfaces are represented with more numerous smaller Gaussians—ray tracing can approach logarithmic scaling in the number of primitives. As such, in the limit of large counts of tiny Gaussians, ray tracing can potentially outperform rasterization. 

\begin{figure}[t]
    \centering
    \includegraphics[width=0.48\linewidth]{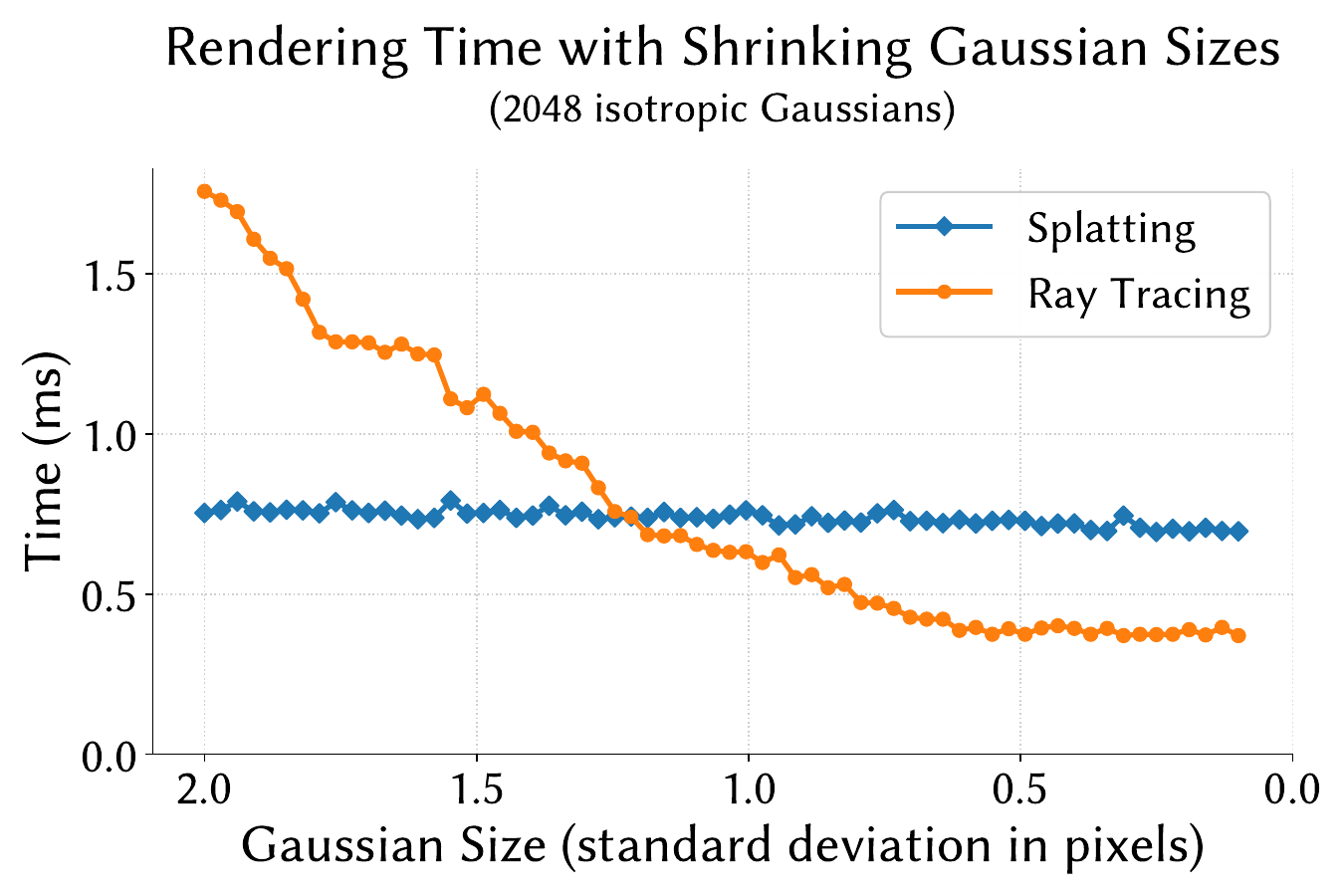}
    \includegraphics[width=0.48\linewidth]{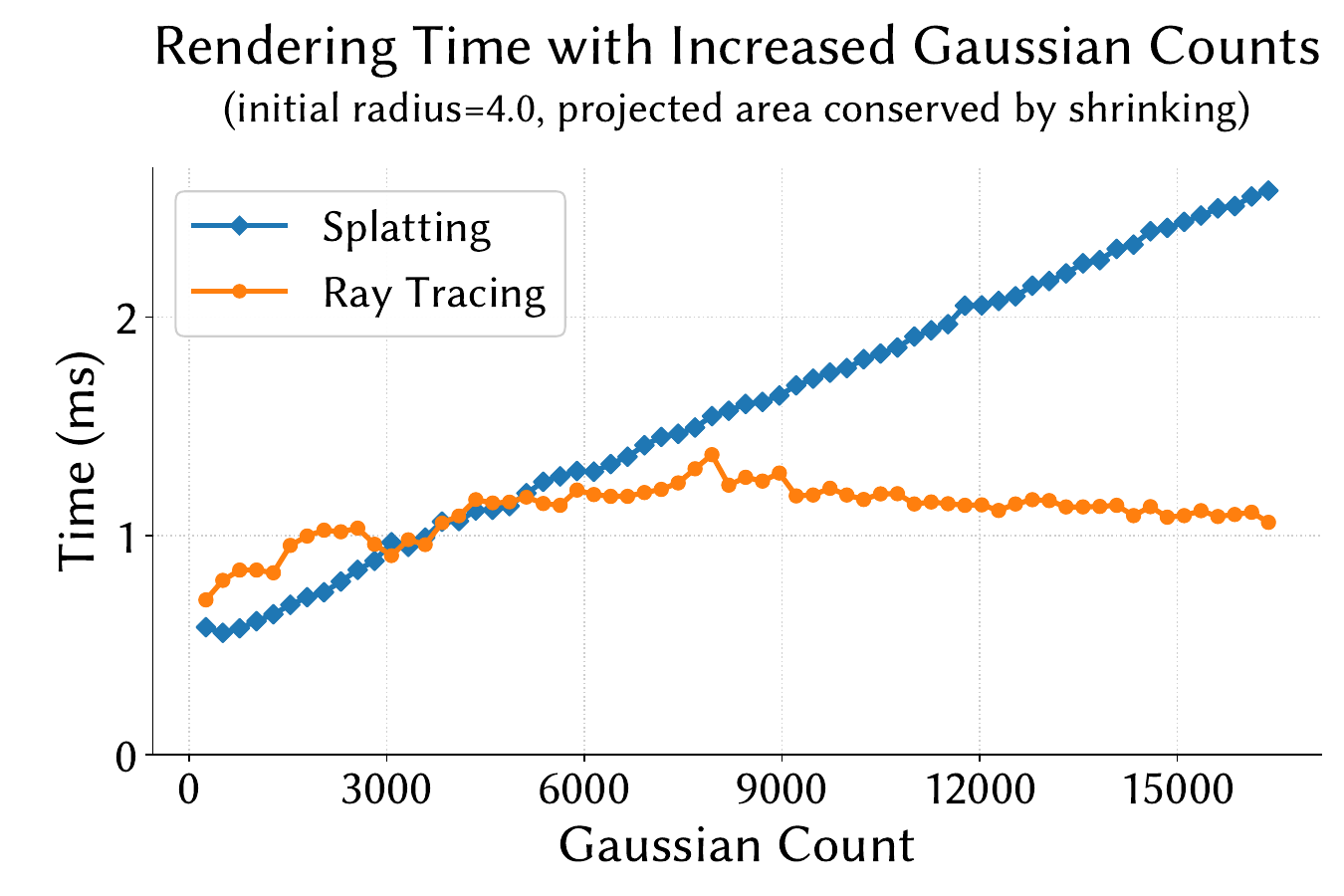}
    \caption{Ray tracing can scale better than splatting when the rendered primitives are small. This test renders a single $16 \times 16$ pixel tile at different Gaussian counts and scales. Left: as Gaussians shrink, ray tracing speeds up to eventually overtake splatting which stays nearly constant. Right: when more Gaussians are added without increasing their total projected area, ray tracing exhibits logarithmic scaling. }
    \Description{Left: a plot showing the rendering time with shrinking Gaussians sizes (radius going down from 2.5 to 0.0), where splatting stays nearly constant around 0.8-0.9 while ray tracing starts at 2.5 but decreases to around 0.4. Right: a plot showing the rendering time with increasing Gaussian counts while projected area is conserved, splatting increases linearly from around 0.5 to 2.5 as the Gaussian count grows from 0 to 17.5k, while ray tracing exhibits logarithmic growth, starting at around 0.8 but plateauing near 1.25. }
    \label{fig:rt-outscales}
\end{figure}

\paragraph{Ray Tracing Speeds Up When Gaussians Shrink} To show this difference in computational complexity clearly, we rendered a single $16 \times 16$ pixel tile with both 3DGS and our ray tracer GRay, which we introduce later in this paper, and progressively reduced Gaussian sizes. The toy scene consists of a grid of 2048 isotropic Gaussians, with a $90^\circ$ field of view camera looking down on the grid. We rendered a first image with Gaussians at a standard deviation of 2.0~pixels, and then progressively reduced their standard deviation down to 0.1~pixels. Results are shown in the left of \cref{fig:rt-outscales}. While ray tracing initially renders much slower, as the Gaussians shrink, it can eventually overtake splatting since it benefits from smaller Gaussian sizes. Which method is fastest depends on the number of Gaussians, but such trends are consistent across counts; see \cref{fig:toy-sweeps} for details. 

\paragraph{Ray Tracing Can Scale Logarithmically} Next, we tested scaling in the number of primitives. We rendered the same toy scene at one Gaussian per pixel with standard deviations of 4.0~pixels and progressively increased Gaussian counts up to 64 per pixel. While adding Gaussians, we progressively shrunk their size such as to conserve the same projected surface area; for instance, quadrupling the Gaussian count involved halving their standard deviations. Results are shown in the right of \cref{fig:rt-outscales}: when Gaussian size is controlled for, ray tracing exhibits logarithmic scaling under increased primitive counts. Hence, depending on the size of its primitives, ray tracing can eventually surpass splatting as Gaussian counts increase. Which method is fastest in practice is sensitive to the initial standard deviation, but the scaling behavior is unchanged by this initial size; see \cref{fig:toy-sweeps} for details. \changed{Also note that although BVH construction remains $O(n \log n)$, update cost is small in practice; this is also explained in \cref{sec:bvh-details}}.

\section{Algorithmic Complexity Matters in Practice}
To test whether the difference in algorithmic complexity exposed in \cref{sec:outscale} is relevant in practice, we employ the well-known ray tracer 3DGRT~\cite{3dgrt} and switch its configuration to use dense initialization (DI)~\cite{edgs}, a realistic scenario where Gaussians are both smaller and more numerous.

\begin{figure}[h!]
\centering
\resizebox{\linewidth}{!}{%
\setlength{\tabcolsep}{1pt}
\begin{tabular}{lccc}
 & $\mathrm{3DGRT}_{\mathrm{SI}}$ & $\mathrm{3DGRT}_{\mathrm{SI}}^{*}\ (\text{matches 3DGS})$ & $\mathrm{3DGRT}_{\mathrm{DI}}$ \\

 & 
\includegraphics[width=0.33\linewidth]{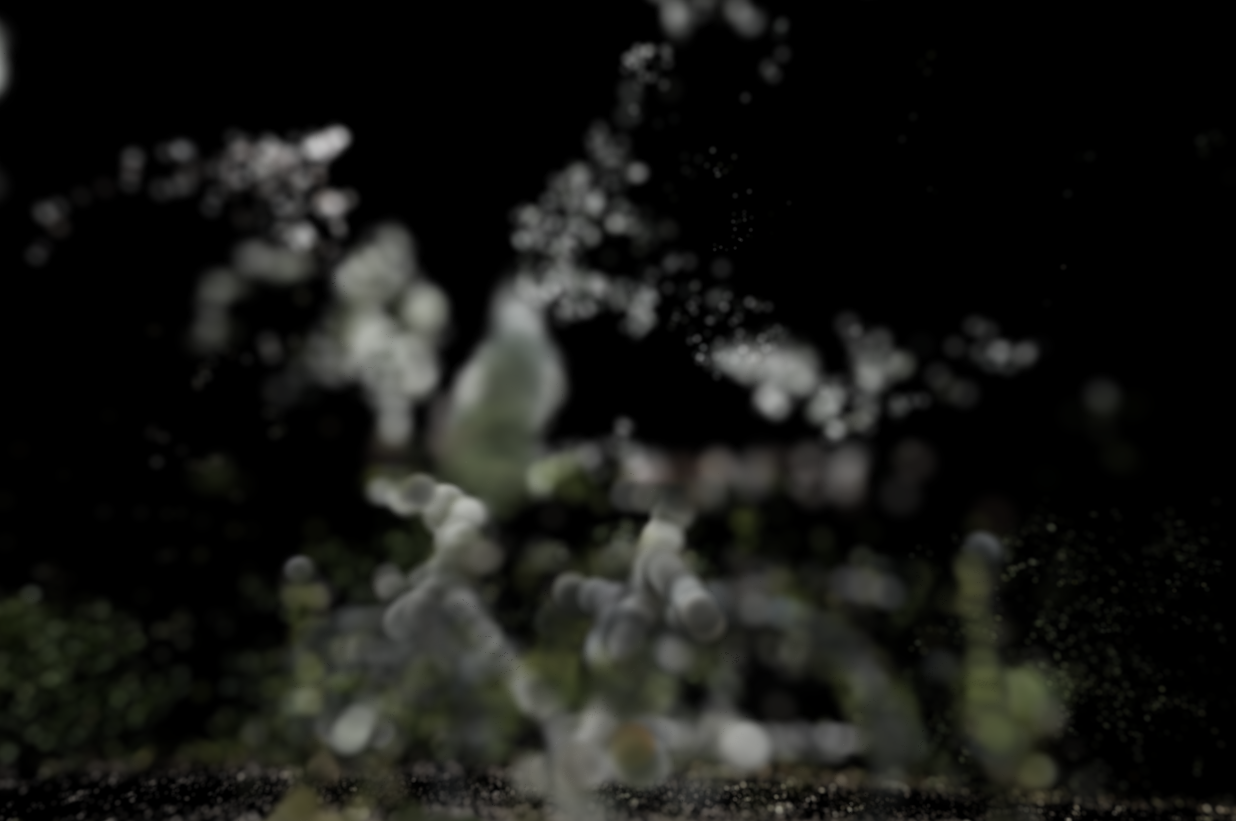} &
\includegraphics[width=0.33\linewidth]{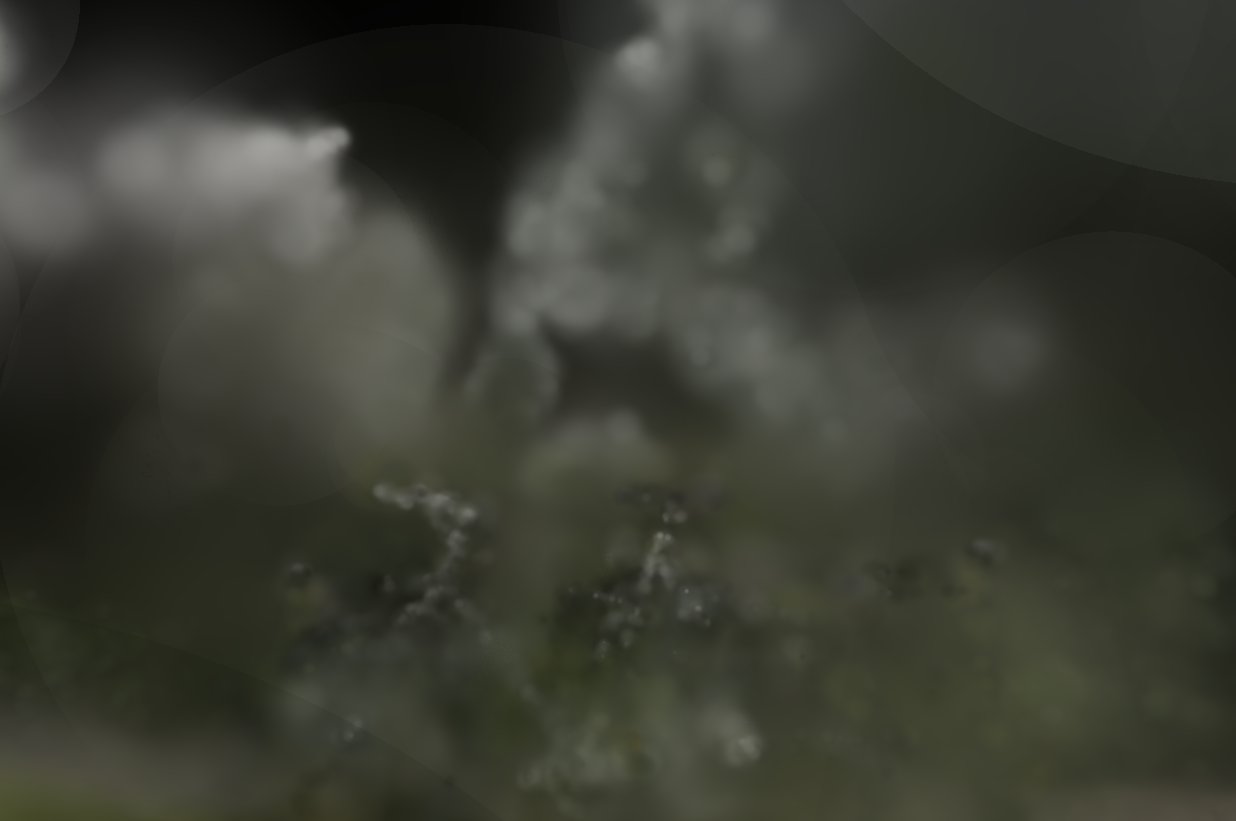} &
\includegraphics[width=0.33\linewidth]{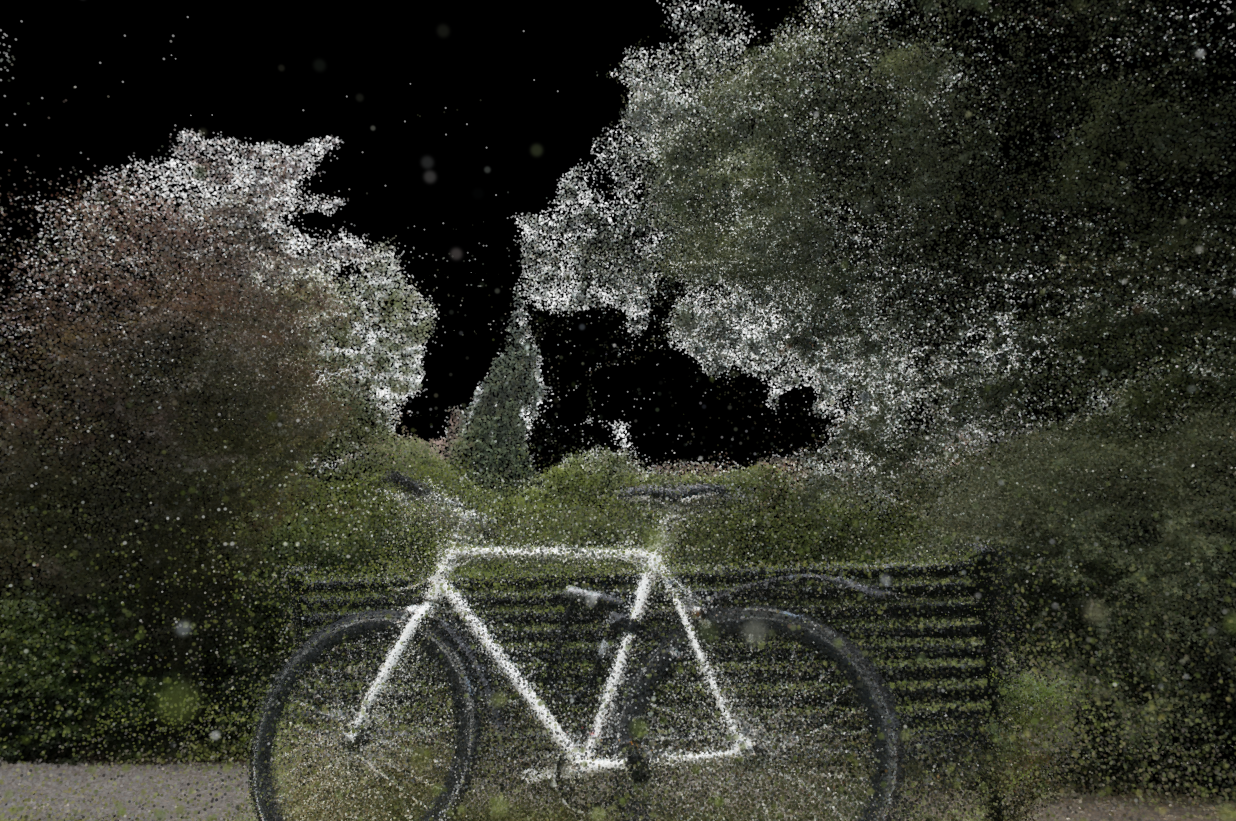} \\
\end{tabular}
}
\caption{Different Gaussian initializations visualized with 3DGRT. From left to right: sparse (SfM) initialization, sparse initialization with Gaussian scales matching 3DGS exactly (note how primitives are very large, which is unfavorable for ray-tracing), and dense initialization.}
\Description{Rendering of the Bicycle scene showing that sparse initialization starts with fewer large Gaussians, while dense initialization starts with many more tiny Gaussians showing a render the scene directly at initialization. }
\label{fig:init-viz}
\end{figure}
\paragraph{DI Slows Down Splatting but Speeds Up Ray Tracing} We ran both 3DGS and 3DGRT with and without DI for just 500 steps (stopping before any densification or pruning takes place \changed{so Gaussian counts stay constant}). This lets us accurately measure the performance impact of switching to DI and compare both rendering methods at equal Gaussian counts. \Cref{tbl:dense-init-speed-up} gives the results, where $\cdot_\text{SI}$ denotes a method with sparse (SfM) initialization while $\cdot_\text{DI}$ denotes a method with dense initialization. $\textsc{3DGRT}\textsubscript{SI}^{*}$ further denotes 3DGRT\textsubscript{SI} with initialization adjusted to match 3DGS\textsubscript{SI} exactly\footnote{While 3DGS uses the average distance to the approximate 3 nearest neighboring points to scale the initial Gaussians, 3DGRT instead uses the distance to the nearest camera.}; these different initialization types are displayed in \cref{fig:init-viz}. As expected, switching to DI slows down 3DGS’s optimization by over $3\times$ while halving its frame rate, which is explained by the Gaussian count increase from around $100\mathrm{K}$ to nearly $4\mathrm{M}$. Yet strikingly, with the same Gaussian counts, dense initialization \emph{speeds up} ray tracing instead of slowing it down, both in terms of optimization time and FPS. 

\begin{table}[h!]
\centering
\small
\caption{Speed of the first 500 iterations for 3DGS and 3DGRT for both initialization types, averaged over all scenes. Speedups are measured within each family using \textsc{3DGS}\textsubscript{SI} and \textsc{3DGRT}\textsubscript{SI} as baselines. \changed{FPS is measured at iteration 500.}}
\label{tbl:dense-init-speed-up}
\begin{tabular}{@{}lcccc@{}}
\toprule
Method & \#Gaussians$_\downarrow$ & Opt Time$_\downarrow$ & FPS$_\uparrow$ \\
\midrule
\textsc{3DGS}\textsubscript{SI}   & \textbf{0.11M} & \textbf{00{:}16} & \textbf{660} \\
\textsc{3DGS}\textsubscript{DI}   & 3.97M & 00{:}50 {\scriptsize(0.32$\times$)} & 282 {\scriptsize(0.43$\times$)} \\
\midrule
\textsc{3DGRT}\textsubscript{SI}  & \textbf{0.11M} & 00{:}31 & 96 \\
$\textsc{3DGRT}\textsubscript{SI}^{*}$ & \textbf{0.11M} & 00{:}30 & 98 \\
\textsc{3DGRT}\textsubscript{DI}  & 3.97M & \textbf{00{:}28} {\scriptsize(1.11$\times$)} & \textbf{140} {\scriptsize(1.46$\times$)} \\
\bottomrule
\end{tabular}
\end{table}

\begin{figure*}[h]
    \centering
    \includegraphics[width=\textwidth]{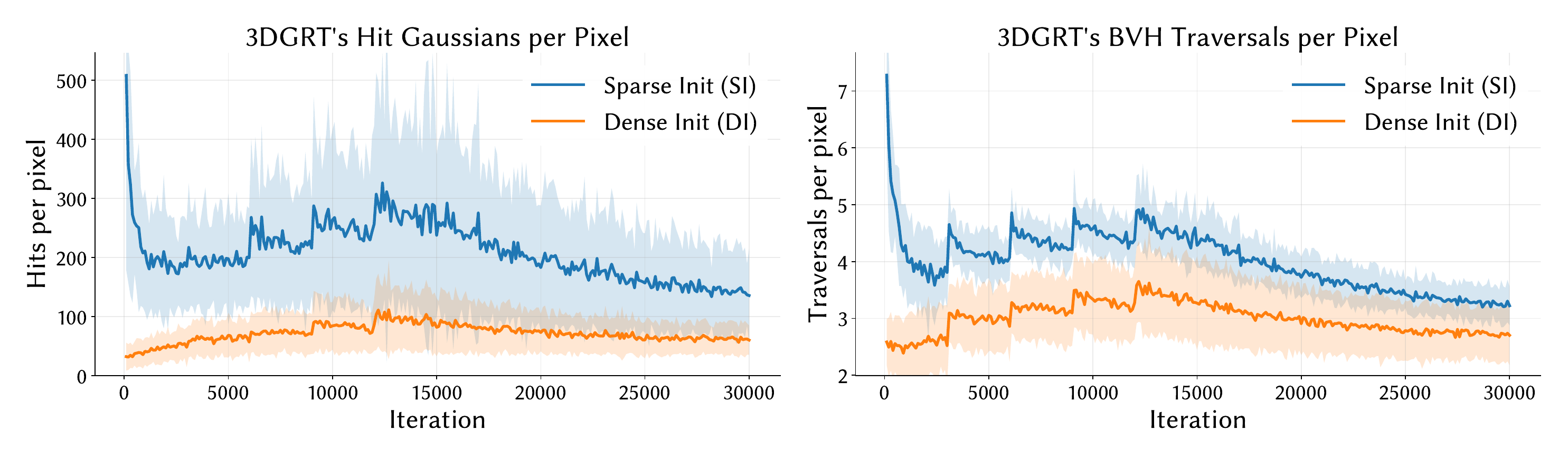}
    \caption{Per-pixel ray–Gaussian intersection statistics over training with sparse (SI) and dense (DI) initialization. Mean hits per pixel (left) and mean BVH traversals per pixel (right), averaged across all scenes, with shaded $\pm 1$ standard deviation bands. }
    \Description{Plots going from iteration 0 to 30000 showing that the number of ray-Gaussian intersections remains higher on average for SI throughout training, and idem for the number of BVH traversals.}
    \label{fig:stats-per-pixel} 
\end{figure*}

\paragraph{DI Reduces the Ray Tracing Workload Despite Increased Gaussian Counts} We further analyzed the ray tracing workload induced by both initialization types in more detail. Our experiments show that switching to DI \emph{reduces} the ray tracing workload---even at the start of training, when the Gaussian counts are ${>}35{\times}$ larger. \Cref{fig:stats-per-pixel} compares the number of hit Gaussians per ray across training alongside the required number of BVH traversals\footnote{A hit is counted for each ray intersection with a Gaussian bounding box; hits and traversals are counted for the forward pass only, and so the effective counts are doubled when accounting for the backward pass.}: DI reduces the number of hit Gaussians by about $3.5{\times}$ ($209.2 \rightarrow 60.2$ per pixel) and the number of traversals by around $2.75{\times}$ ($4.01 \rightarrow 1.45$ per pixel) on average across iterations. We conjecture that smaller Gaussians model surfaces more precisely, with less ``fuzz'' and without layering numerous mid-to-large Gaussians within object volumes. In any case, DI appears unequivocally more efficient than SI when working with ray tracing.

\section{GRay: A Fast Ray-Tracing Method for 3D Gaussians}
\label{sec:method}

Based on these insights, we develop a novel method designed to exploit ray tracing's algorithmic scaling. We build our method on 3DGRT~\cite{3dgrt}, EPBRR~\cite{epbrr} and EDGS~\cite{edgs} by curating their techniques and modifying the training regimen to better exploit large counts of smaller Gaussians. We validated our design choices through detailed ablations and analysis presented in \cref{additional-analysis,sec:detailed-ablations}; highlights from these experiments are also presented here directly. All metrics report averages over all thirteen 3DGS standard benchmarking scenes.\footnote{All experiments were performed on a single NVIDIA GeForce RTX 4090.} Code for our method will be open sourced. 

\paragraph{Dense Initialization (DI)}
We use the DI method proposed by EDGS directly and with their recommended parameters: $25{\mkern1mu,\mkern1mu}000$ matches per reference frame, a total of 180 reference frames, and 3 neighbors per reference. We briefly explored alternatives but did not find important differences in quality; notably, adding more Gaussians did not help. Likewise, we initialize Gaussian scales isotropically to $0.0005$ times the distance to the reference camera which produced the Gaussian. 

\paragraph{Restricting Gaussian Kernel Support}
3DGRT proposed reducing the Gaussians’ footprints with two techniques. First, by replacing the Gaussian kernel with a generalized version like:
$$
\exp\big[-\changed{\frac{1}{2n_{\mathrm{expo}}}} \big( (x-\mu)^{T} \Sigma^{-1} (x-\mu) \big)^{n_{\mathrm{expo}}}\big] \,,
$$
where increasing the exponent $n_{\mathrm{expo}}$ makes Gaussians ``bulkier''. Second, with adaptive clamping, i.e., by defining a Gaussian’s support in terms of its alpha value after opacity multiplication, such that all alpha values falling below a threshold $\alpha_{\min}$ are zeroed. Both of these techniques decrease superfluous ray-Gaussian intersections. We employ these two techniques with their suggested hyperparameters ($n_{\mathrm{expo}}=2$ and $\alpha_{\mathrm{min}}=0.01$) since we aim to stay near 3DGRT’s quality. \changed{However, 3DGRT uses $\frac{4.5}{3^{2 n_{\mathrm{expo}}}}$ instead of $\frac{1}{2n_{\mathrm{expo}}}$; opting for the latter makes initialization less sensitive to the setting of $n_{\mathrm{expo}}$ and results in smaller initial Gaussians in practice.}

Interestingly, while dynamic clamping leads to visible artifacts for large Gaussians, they can be less apparent when working with DI since most Gaussians stay small.

\paragraph{Oriented Bounding Boxes (OBBs)}
We found OBBs essential when working with DI since the Gaussian counts start much higher, making BVH update times more important. 3DGRT initially proposed using triangle icosahedrons to bound Gaussians, but they have slow BVH updates. This becomes problematic when dealing with millions of tiny Gaussians, since updating them at each training iteration requires transforming 12 vertices for each icosahedron. EPBRR and 3DGRT’s code release replaced polygonal cages with oriented bounding boxes (OBBs), which allow significantly faster updates, albeit with less accurate bounding. Although OBBs are not natively supported in OptiX, they can be emulated efficiently through two-level instancing~\cite{using-hardware-ray-transforms}, \changed{where a top-level acceleration structure (TLAS) transforms instanced copies of a bottom-level acceleration structure (BLAS) which consists of a single unit bounding box.}

In \Cref{tbl:bvh-update-time-highlight} we compare the BVH incremental update times at initialization for both icosahedron bounds and emulated OBBs in 3DGRT. We reported averages over all test views and over all scenes, run $100$ times and averaged. While update times remain negligible at ${<}5\mathrm{ms}$ with SI, they grow to over $125\mathrm{ms}$ under DI becoming a substantial part of the training cost. Because of this, training with OBBs becomes two times faster than icosahedrons under DI ($01{:}22{:}12 \rightarrow  40{:}05$), while FPS is slightly lower ($145 \rightarrow 131$).

\begin{table}[h!]
\centering
\small
\caption{BVH update times become critical under DI.}
\label{tbl:bvh-update-time-highlight}
\begin{tabular}{@{}lcccc@{}}
\toprule
Method & BVH Update$_\downarrow$ & \#Gaussians$_\downarrow$ \\
\midrule
\textsc{3DGRT}\textsubscript{SI}  & \textbf{0.9\,ms} & 0.11M \\
\textsc{3DGRT}\textsubscript{SI+Icosahedrons}  & 4.5\,ms & 0.11M \\
\midrule
\textsc{3DGRT}\textsubscript{DI}  & \textbf{11.4\,ms} & 3.97M \\
\textsc{3DGRT}\textsubscript{DI+Icosahedrons}  & 125.7\,ms & 3.97M \\
\bottomrule
\end{tabular}
\end{table}

\begin{figure}[t]
\centering
\includegraphics[width=\textwidth]{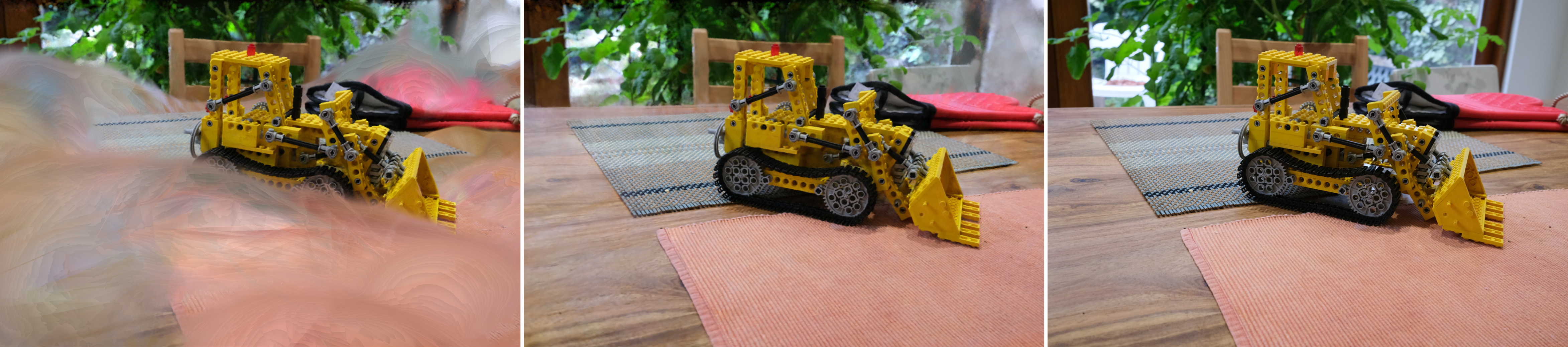}
\caption{Test view after training for 1000 iterations with a very high truncation threshold $\tau_{\mathrm{min}} = 0.1$ without (left) and with DHT (center), compared to the ground truth (right).  }
\Description{Renderings of the Counter scene showing that training without DHT results in large overly expanded Gaussians, producing a blurry appearing in large sections of the scene.}
\label{fig:no-dht-crash}
\end{figure}

\paragraph{Detached Hybrid Transparency (DHT)}
To avoid evaluating occluded Gaussians, 3DGRT proposed early ray termination: accumulation terminates once transmittance falls below a threshold $\tau_{\mathrm{min}}$, skipping the remaining Gaussians along the ray. While this is safe at runtime (i.e., rendering after optimization has finished), their naive implementation destabilizes training and requires different threshold values at inference ($\tau_{\mathrm{min}}=0.03$) than during training ($\tau_{\mathrm{min}}=0.001$). To address this, RaySplats and EPBRR approximate the contribution of the skipped Gaussians to reduce the gradient’s bias. We further propose using unordered alpha blending~\cite{meshkin,weighted-blended-oit} as an estimator and rename this technique Detached Hybrid Transparency~\cite{hybrid-transparency,htgs} for clarity. DHT stabilizes training under early termination, enabling early ray termination to be used safely during training. This is exemplified in \cref{fig:no-dht-crash} and further explained in \cref{sec:dht-convergence}.

More specifically, during BVH traversal we compute the exact final transmittance $\tau_{\mathrm{exact}}$ (which can be done out of order). Then, during accumulation, after reaching a transmittance value $\tau < \tau_{\mathrm{min}}$, we approximate the color of the ``tail'' (the skipped Gaussians) with
$$
\boldsymbol{c}_{\mathrm{tail}} = \frac{1}{\alpha_\mathrm{tail}} \sum_{i\in G_{\mathrm{tail}}} \alpha_i \boldsymbol{c}_i \,,
$$
given $\alpha_\mathrm{tail} = \sum_{i\in G_{\mathrm{tail}}} \alpha_i$, where $G_{\mathrm{tail}}$ are the skipped Gaussians.\footnote{We define $\boldsymbol{c}_{\mathrm{tail}} = \mathbf{0}$ if $\alpha_\mathrm{tail}=0$.} The final color $\boldsymbol{c}$ for that pixel becomes
$$
\boldsymbol{c} = \boldsymbol{c}_{\mathrm{head}} + \operatorname{sg}\big((\tau - \tau_{\mathrm{exact}}) \boldsymbol{c}_{\mathrm{tail}}\big) \,,
$$
where $\boldsymbol{c}_{\mathrm{head}}$ is the accumulated color for all sorted Gaussians and $\operatorname{sg}(x)$ denotes the ``stop-gradient'', operator which detaches this value, treating it as constant during gradient computation. 

\Cref{tbl:dht-highlight} shows our method trains stably at high transmittance thresholds ($\tau_{\mathrm{min}}{=}0.03$) while skipping the accumulation of $40\%$ of hit Gaussians. Interestingly, it reaches slightly better LPIPS than when early termination is disabled ($\tau_{\mathrm{min}} = 0.0$). Since these Gaussians are further skipped during the backward pass, optimization time drops significantly ($6{:}33 \rightarrow 5{:}40$). Finally, running our method with DHT disabled (\textsc{GRay}$_{\mathrm{NoDHT}}$) shows rapid deterioration of quality metrics when truncation levels increase.

\begin{table}[h!]
\centering
\small
\caption{DHT stabilizes early ray termination during training ($\tau_{\mathrm{min}}=0.03$).}
\label{tbl:dht-highlight}
\begin{tabular}{@{}l|cccc@{}}
\toprule
Variant & PSNR$_\uparrow$ & LPIPS$_\downarrow$ & Opt Time$_\downarrow$ & \% Skipped \\
\midrule
\textsc{GRay} & \textbf{26.47} & \textbf{0.236} & 05{:}40 & 40.28 \\
\textsc{GRay}$_{\mathrm{NoEarlyTerm.}}$ & 26.46 & 0.241 & 06{:}33 & 0.00 \\
\textsc{GRay}$_{\mathrm{NoDHT}}$ & 24.54 & 0.284 & \textbf{05{:}34} & 54.53 \\
\bottomrule
\end{tabular}
\end{table}

\paragraph{Per-Pixel Linked List Replay Buffers}
We found per-pixel linked lists~\cite{ppll} (PPLLs) to still be beneficial even under increased Gaussian counts. Because per-pixel Gaussian arrays do not fit in register memory, 3DGRT uses a collection strategy that processes them in fixed-size batches of $K$ using multiple BVH traversals to avoid register spilling. Unfortunately, the overhead of performing multiple traversals is substantial. EPBRR instead performs a single BVH traversal and stores all intersected Gaussians in a PPLL replay buffer, deliberately trading memory for speed. The buffer is then replayed to collect batches of Gaussians without additional traversals, improving performance overall. EPBRR also stored temporary data for the backward pass in PPLLs since they allow for coalesced memory access. 

We adopt PPLLs for our ray tracer for both the forward and backward pass, but we found EPBRR's implementation suboptimal, as too many attributes were stored in lists even when they can be recomputed efficiently. Our implementation uses lightweight PPLLs which only store Gaussian IDs, hit distances, and alpha values.

We validated that PPLLs are still worth using since DI reduces the required BVH traversals when using 3DGRT's Gaussian collection strategy. \Cref{tbl:ppll-vs-repeated-traversals-highlights} shows that PPLLs are still much faster even in the DI regime, increasing frame rates by around $40\%$ ($175 \rightarrow 248$).\footnote{Note that we disabled DHT for the multiple  traversals used in 3DGRT since the list of unprocessed Gaussians isn't readily available; implementing DHT for it would require an additional traversal, further decreasing speed.} 
% Also note that our method uses PPLLs for the backward pass as well; although we did not measure it, the performance difference during training should be larger than at inference.

\begin{table}[h!]
\centering
\small
\caption{PPLLs still outmatch multiple traversals under DI.}
\label{tbl:ppll-vs-repeated-traversals-highlights}
\begin{tabular}{@{}l c|cccccc@{}}
\toprule
Variant & {Collection Method} & FPS$_\uparrow$ \\
\midrule
\textsc{GRay}\textsubscript{MultipleTraversal} & {Multiple BVH Traversals} & 175 \\
\textsc{GRay}                         & {Per-Pixel Linked Lists}                & \textbf{248} \\
\bottomrule
\end{tabular}
\end{table}

\paragraph{Weight-Based Pruning}
We propose changing the pruning approach to further cut down on bloated Gaussian counts. Several recent approaches, including EPBRR, explore pruning unimportant Gaussians~\cite{spotlesssplats,pup3dgs,speedy-splat,radsplat,eagles,mini-splatting,reducing-the-memory-footprint,epbrr}. Inspired by these ideas, we adopt a simple weight-based pruning criterion, that is to prune according to the average accumulated weight of a Gaussian over views that observe it:
$$
\left[ \frac{1}{|V_i|\,|v|}\sum_{v\in V_i}\sum_{r\in v} w_i(r) \right] < \gamma_{\text{prune}}\,.
$$
Here, $v$ is a training view from the set of views $V_i$ that see Gaussian $i$, $r$ is a ray in that view, and $w_i(r)$ is the Gaussian’s accumulation weight for that ray. This is done every $500$ iterations.

\Cref{tbl:pruning-highlight} shows the impact of our pruning method. Our selected value $\gamma_{\textrm{prune}}=10^{-7}$ reduces the Gaussian count to less than half ($3.27\text{M} \rightarrow 1.52\text{M}$) while \emph{improving} PSNR, and only worsening LPIPS by small amounts ($0.234 \rightarrow 0.236)$. This results in substantial improvement to FPS ($189 \rightarrow 248$) and to optimization times ($7{:}26 \rightarrow 5{:}40$). Our weight-based pruning significantly outperforms opacity-based pruning, which we show in \cref{tbl:opacity-pruning-sweep} with extensive parameter sweeps.

\begin{table}[h!]
\centering
\small
\caption{Aggressive weight-based pruning maintains good quality ($\gamma_{\mathrm{prune}}=10^{-7}$).}
\label{tbl:pruning-highlight}
\begin{tabular}{@{}l|ccccc@{}}
\toprule
Variant & PSNR$_\uparrow$ & LPIPS$_\downarrow$ & Opt. Time$_\downarrow$ & FPS$_\uparrow$ & Final \#Gaussians$_\downarrow$ \\
\midrule
$\mathrm{GRay}$ & \textbf{26.47} & 0.236 & \textbf{05{:}40} & \textbf{248} & \textbf{1.52M} \\
$\mathrm{GRay}_{\mathrm{NoPruning}}$ & 26.42 & \textbf{0.234} & 07{:}26 & 189 & 3.27M \\
\bottomrule
\end{tabular}
\end{table}

\paragraph{Halving Training Iterations}

One of EDGS's main claims is that DI converges more rapidly than SI~\cite{edgs}. Thus, benefiting from DI’s faster convergence, we train for $15\mathrm{K}$ iterations instead of $30\mathrm{K}$. Of course, training at reduced iteration counts can lead to reduced quality. To compensate, we adjust learning rates by adding schedules for scale, rotation, and SH DC coefficients. These schedules assign higher values early during training while decaying them down progressively to standard values. We also increase the SH higher-order coefficient learning rate $5 \times$ to $0.000625$. The exact schedules used are described in \cref{sec:lr-schedules}.

\Cref{tbl:reduced-iter-highlight} shows that it is in fact possible to train at $15\mathrm{K}$ iterations with slightly better LPIPS score than $30\mathrm{K}$ iterations ($0.243 \rightarrow 0.236$) and only a slight decrease in PSNR ($26.51 \rightarrow 26.47$), leading to halved training times ($10{:}24 \rightarrow 05{:}40$) almost for free. In \cref{tbl:reduced-iter-count}, additional configurations are presented and the learning rate schedules are ablated.

\begin{table}[h!]
\centering
\small
\caption{Fewer iterations can still converge fully.}
\label{tbl:reduced-iter-highlight}
\begin{tabular}{@{}lc|ccc@{}}
\toprule
Variant & Iterations & PSNR$_\uparrow$ & LPIPS$_\downarrow$ & Opt. Time$_\downarrow$ \\
\midrule
\textsc{GRay}\textsubscript{30K} & 30000 & \textbf{26.51} & 0.243 & 10{:}24 \\
\textsc{GRay}                   & 15000 & 26.47 & \textbf{0.236} & \textbf{05{:}40} \\
\bottomrule
\end{tabular}
\end{table}

\paragraph{Scale Decay} To further encourage smaller Gaussians, we included a scale decay. At each iteration, Gaussians are scaled down by a factor of $\eta_{\textrm{decay}}=0.999875$. We found this regularizer highly effective at controlling final Gaussian sizes while also encouraging the pruning of occluded or out-of-view Gaussians, which increases final FPS by noticeable amounts while only impacting quality minimally.

\Cref{tbl:scale-decay-highlight} shows that increasing decay improves FPS by large amounts ($157 \rightarrow 248$), which is explained by both reduced final Gaussian average scales ($0.0211 \rightarrow 0.0166$) and an increase in pruning ($1.73\textrm{M} \rightarrow 1.52\textrm{M}$ final Gaussian count). Since decay also reduces quality slightly, we conclude that scale decay is a highly effective lever for balancing quality and speed. An extensive sweep showing this is provided in \cref{tbl:scale-decay-sweep}.

\begin{table}[h!]
\centering
\small
\caption{Scale decay efficiently trades quality for speed ($\eta_{\mathrm{decay}} = 0.999875$).}
\label{tbl:scale-decay-highlight}
\begin{tabular}{@{}l|ccccc@{}}
\toprule
Variant & PSNR$_\uparrow$ & LPIPS$_\downarrow$ & FPS$_\uparrow$ & Final \#Gaussians$_\downarrow$ & Final Avg Scale \\
\midrule
\textsc{GRay} & 26.47 & 0.236 & \textbf{248} & \textbf{1.52M} & \textbf{0.0166} \\
\textsc{GRay}$_{\mathrm{NoDecay}}$ & \textbf{26.60} & \textbf{0.231} & 157 & 1.73M & 0.0211 \\
\bottomrule
\end{tabular}
\end{table}

\paragraph{Initialization Binning} DI densely covers surfaces with Gaussians, but can result in highly redundant Gaussians when the selected reference views are similar. We propose using binning to remove duplicate Gaussians at initialization. In practice, we voxelize all Gaussians positions at a fixed grid of size $\psi_{\mathrm{bin}} r_\mathrm{scene}$, where $\psi_{\mathrm{bin}}=0.04$ is the grid cell size and the scene radius $r_\mathrm{scene}$ is defined as $1.1$ times the maximum distance from the average camera center to any other camera center~\cite{nerfplusplus}\footnote{\changed{A single value $\psi_{\mathrm{bin}}=0.04$ worked well in the 3DGS benchmark scenes but might require adjustment for larger scenes.}}. The position, color and scale properties of each Gaussians within a voxel are averaged. 

\begin{table}[h!]
\centering
\small
\caption{Many initial DI Gaussians are useless ($\psi_{\mathrm{bin}}=0.04$).}
\label{tbl:init-binning-highlight}
\begin{tabular}{@{}l|ccc@{}}
\toprule
Variant & PSNR$_\uparrow$ & LPIPS$_\downarrow$ & Init \#Gaussians$_\downarrow$ \\
\midrule
$\mathrm{GRay}$ & 26.47 & 0.236 & \textbf{3.27M} \\
$\mathrm{GRay}_{\mathrm{NoBinning}}$ & \textbf{26.49} & \textbf{0.234} & 3.97M \\
\bottomrule
\end{tabular}
\end{table}

Our experiments show that this practice maintains visual quality nearly identical while reducing initial Gaussian counts significantly.
\Cref{tbl:init-binning-highlight} gives our final setting $\psi_{\textrm{bin}}=0.04$, which removes over $15\%$ of the initial Gaussians at limited cost to PSNR ($26.49 \rightarrow 26.47$) and LPIPS ($0.234 \rightarrow 0.236$). A parameter sweep is also provided in \cref{tbl:init-binning-sweep}.

\paragraph{Additional Changes} \changed{Unlike 3DGS and 3DGRT, we do not use opacity resets, and unlike 3DGS, do not prune Gaussians based on size. Remaining small performance differences may stem from implementation improvements. All differences are detailed in \cref{tbl:method-features}.}

\deleted{As compared to EPBRR, we do not fuse the forward and backward pass to allow for the SSIM loss and easy addition of other custom losses. We also use a CUDA-side Adam optimizer that launches a separate thread for every spherical harmonic coefficient instead of just launching one per Gaussian.}

\deleted{We use the fused SSIM loss proposed in Taming 3DGS~\cite{taming-3dgs} as do most existing works. In supplemental experiments, we also use the lazy stepping for non-DC Spherical Harmonic (SH) bands they propose, although we leave it disabled by default; this technique consists of only taking optimization steps every $K_{\mathrm{lazy}}$ iterations (Taming 3DGS uses $K_{\mathrm{lazy}} = 12$).}

\deleted{Unlike 3DGS and 3DGRT, we do not use opacity resets; unlike 3DGS we do not prune out large Gaussians based on their viewport or world-space size; and unlike EDGS, we do not decay opacity or modify the positional learning rate schedule. These details are further clarified in the supplemental.}

\section{Experiments}

In this subsection, we compare our ray tracer GRay to 3DGRT and 3DGS under various configurations, and also run EDGS's original method for reference. 

More specifically, we extracted the dense initialization code from EDGS’s codebase and incorporated it into 3DGS, 3DGRT, and our ray tracer GRay while dispensing of any other changes.\footnote{The original EDGS implementation includes other minor changes like an opacity decay and different LR schedule, which improve quality slightly but make precise comparisons harder. For this reason we exclude such changes and apply DI directly to 3DGS. This is \changed{further} discussed in \cref{sec:additional-hyperparams}.} We used the indoor RoMa model for all indoor scenes and the outdoor one for all outdoor scenes. For simplicity we ran dense initialization on the full-resolution images, but worked with downsized images, as was done in the 3DGS paper. \changed{We also compared to RayGaussX, a slower high quality ray tracing method.} We standardized downsizing across all methods, using Pillow’s Lanczos filter. We reported averages over all thirteen Mip-NeRF360, Tanks \& Temples, and Deep Blending scenes which are the standard 3DGS benchmark introduced in the original paper~\cite{3dgs}. 

For 3DGS, we used the improved optimizer proposed by Taming 3DGS~\cite{taming-3dgs}; for 3DGRT, we did not use Sparse Adam in most experiments, since it reduced quality slightly without providing a significant performance boost. We ran 3DGRT with oriented bounding boxes unless otherwise specified. We measured FPS over all test views with a warm start\footnote{To ensure caches are filled, we run the FPS measurement twice and retain the second result.} \changed{and averaged their values directly}. We also reported PSNR, SSIM~\cite{ssim}, and LPIPS~\cite{lpips} as usual. Note that the original 3DGS implementation computed slightly erroneous LPIPS scores which are unfortunately reported in numerous works~\cite{revisiting-densification}; we reported accurate LPIPS for all methods. 

Rendered videos and captured interactive sequences are also provided alongside the paper. Unfortunately, dense initialization seems prone to floaters, especially in outdoor scenes with vegetation---this applies to all methods tested with DI. As such, for \changed{all qualitative results} we masked floaters with manually configured near clipping planes \changed{with distances indicated in \cref{tbl:znear-values}}. The same clipping planes were used for all methods; \changed{they were not used when measuring quality.} 

\Cref{tbl:main-results} contains our main comparisons; 3DGS was run with Taming 3DGS's techniques unless marked with $\cdot_{\mathrm{RegularAdam}}$ while 3DGRT was run with the regular optimizer unless marked with $\cdot_{\mathrm{SparseAdam}}$. We ran all experiments on a single NVIDIA GeForce RTX 4090.

% \todo{Clarify: which parts of Taming 3DGS are included in 3DGS's accelerated variant? And what does "Sparse Adam" do for 3DGRT exactly?}

\begin{table}[t]
\caption{Reconstruction quality and performance between methods and configurations.}
\label{tbl:main-results}
\centering
\resizebox{\linewidth}{!}{%
\begin{tabular}{@{}lccccccccc@{}}
\toprule
Method & PSNR$_\uparrow$ & SSIM$_\uparrow$ & LPIPS$_\downarrow$ & Init Time$_\downarrow$ & Opt Time$_\downarrow$ & FPS$_\uparrow$ & Init \#G$_\downarrow$ & Final \#G$_\downarrow$ & \#Iterations \\
\midrule
$\mathrm{3DGS}_{\mathrm{SI}}$          & 27.10 & 0.831 & 0.262 & 00{:}00 & 06{:}18 & 253 & 0.11M  & 2.25M  & 30000 \\
$\mathrm{3DGS}_{\mathrm{DI}}$          & 26.97 & 0.827 & 0.226 &  01{:}58 & 10{:}09 & 241 & 3.97M  & 3.28M  & 30000 \\
$\mathrm{3DGS}_{\mathrm{SI+15Kiters}}$  & 26.55 & 0.814 & 0.296 & 00{:}00 & 02{:}47 & 299 & 0.11M  & 1.87M  & 15000 \\
$\mathrm{3DGS}_{\mathrm{DI+15Kiters}}$  & 26.68 & 0.830 & 0.231 &  01{:}58 & 05{:}37 & 235 & 3.97M & 3.39M  & 15000 \\
$\mathrm{3DGS}_{\mathrm{SI+RegularAdam}}$  & 27.28 & 0.836 & 0.253 & 00{:}00 & 18{:}13 & 190 & 0.11M & 2.49M & 30000 \\
$\mathrm{3DGS}_{\mathrm{DI+RegularAdam}}$  & 27.29 & 0.838 & 0.211 &  01{:}58 & 23{:}56 & 157 & 3.97M & 3.50M & 30000 \\
\midrule
$\mathrm{EDGS}_{\mathrm{RegularAdam}}$  & 27.55 & 0.850 & 0.204 & 00{:}00 & 30{:}38 & 173 & 3.98M & 2.26M & 30000 \\
$\mathrm{EDGS}_{\mathrm{RegularAdam+15Kiters}}$  & 27.26 & 0.851 & 0.212 & 00{:}00 & 20{:}07 & 167 & 3.98M & 2.16M & 15000 \\
\midrule
$\changed{\mathrm{RayGaussX}}$ & \changed{28.14} & \changed{0.856} & \changed{0.221} & \changed{00{:}00} & \changed{56:18} & \changed{39} & \changed{0.11M} & \changed{3.28M} & \changed{30000} \\
\midrule
$\mathrm{3DGRT}_{\mathrm{SI}}$  & 26.77 & 0.828 & 0.258 & 00{:}00 & 55{:}01 & 68  & 0.11M  & 3.24M & 30000 \\
$\mathrm{3DGRT}_{\mathrm{DI}}$  & 26.46 & 0.824 & 0.229 &  01{:}58 & 40{:}05 & 131 & 3.97M & 2.64M & 30000 \\
% \midrule
$\mathrm{3DGRT}_{\mathrm{SI+15Kiters}}$  & 26.05 & 0.818 & 0.268 & 00{:}00 & 24{:}04 & 50  & 0.11M  & 3.39M & 15000 \\
$\mathrm{3DGRT}_{\mathrm{DI+15Kiters}}$  & 26.28 & 0.830 & 0.230 &  01{:}58 & 20{:}13 & 120 & 3.97M & 2.62M & 15000 \\
% \midrule
$\mathrm{3DGRT}_{\mathrm{SI+SparseAdam}}$  & 26.61 & 0.825 & 0.264 & 00{:}00 & 53{:}37 & 52  & 0.11M  & 3.32M & 30000 \\
$\mathrm{3DGRT}_{\mathrm{DI+SparseAdam}}$  & 26.16 & 0.815 & 0.239 &  01{:}58 & 33{:}43 & 136 & 3.97M & 2.69M & 30000 \\
\midrule  % separator between 3DGS and the ray tracing methods
$\mathrm{GRay}$  & 26.47 & 0.819 & 0.236 &  01{:}58 & 05{:}40 & 248 & 3.27M & 1.52M & 15000 \\
\bottomrule
\end{tabular}
}
\end{table}

\paragraph{GRay Runs as Fast as 3DGS Near 3DGRT’s Quality}
Our main results are shown directly in \Cref{fig:teaser}: GRay roughly matches the optimization times and FPS of 3DGS while maintaining a quality level close to 3DGRT (worse PSNR and SSIM, but better LPIPS). We credit the majority of this improvement to DI. Note, however, that DI takes roughly two minutes to run, and so the total training times (initialization + optimization) remain in favor of 3DGS, along with the overall visual quality. FPS are also nearly $4\times$ faster for our method compared to 3DGRT ($68 \rightarrow 248$). Finally, despite starting at highly inflated Gaussian counts, our weight-based pruning results in lower final counts ($3.24\textrm{M} \rightarrow 1.52\textrm{M})$ which ultimately saves memory. 

\paragraph{Splatting Fails to Exploit DI for Speed} 
Looking at overall optimization times and frame rates, we can see that optimization speed (on a per-iteration basis) is consistently worse for 3DGS under DI ($06{:}18 \rightarrow 10{:}09$), and that the frame rates after optimization completes are also slightly worse ($253 \rightarrow 241$). In fact, optimization times after halving iteration count with DI do not even improve much over regular training ($\mathrm{3DGS}_{\mathrm{DI+15Kiters}}@05{:}37 \approx \mathrm{3DGS}_{\mathrm{SI}}@06{:}18$). In great contrast, ray tracing with 3DGRT obtains a marked $2\text{–}3\times$ FPS increase when switching to DI in all tested configurations, with improved optimization times as well ($55{:}01 \rightarrow 40{:}05$ in the main configuration). We conclude that splatting \deleted{cannot} fails to leverage DI for speed. This means our ray tracer's speed remains comparable to 3DGS even when the latter is also switched to DI.

\paragraph{DI can Lower PSNR/SSIM but Improves LPIPS} 
While the original EDGS method improved PSNR and SSIM over 3DGS, they also included a modified learning rate schedule and opacity decay which improved visual quality irrespective of DI. When switching 3DGS or 3DGRT to DI with no other changes applied, we found that PSNR and SSIM tend to worsen slightly (most notably PSNR $26.77 \rightarrow 26.46$ for 3DGRT), while LPIPS improved considerably (for 3DGS $0.262 \rightarrow 0.226$; for 3DGRT $0.258 \rightarrow 0.229$). While LPIPS is arguably a better assessment of human perceived perceptual quality~\cite{lpips,preception-distortion-tradeoff,comparison-of-iqa}, in practice DI performs better in certain types of scenes and worse in others, which we discuss below.

\section{Discussion}\label{sec:discussion}

We showed that ray tracing can be greatly accelerated by exploiting its synergy with dense initialization, which we attribute to its advantage in computational scaling over splatting. We discuss below the implications and main limitations of our work.

\begin{figure}[t]
\centering
\resizebox{\linewidth}{!}{%
\setlength{\tabcolsep}{0.5pt}
\begin{tabular}{lcccc}
 & $\mathrm{3DGS}_{\mathrm{SI}}$ & $\mathrm{3DGRT}_{\mathrm{SI}}$ & $\mathrm{3DGRT}_{\mathrm{DI}}$ & \textsc{GRay} (Ours) \\

\rotatebox{90}{\textsc{Flowers} {\scriptsize (easy)}} &
\includegraphics[width=0.24\linewidth]{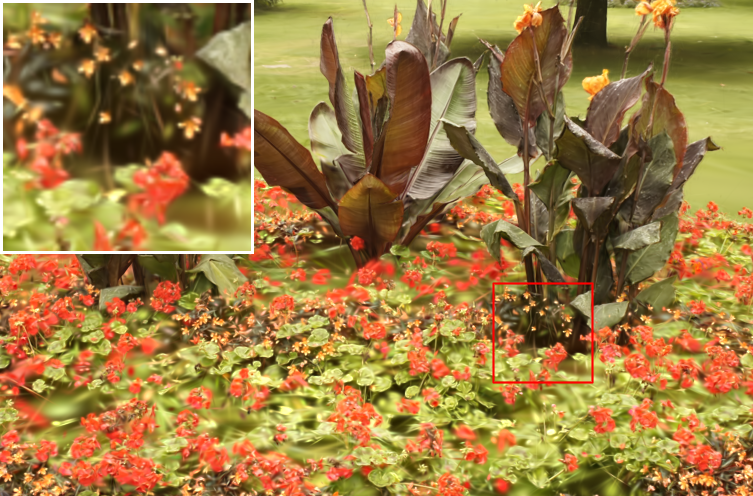} &
\includegraphics[width=0.24\linewidth]{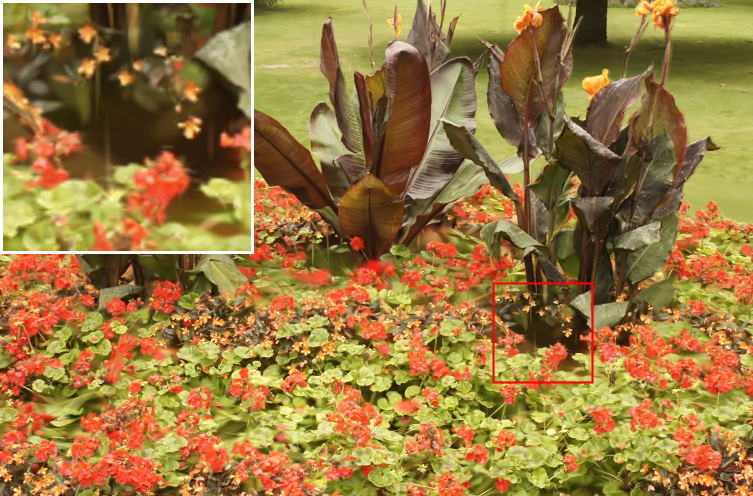} &
\includegraphics[width=0.24\linewidth]{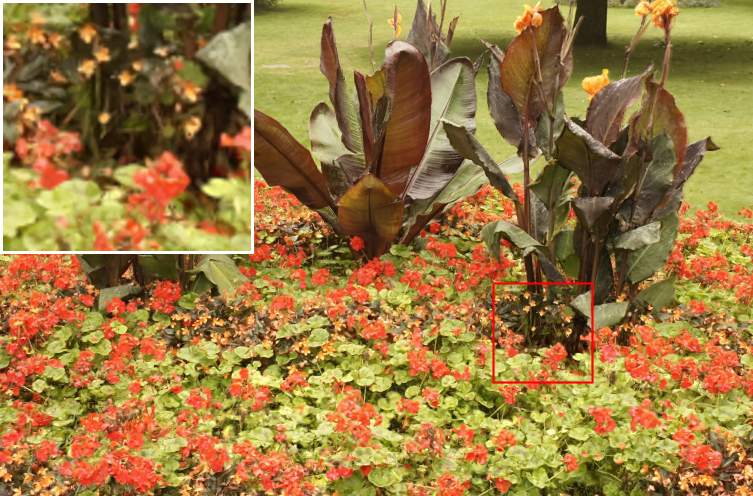} &
\includegraphics[width=0.24\linewidth]{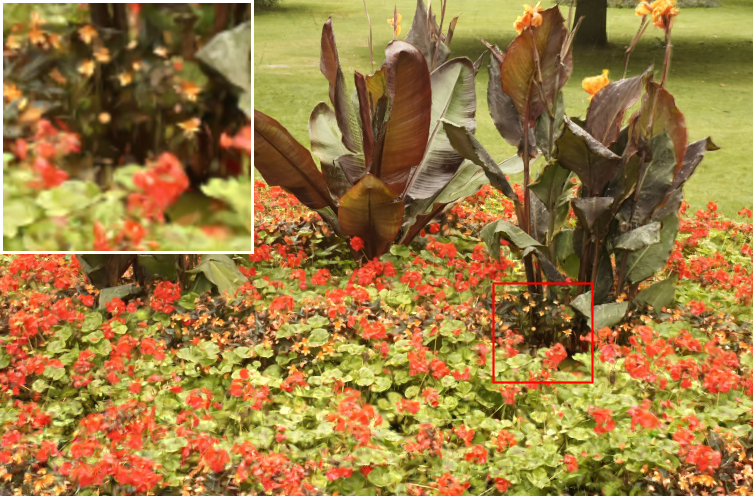} \\

\rotatebox{90}{\textsc{\changed{Bicycle}} {\scriptsize (med.)}} &
\includegraphics[width=0.24\linewidth]{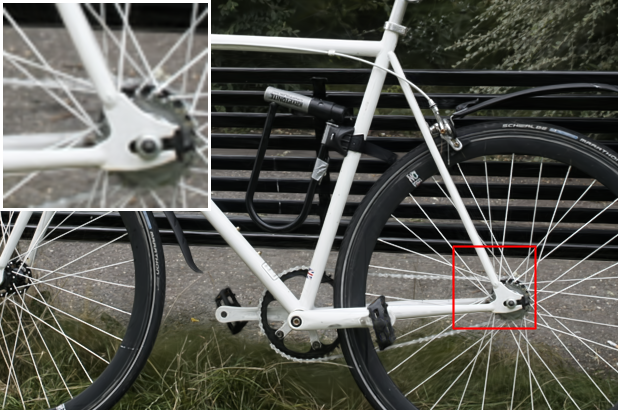} &
\includegraphics[width=0.24\linewidth]{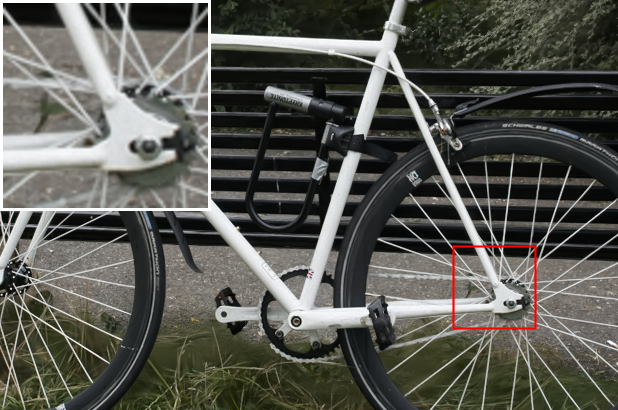} &
\includegraphics[width=0.24\linewidth]{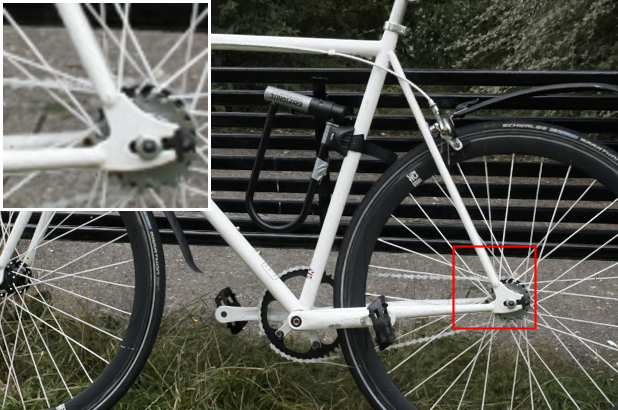} &
\includegraphics[width=0.24\linewidth]{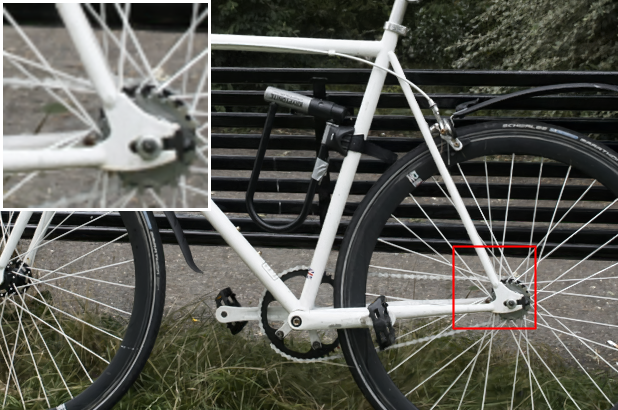} \\

\rotatebox{90}{\textsc{Room} {\scriptsize (hard)}} &
\includegraphics[width=0.24\linewidth]{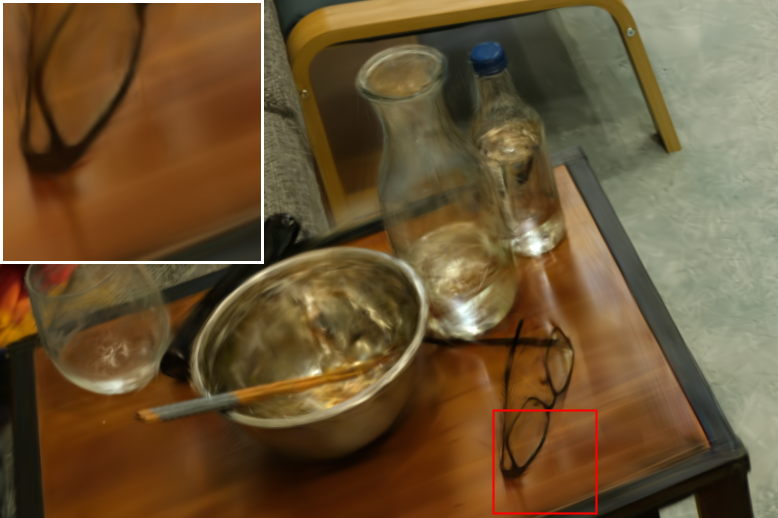} &
\includegraphics[width=0.24\linewidth]{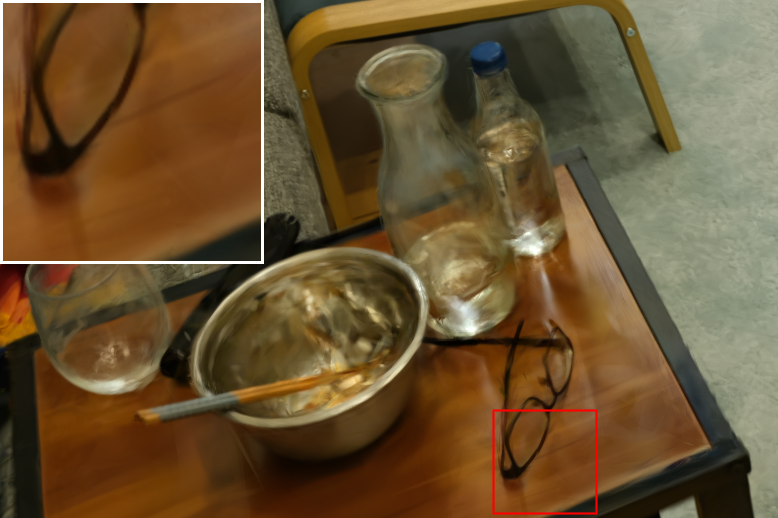} &
\includegraphics[width=0.24\linewidth]{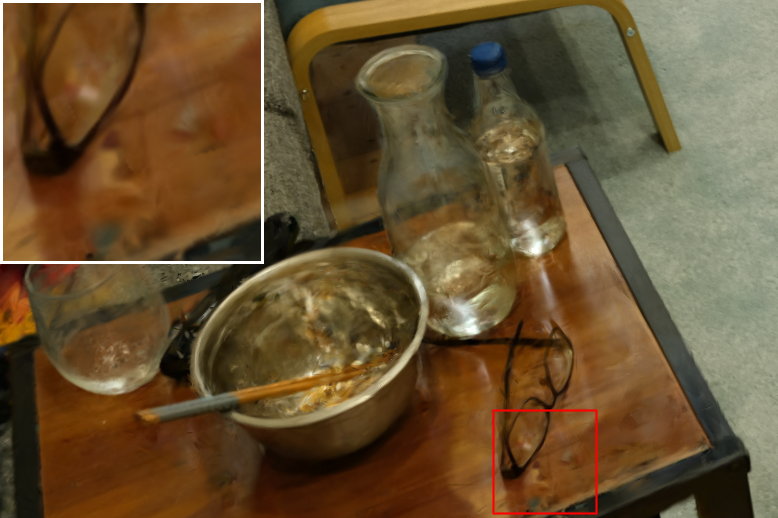} &
\includegraphics[width=0.24\linewidth]{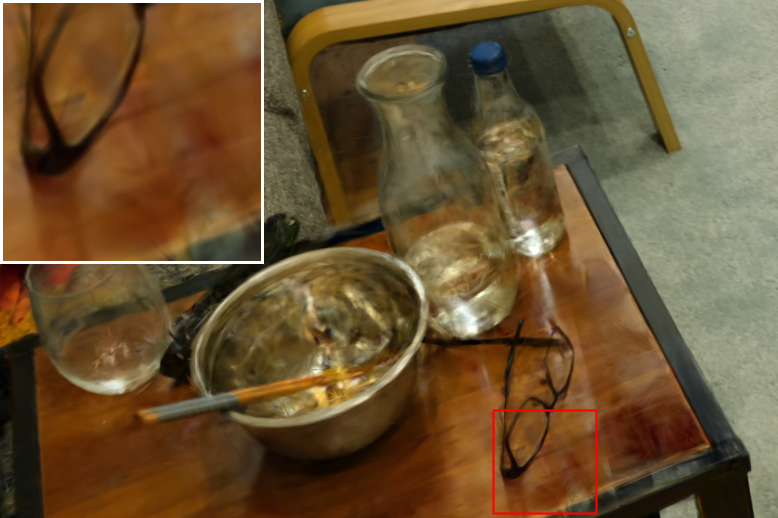} \\
\end{tabular}
}
\caption{Novel view rendering on \textsc{Flowers} (an easy case for dense initialization), \textsc{Bicycle} (an average case), and \textsc{Room} (a hard case for dense initialization).}
\Description{Top row: renderings of the Flowers scene showing that DI results is better reconstruction of finer details (SI contains several blurry regions). Bottom row: renderings of the Room scene showing increased artifacts around reflective objects like a metallic blow and glass objects. }
\label{fig:di-quality-profile}
\end{figure}

\paragraph{DI's Quality Profile} Most of the quality difference between our method and 3DGRT is explained by the switch to DI. DI exhibits a notably different quality profile: it appears much better at reconstructing high-frequency diffuse details like vegetation (as already discussed in EDGS), but appears notably worse at reconstructing highly reflective, smooth objects. We hypothesize that this is because DI struggles to initialize Gaussians at their correct location when objects are highly reflective, and/or because larger Gaussians tend to smooth out details which leaves fewer artifacts. \Cref{fig:di-quality-profile} shows such cases for all three methods.

\paragraph{Performance vs.\ Splatting}
While our method is significantly faster than 3DGRT and near its quality, splatting continues to outperform ray tracing, achieving better reconstruction quality at comparable frame rates, or higher frame rates at equal quality. For these reasons, rasterization is unlikely to be fully replaced. Nonetheless, ray tracing proves highly effective in the dense initialization regime: our results demonstrate that the performance gap between splatting and ray tracing can be closed. Moreover, ray tracing naturally supports light-transport simulation; producing robust, high-quality, relightable 3D Gaussian scenes remains an open challenge, for which fast ray tracing is likely to become a crucial enabling tool.

\begin{figure}[h]
    \centering
    \includegraphics[width=0.5\linewidth]{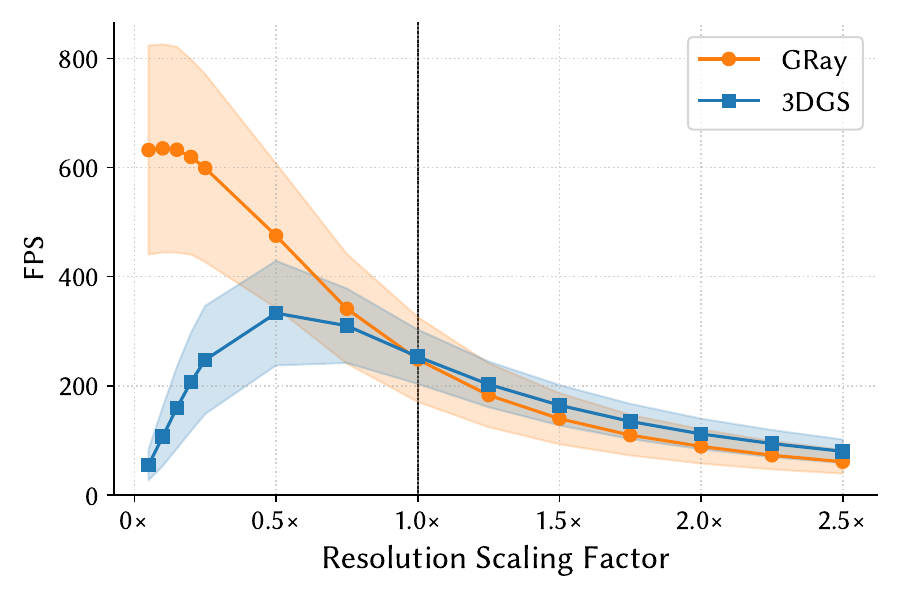}
    \caption{Frame rates at different resolutions measured after training, for GRay and 3DGS. We rescaled all scene resolutions by a constant scaling factor and report averages over all scenes, with shaded regions indicating $\pm$ one standard deviation.  }
    \Description{Plot showing that the frame rates of splatting reduce more slowly than ray tracing as resolution increases, but after increasing initially, start decreasing sharply as resolutions become smaller. In contrast, ray tracing's frame rates increase to nearly triple as resolutions become very small.}
    \label{fig:fps_vs_resolution}
    \vspace{-1em}

\end{figure}

Moreover, our results suggest that ray tracing can scale more efficiently to high Gaussian densities relative to the pixel density, while splatting may scale better to higher resolution unless more Gaussians are added. To test this, we re-rendered all trained scenes at different resolutions, by scaling the existing resolution up and down by a constant factor. We start at the resolutions from the original 3DGS evaluation, which average around 1.2 megapixels. Results are shown in \cref{fig:fps_vs_resolution}: as expected, our ray tracing method appears slower than 3DGS at larger resolutions, but several times faster at smaller resolutions when the same Gaussians are rendered. Interestingly, splatting slows down significantly at such low resolutions.\footnote{Do note however that ray tracing only intersects the Gaussians at the exact center of each pixel; at very small resolutions Gaussians can fall in-between pixels centers and be skipped, which can explain the performance difference beyond algorithmic complexity.}

\paragraph{Compatibility with Splatting}
Ray tracing also addresses two key limitations of 3D Gaussian Splatting: popping artifacts and the affine approximation. Popping occurs because 3DGS sorts Gaussians per primitive rather than per pixel; as the camera or scene changes, small variations in visibility ordering can lead to abrupt changes. By contrast, ray tracing naturally sorts Gaussians along each viewing ray, ensuring a consistent compositing order that eliminates such pops. The affine approximation, on the other hand, arises from the projection model used in 3DGS, where each Gaussian is approximated by an affine transform in screen space. This assumption breaks under strong perspective effects, leading to visible distortion. Ray tracing resolves this issue entirely by evaluating each Gaussian exactly in 3D under perspective projection.

While these differences mean the two rendering paradigms are not directly compatible, prior work has demonstrated promising solutions. In particular, \RM{\textit{Efficient Perspective-Correct 3D Gaussian Splatting Using Hybrid Transparency}}\GDchange{Hahlbohm et al. }~\cite{htgs} introduced a rasterizer capable of evaluating Gaussian kernels exactly without relying on the affine approximation, while also mitigating popping through hybrid transparency.\footnote{Their hybrid-transparency method includes gradients and truncates at a fixed primitive count, whereas ours truncates at a specific transmittance threshold, but the ideas are otherwise very similar.} Similarly, \textit{Stop the Pop}~\cite{stop-the-pop} showed that the sorting discrepancy between splatting and ray tracing can be reconciled. Finally, if popping is tolerated, ray tracing can emulate per-primitive sorting without issue. These insights suggest that it should be possible to design a rasterizer–ray tracer pair compatible with our approach, unifying both under a shared framework, even though such an implementation does not yet exist. Note that 3DGUT~\cite{3dgut} implements a rasterizer that is nearly compatible with 3DGRT, although achieving this required more substantial modifications.

\paragraph{Limitation: Higher Memory Use}
Our method uses more memory than 3DGRT, as a consequence from choosing PPLLs. To verify this, we profiled the \textsc{Bicycle} scene and measured its peak memory consumption. Our method peaks at 16.52~ GB of VRAM, whereas 3DGRT only peaks at 12.42~GB. For reference, 3DGS peaks at 14.02~GB with SI, and 11.62~GB with DI. All scenes ran successfully with roughly 24~GB of VRAM, which corresponds to the recommended minimum value in 3DGS’s official implementation. 

Here is a breakdown of our memory consumption: we ran all scenes with a preallocated forward pass PPLL of size 300M (which uses over $4.8\,\mathrm{GB}$ of VRAM) and a preallocated backward PPLL of size 120M (which uses over $1.9\,\mathrm{GB}$ of VRAM). We additionally store $1005\,\mathrm{bytes}$ of data per Gaussian; Gaussians take 3.62~GB of VRAM at the start of training for the \textsc{Bicycle} scene, which decreases to 2.23~GB at the end of training due to pruning. We also store all test images in VRAM, which requires 2.36~GB for this scene. In total, 14.5~GB is allocated before training even starts.

Scaling training to higher resolutions would require very large PPLLs, which would take up considerably more memory (linear in the number of pixels). However, this usage could be reduced by rendering tiles instead of the whole image at once.

\section{Conclusion}
Gaussian ray tracing has several advantages, including avoiding the affine approximation and popping artifacts, and is of critical importance for inverse rendering and relighting applications. As of yet however, it has remained significantly slower than splatting. We argue that ray tracing can better handle increasing counts of tiny Gaussians, where, unlike splatting, it can exhibit logarithmic scaling. Our experiments on dense initialization suggest that this effect is of great practical significance. In addition, we have analyzed several important algorithmic components and parameters of Gaussian ray tracing for radiance fields. Of these, per-pixel linked lists and scale decay have the most significant effect on performance. 
Leveraging our analysis, we propose novel ray tracing solution designed to exploit ray tracing's algorithmic scaling and dense initialization. Our ray tracer closes the gap in performance between 3DGRT and 3DGS while maintaining a quality level near 3DGRT's. While quality remains somewhat lower than 3DGS, the significant increase in speed will be a key enabler for future work, especially in cases where light transport simulation is required. Our code will be open sourced on publication.

\begin{acks}
Thanks to Jeffrey Hu for pointing us towards dense initialization and programming help, and to Ishaan Shah for his work on the Gaussian Viewer. This research was co-funded by the European Union (EU) ERC Advanced Grant NERPHYS No 101141721 (see https://project.inria.fr/nerphys). Views and opinions expressed are however those of the author(s) only and do not necessarily reflect those of the EU or the European Research Council. Neither the EU nor the granting authority can be held responsible for them. Experiments presented in this paper were carried out using the Grid’5000 testbed, supported by a scientific interest group hosted by Inria and including CNRS, RENATER and several Universities as well as other organizations (see https://www.grid5000.fr). This research was also supported by NSERC grant RGPIN-2020-04799 and the Digital Research Alliance Canada. The authors are grateful to Adobe and NVIDIA for generous donations. 
\end{acks}

\bibliographystyle{ACM-Reference-Format}  
\bibliography{references}

\clearpage
\appendix
\section*{Supplemental Material}

\begin{figure*}[h]
    \centering
    \includegraphics[width=1.0\textwidth]{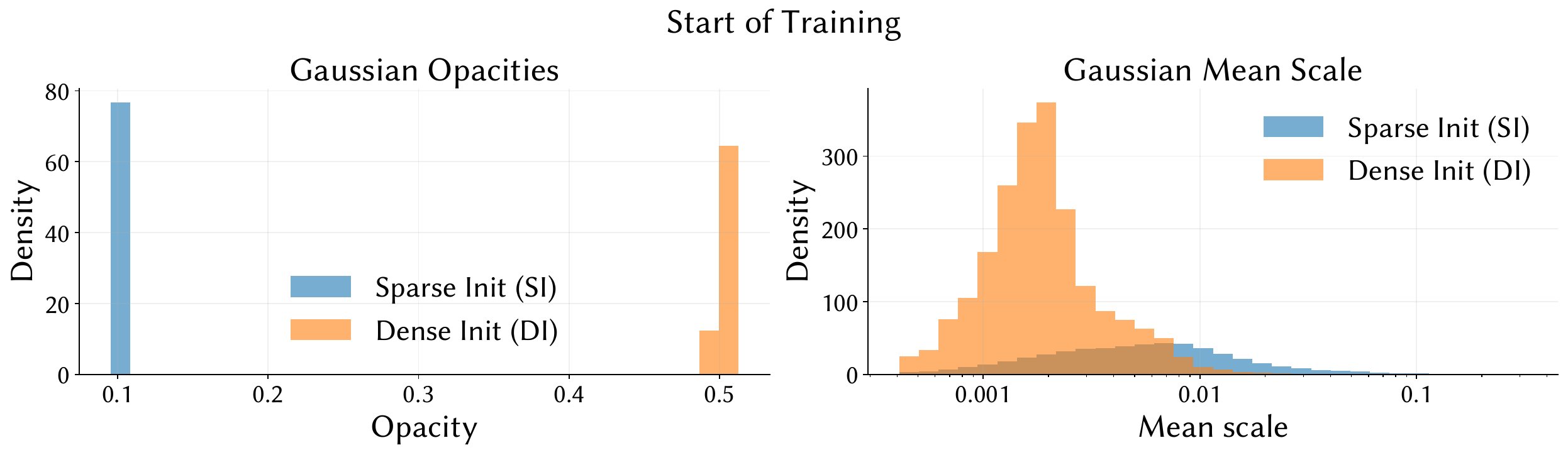}
    \vspace{0.5em}
    \includegraphics[width=1.0\textwidth]{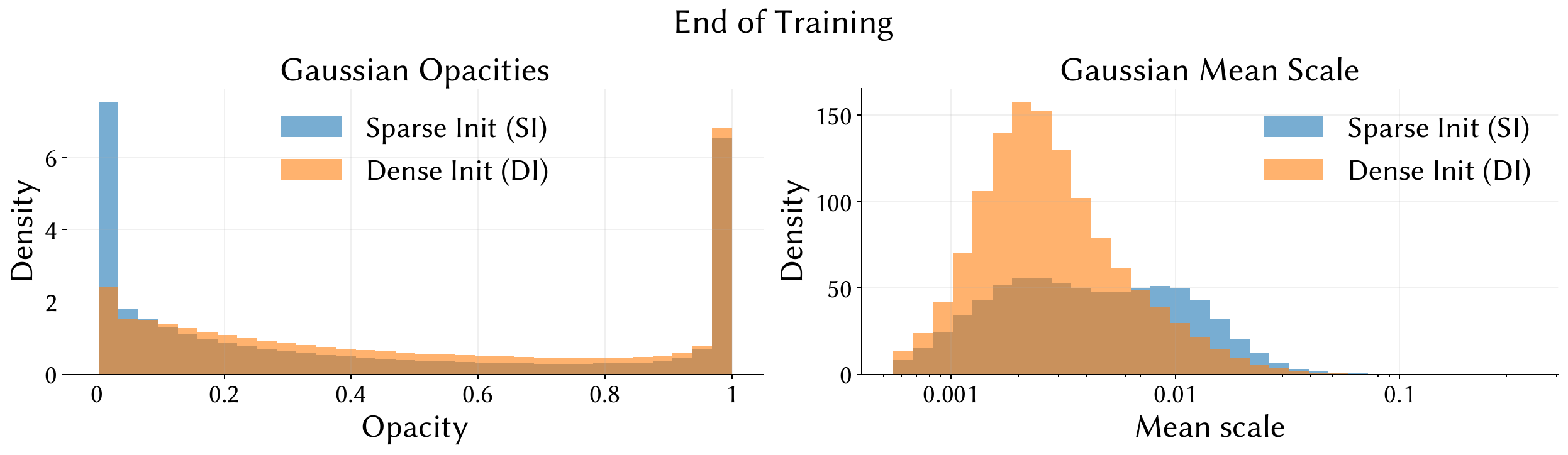}
    \vspace{-0.5em}
    \caption{Comparison of 3DGRT's Gaussian scale and opacity distributions with SI and DI, at the start and at the end of training. Values averaged over all 13 benchmarking scenes.}
    \label{fig:hist-sparse-vs-dense}
\end{figure*}

This supplemental material contains additional analysis, detailed ablations and additional results. Implementation details are then clarified and finally, detailed results for all individual scenes are presented.

\section{Additional Analysis}\label{additional-analysis}

This section contains additional analysis that was not presented in the main text.

\subsection{Gaussian Opacity and Scale Distributions}

We compared the Gaussian opacity and scale distributions between SI and DI in \cref{fig:hist-sparse-vs-dense} using 3DGRT. Interestingly, while SI initializes the Gaussian opacities low (at $0.1$) and DI initializes them high (at $0.5$), their distributions nearly match at the end of training. However, DI results in fewer low-opacity Gaussians. Additionally, while SI's large initial Gaussians tend to shrink a bit during training, once training ends SI still results in larger Gaussians overall.

\begin{figure}[h!]
\centering

\begin{minipage}[t]{0.48\textwidth}
\centering
{\small Rendering Time with Shrinking Gaussian Sizes}

\includegraphics[width=0.8\linewidth]{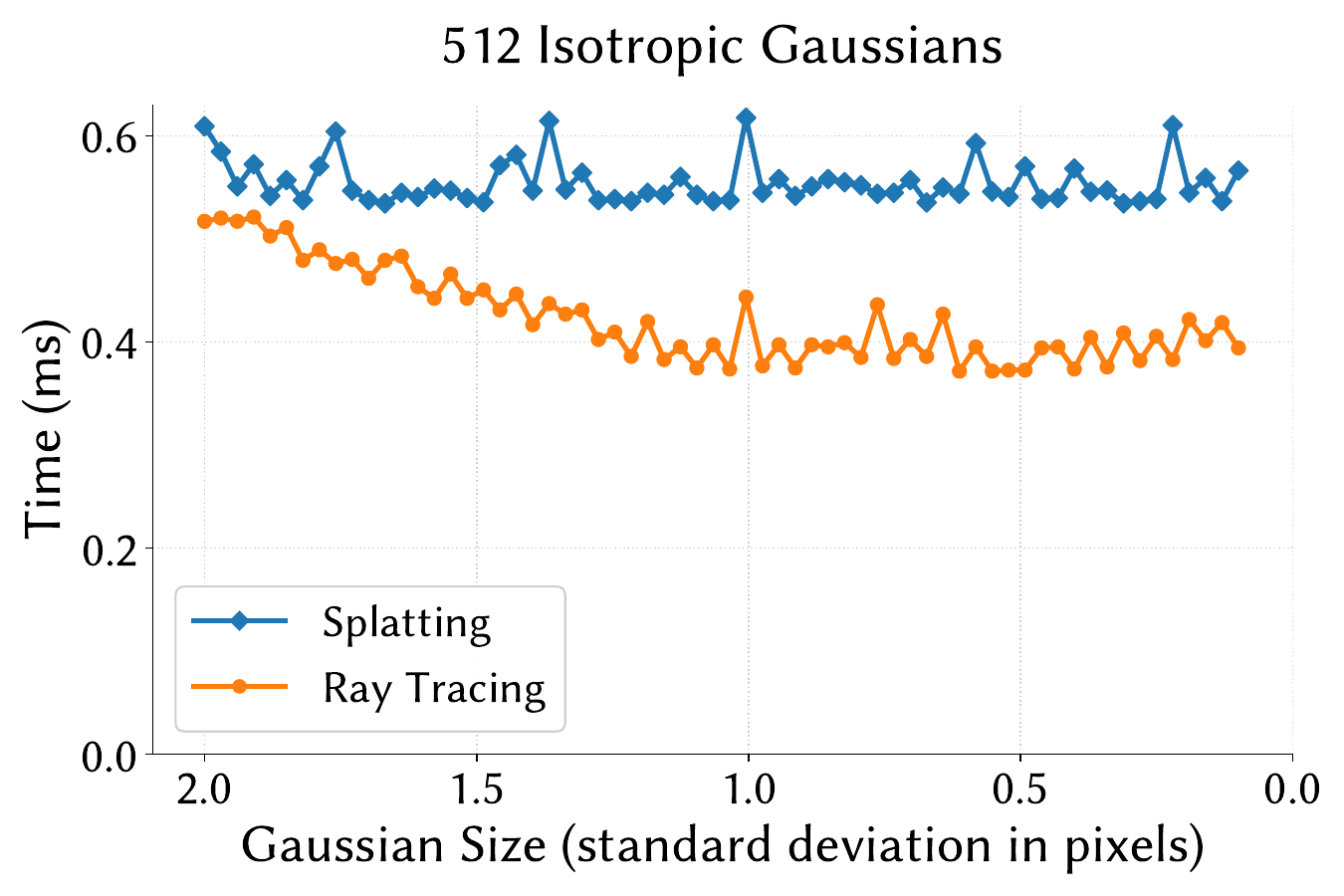}

\includegraphics[width=0.8\linewidth]{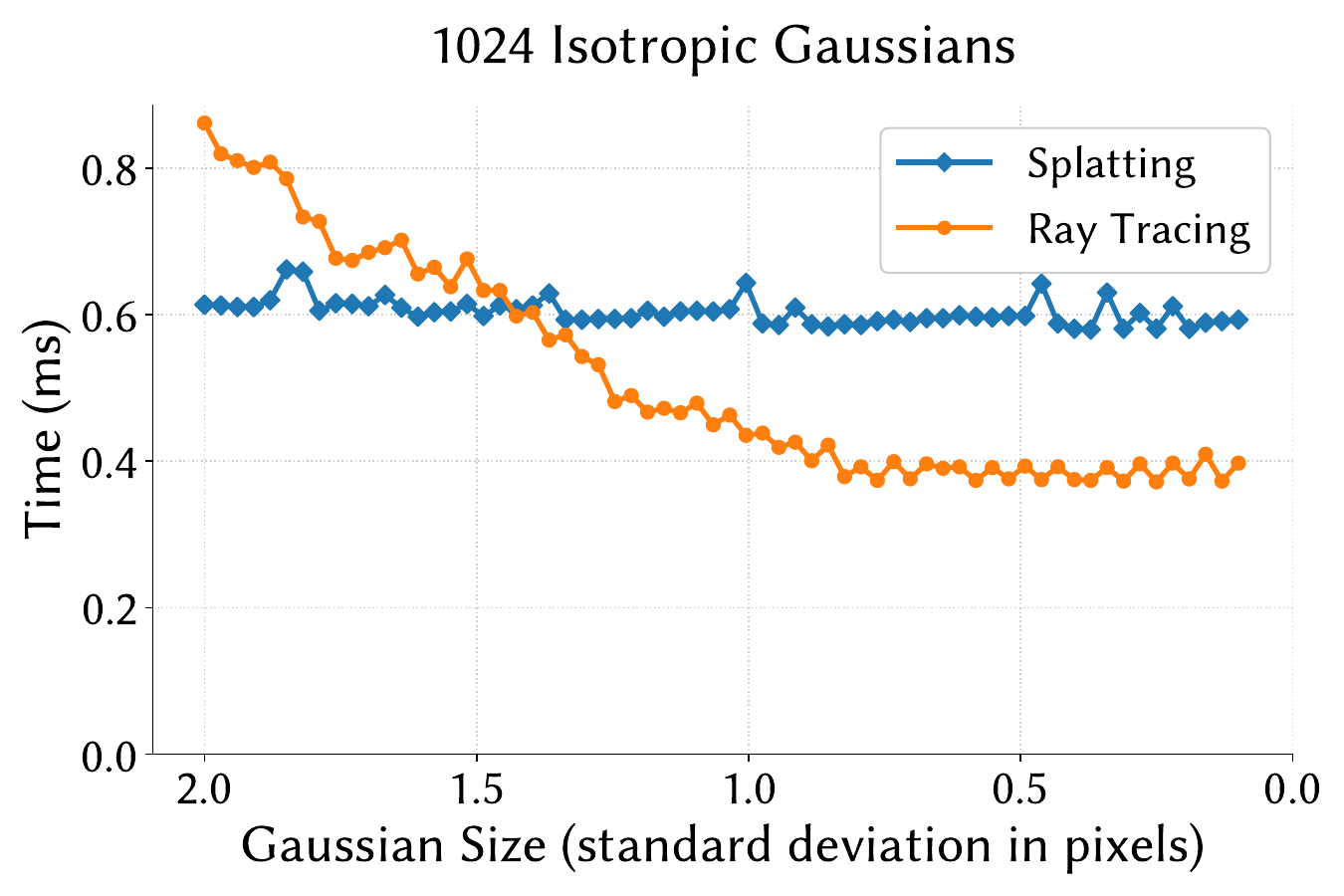}

\includegraphics[width=0.8\linewidth]{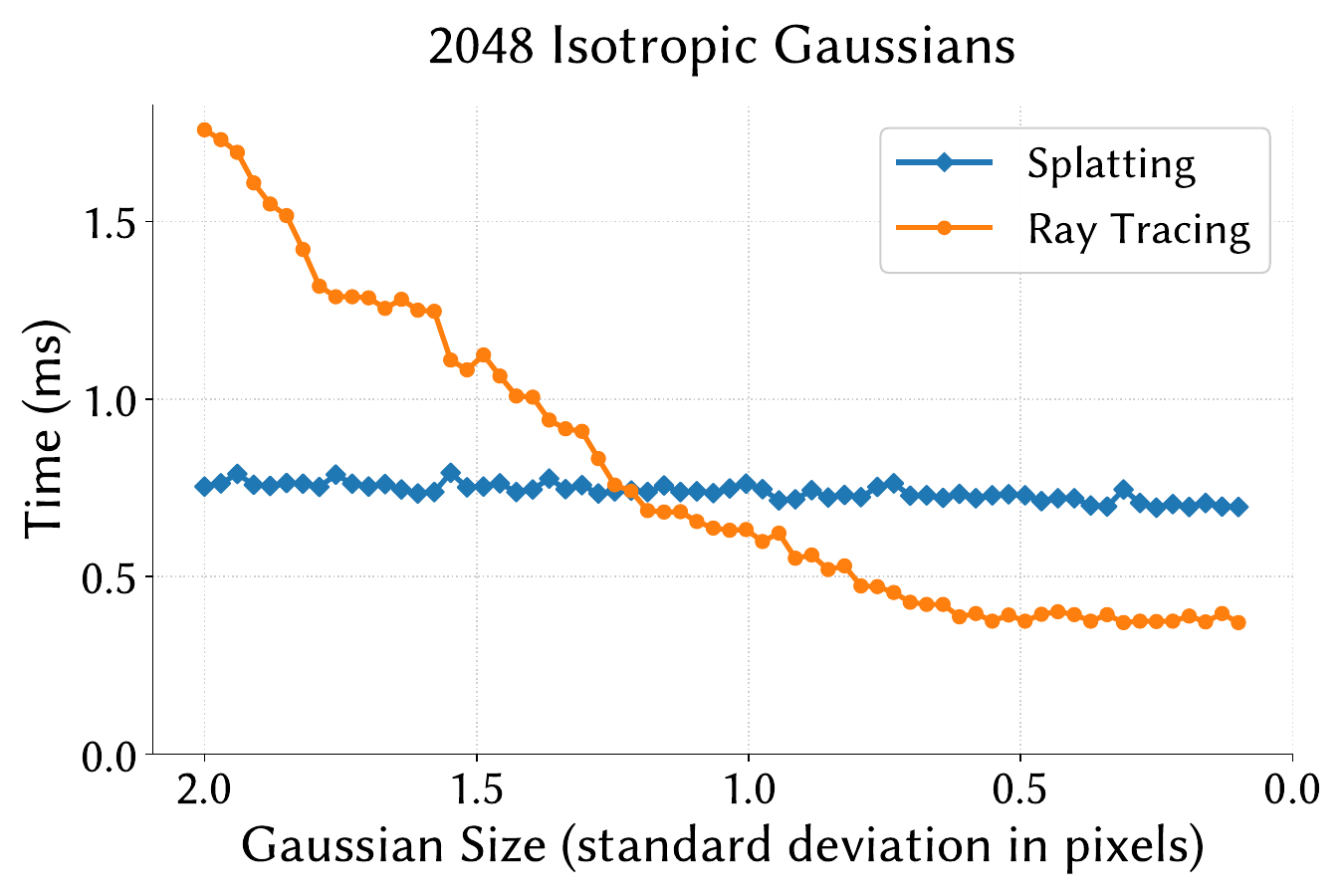}

\includegraphics[width=0.8\linewidth]{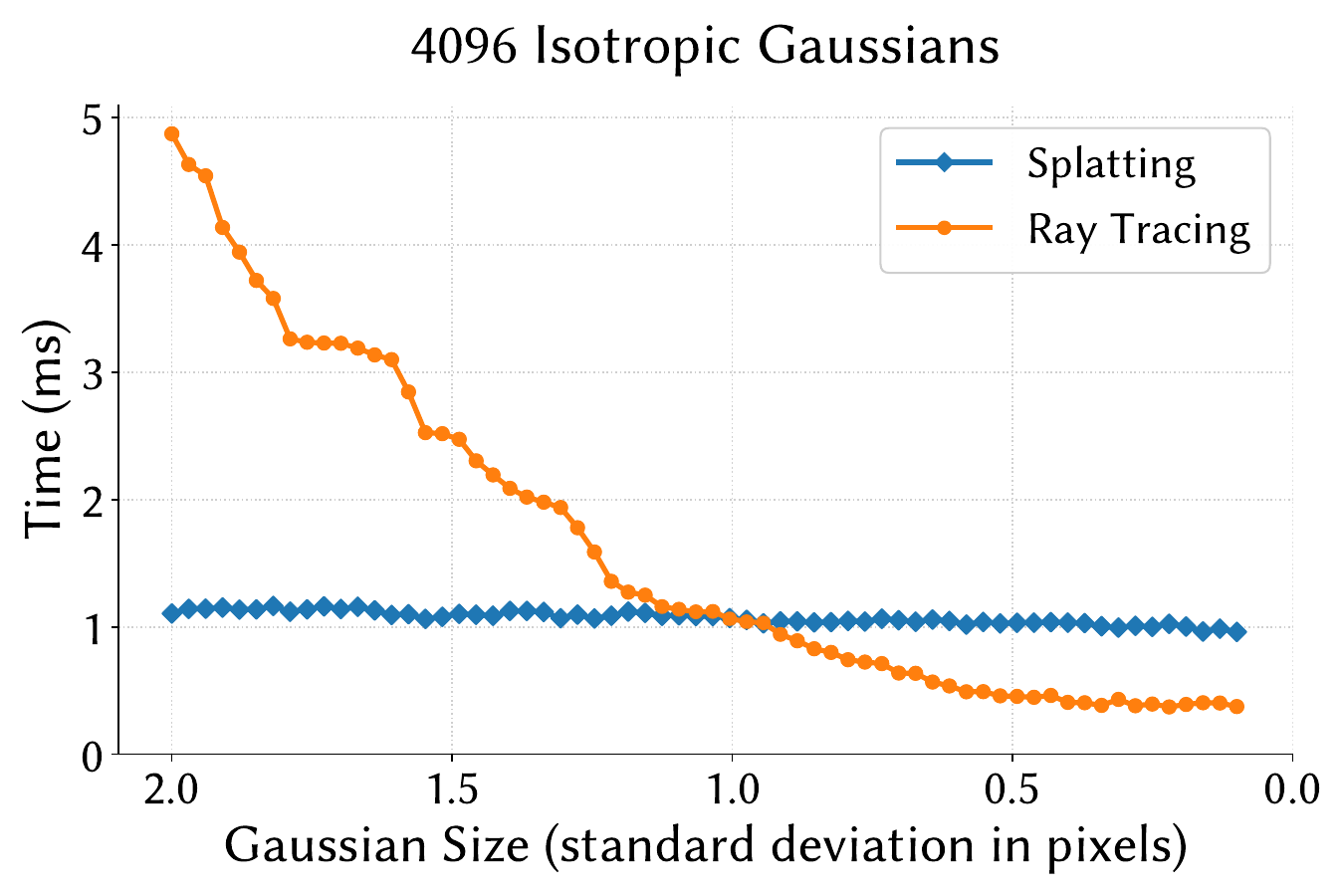}

\includegraphics[width=0.8\linewidth]{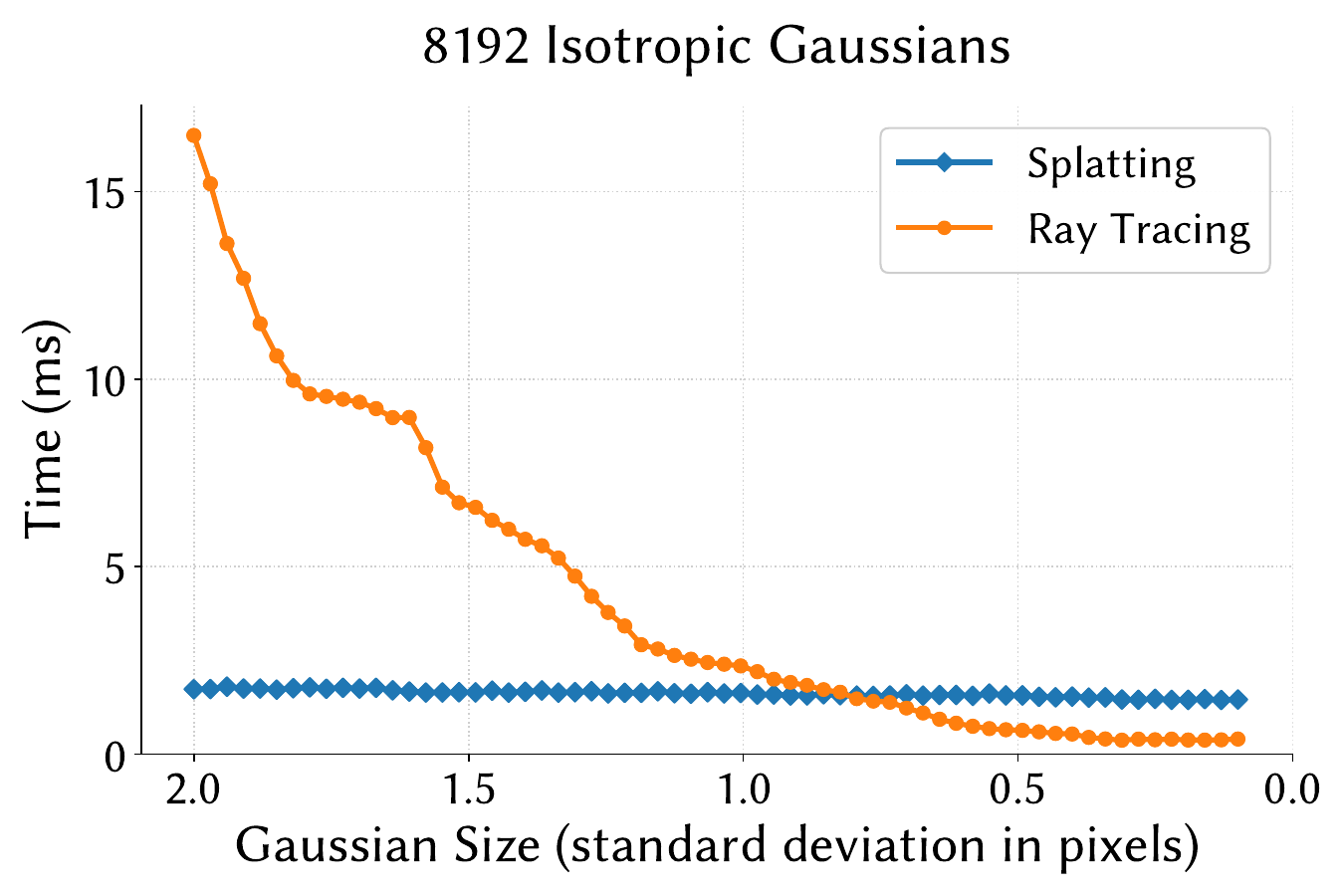}
\end{minipage}
\hfill
\begin{minipage}[t]{0.48\textwidth}
\centering
{\small Rendering Time with Increasing Gaussian Counts}

\includegraphics[width=0.8\linewidth]{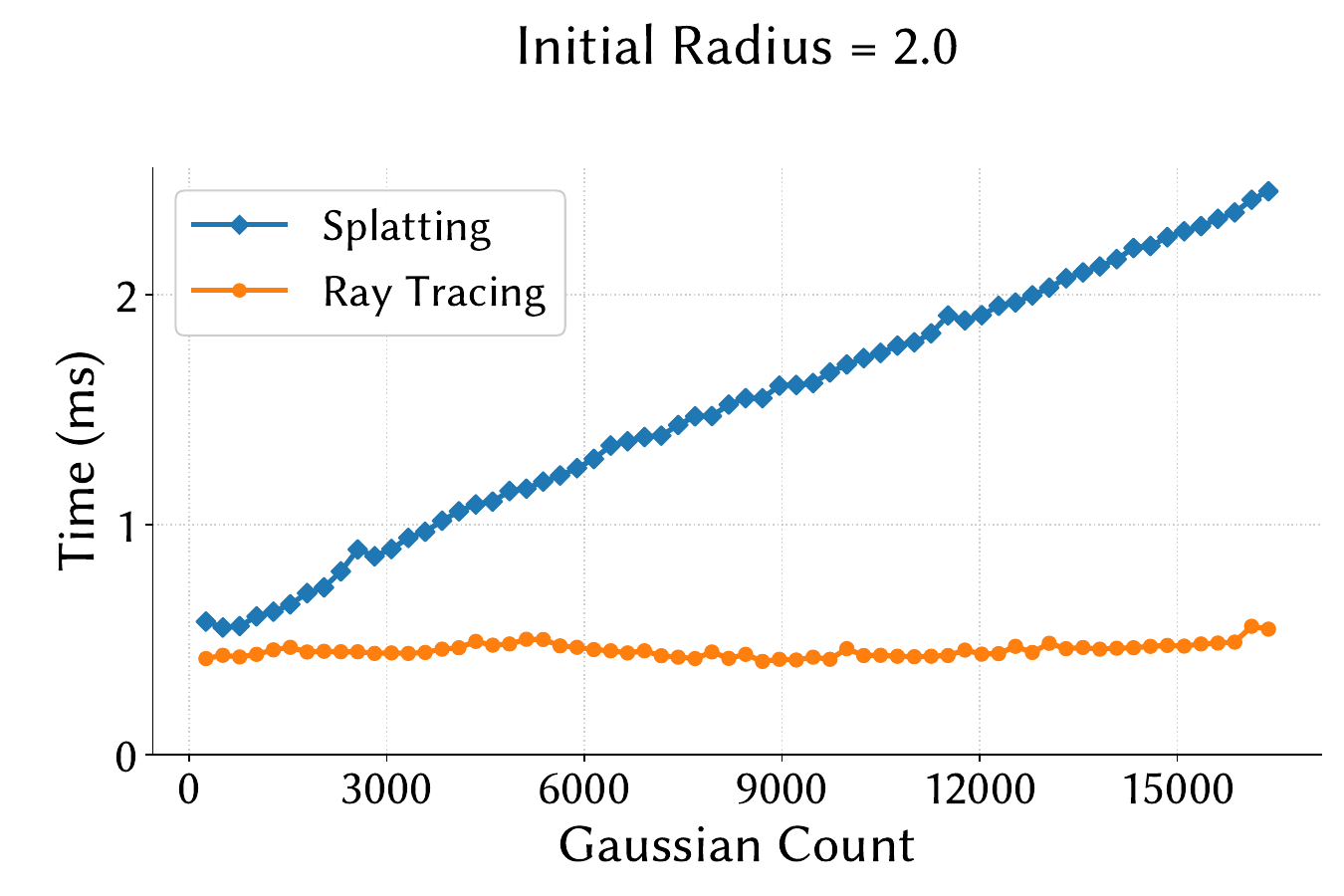}

\includegraphics[width=0.8\linewidth]{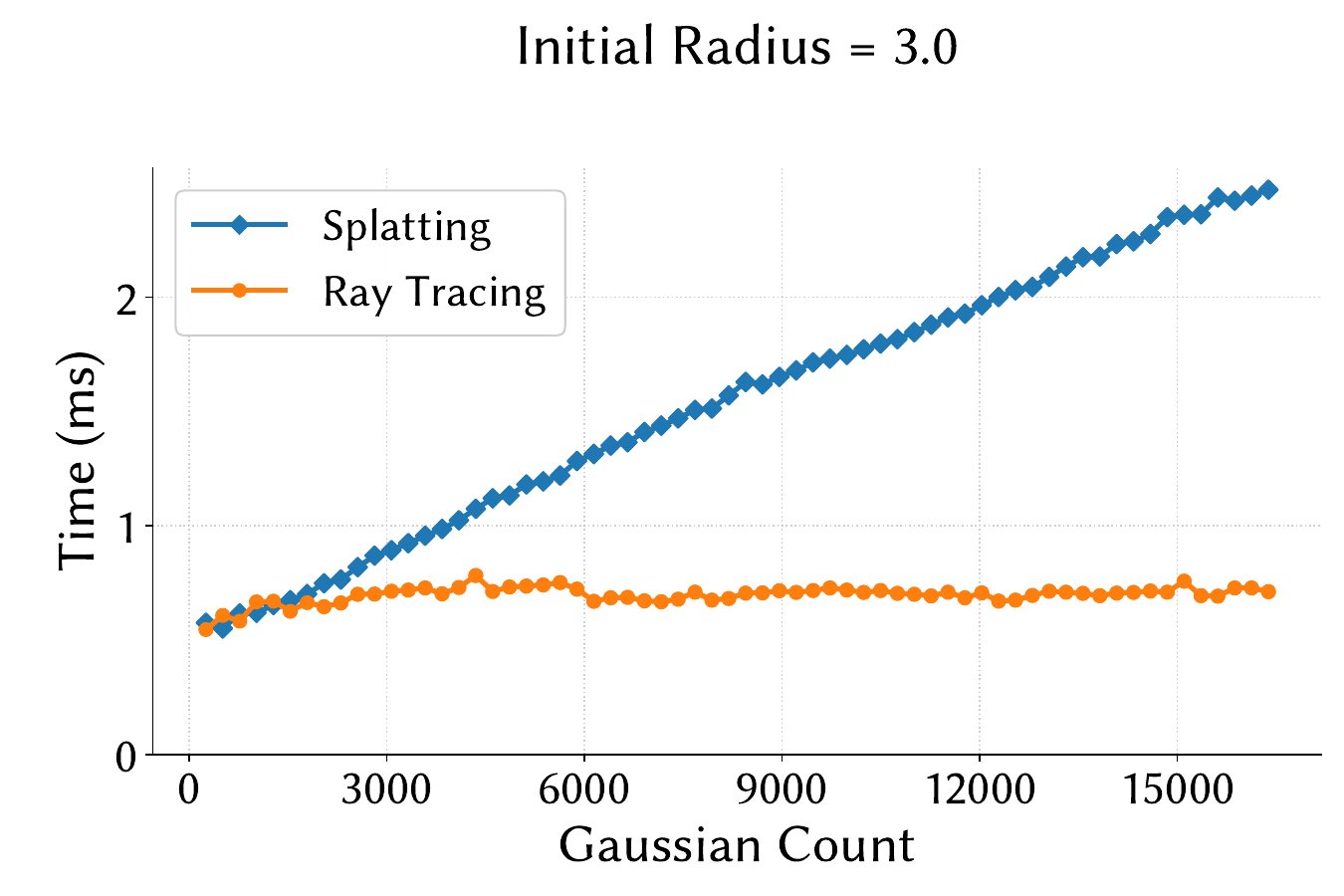}

\includegraphics[width=0.8\linewidth]{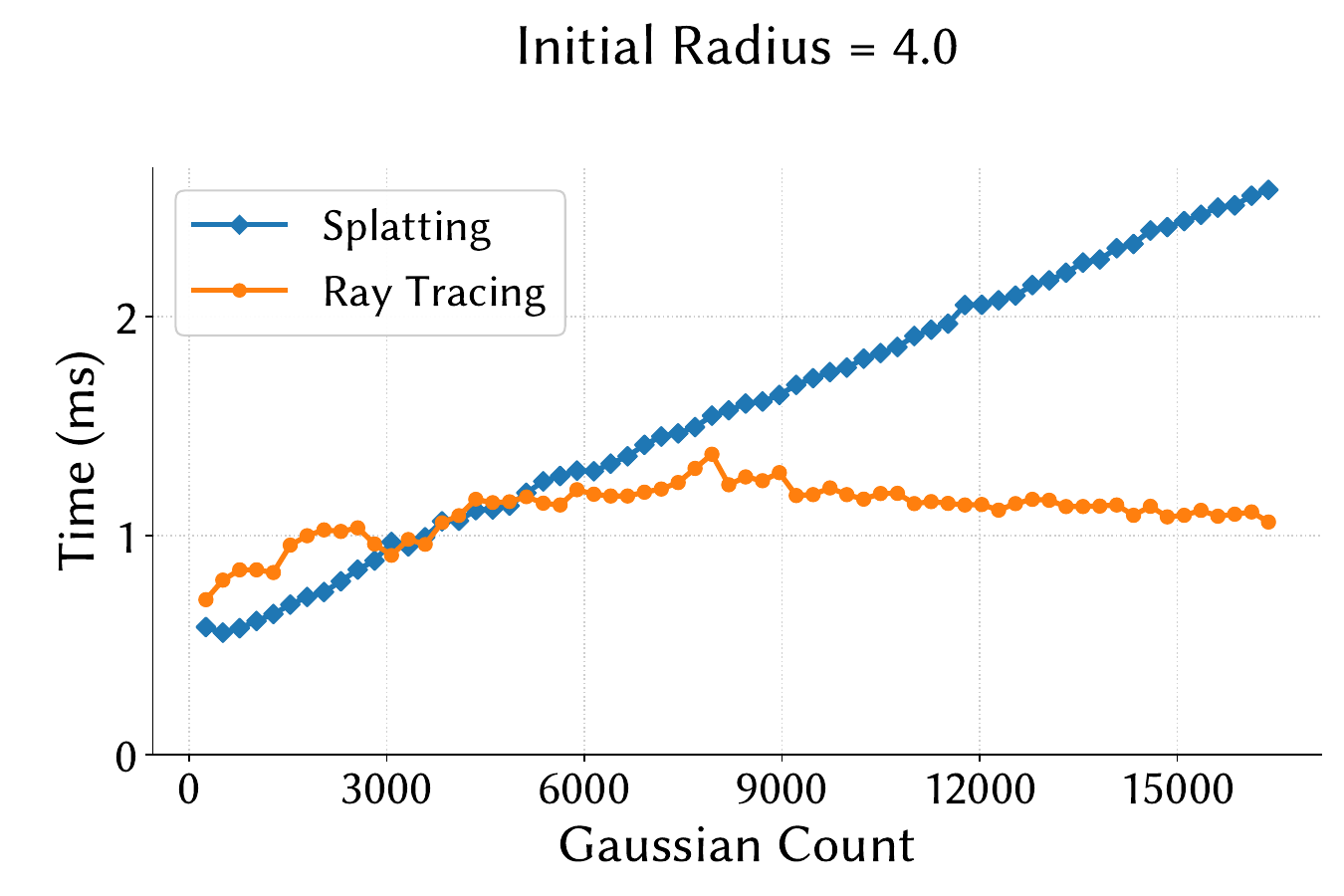}

\includegraphics[width=0.8\linewidth]{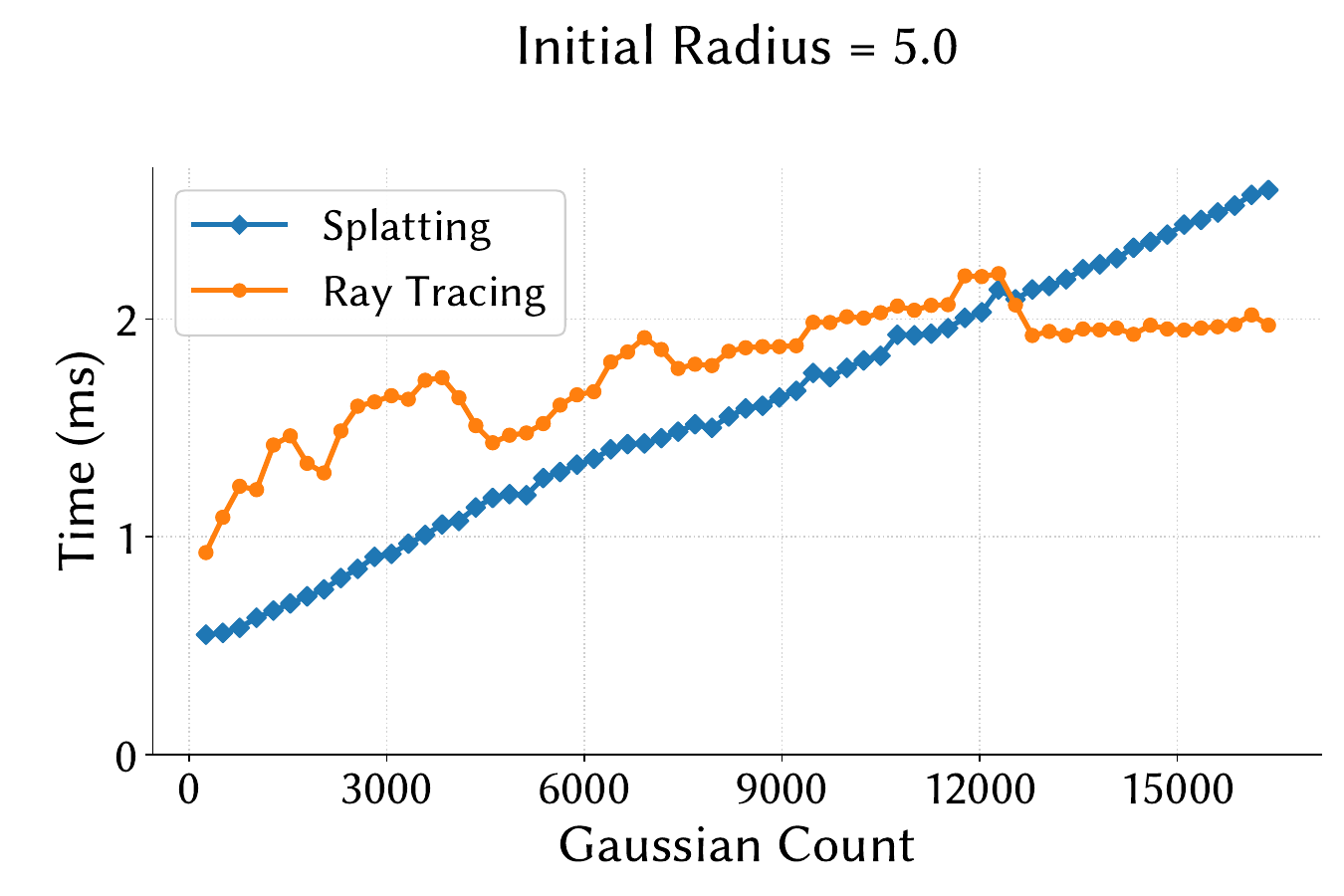}

\includegraphics[width=0.8\linewidth]{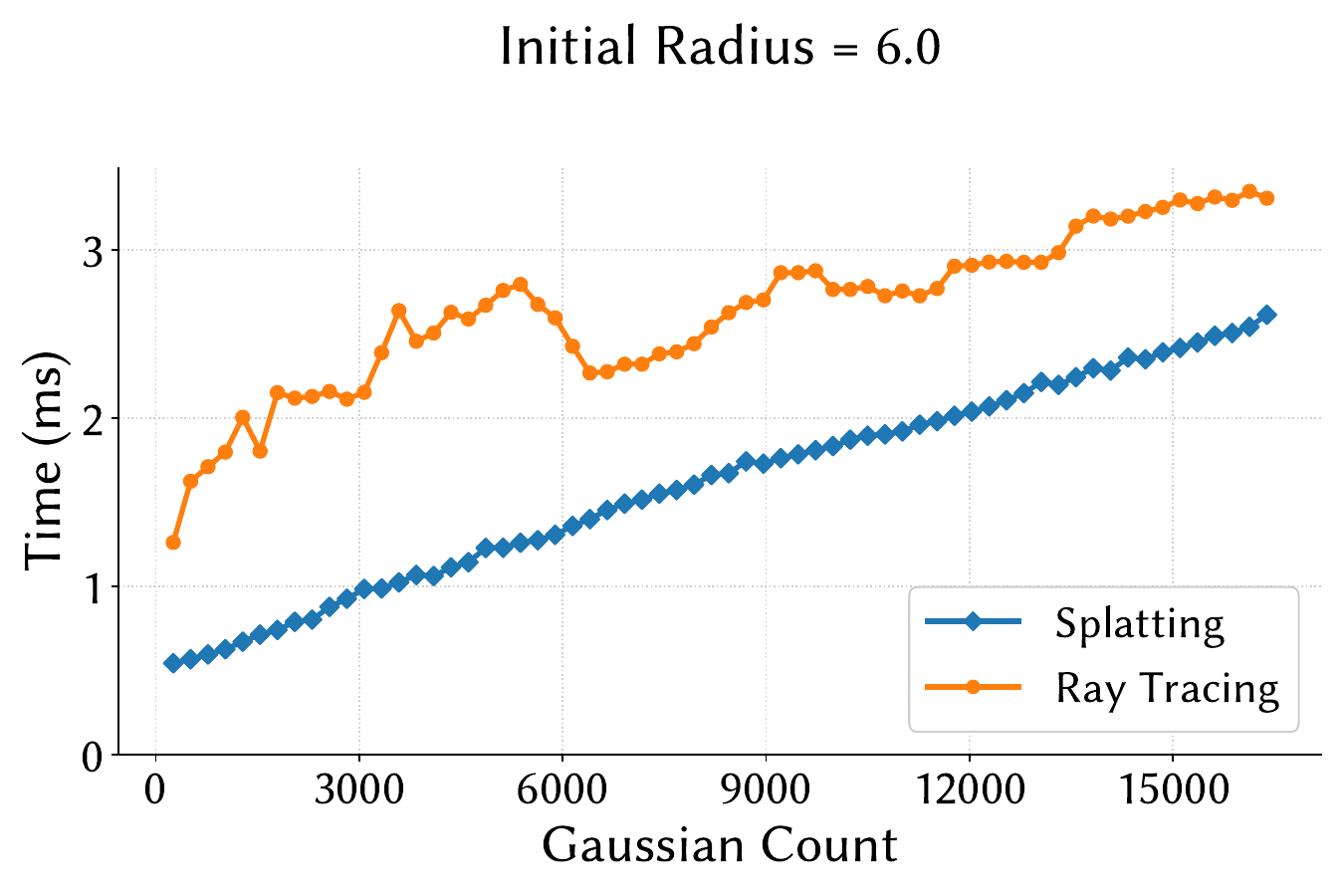}
\end{minipage}

\caption{Toy experiments showcasing ray tracing's ability to benefit from tiny Gaussian sizes, potentially enabling it to scale logarithmically in the Gaussian count.}
\label{fig:toy-sweeps}
\end{figure}

\subsection{Toy Experiments on Computational Complexity}

To demonstrate the difference in computational complexity between splatting and ray tracing, we rendered a toy scene comprising just a few isotropic Gaussians in a single $16 \times 16$ pixel tile. We used a 90~degree camera looking down on the z-plane, positioned at coordinates $(0.0, 0.0, 1.0)$, such that the camera sees a $[-1.0, 1.0] \times [-1.0, 1.0]$ square of the z-plane centered about the origin. We then placed a small number of white, zero rotation and low opacity Gaussians on this plane. We created $k$ isotropic Gaussians per pixel that all share the same scale: as such this scale can be parameterized by a single global standard deviation value $\sigma$. 

To make this experiment robust we had to control for several factors. First, early ray termination can skip many Gaussians evaluations if their opacity is high enough for the ray to go over transmittance threshold. Hence we chose a very low opacity value, 0.01, to avoid this. However, low opacities can benefit ray tracers over splatting because of adaptive clamping. To control for this we set the threshold $\alpha_{\mathrm{min}}=1/255$ to match 3DGS exactly.\footnote{We did keep $n_{\mathrm{expo}}=2$ since we consider it core 3DGRT's proposed approach.} Second, ray tracing only collects Gaussians that intersect the pixel center exactly; when Gaussians are small and far away, they can occasionally fall ``in between'' pixels and not be rendered. To avoid this, rather than placing Gaussians randomly we fixed their positions to match the intersection of camera rays with the $z$ plane: for each pixel, we placed $k$ Gaussians at xy coordinates $\left(x_i, y_j\right)$ with $x_i = 2(i+\tfrac12)/16 - 1$ and $y_j = 2(j+\tfrac12)/16 - 1$ for $i, j \in [0, 15]$. Finally, to prevent the Gaussians from overlapping exactly, we staggered their depth values slightly by setting the z-coordinate of the $n$th Gaussian at pixel $(i, j)$ to $z_n = -n \epsilon$ where $\epsilon = 0.001$. We validated experimentally that each ray always intersects its own $k$ Gaussians, at the very least.  
% todo: Our rendered Gaussian grid is visualized at different settings of $k$ and $\sigma$ in...

In our first experiment, we wanted to show that splatting does not benefit from shrinking Gaussians beyond making them fit inside a single tile. As such, we set $k=8$ to simulate a reasonable workload, yielding a total of 2048 Gaussians, and swept their standard deviation down from $2.0$ all the way to $0.1$. In this supplemental, we repeat the same experiment for $k \in [2, 4, 8, 16, 32]$ for reference. Results are presented in \cref{fig:toy-sweeps}'s left column. Although which method is fastest depends on both the Gaussian sizes and count,
these results clearly show that ray tracing is highly sensitive to Gaussian sizes while splatting is not, unless perhaps they become so large as to intersect multiple tiles.  Do note however that splatting will remain faster at very low Gaussian counts, and ray tracing will require very small Gaussian sizes to surpass splatting at very high Gaussian counts. 

In our second experiment, we wanted to show that ray tracing can exhibit logarithmic scaling in the number of Gaussians, provided they are small enough, while splatting always scales near-linearly. To this end, we selected an initial standard deviation of $\sigma=4.0$ and progressively increased $k$ starting from 1. When increasing $k$, we also decreased the standard deviation such as to approximately conserve the Gaussian's projected surface area; in other words, we used adjust values $\sigma'(k)=\sigma/\sqrt{k}$. This corresponds to halving the standard deviation each time the Gaussian count quadruples. We further validated experimentally that the effective number of ray-Gaussian intersections stayed roughly constant as the Gaussian count increases, which is what we desired. 

In this supplemental, we repeat the same experiment with the initial $\sigma \in [2.0, 3.0, 4.0, 5.0, 6.0]$ for reference. Results presented in \cref{fig:toy-sweeps}'s right column. These results show that, while relative performance is highly sensitive to the Gaussian sizes, if Gaussian sizes are controlled for then ray tracing can exhibit log-like scaling while splatting stays linear. Bear in mind that ray tracing will remain categorically slower if the initial radius is very high, and will remain categorically faster if the initial radius is very small. 

\subsection{\changed{BVH Details and Impact on Performance}}\label{sec:bvh-details}
\changed{While ray tracing small primitives can potentially approach logarithmic complexity, BVH updates remain $O(n \log n)$. As such, our method's efficiency relies on BVH updates having a small constant factor. To keep the update times low, we use OptiX's support for incremental updates,\footnote{Incremental updates refit the BVH internal nodes, while leaving the tree's topology unchanged.} and only perform full rebuilds every $500$ steps.}

\begin{figure}[h!]
    \centering
    \includegraphics[width=1.0\textwidth]{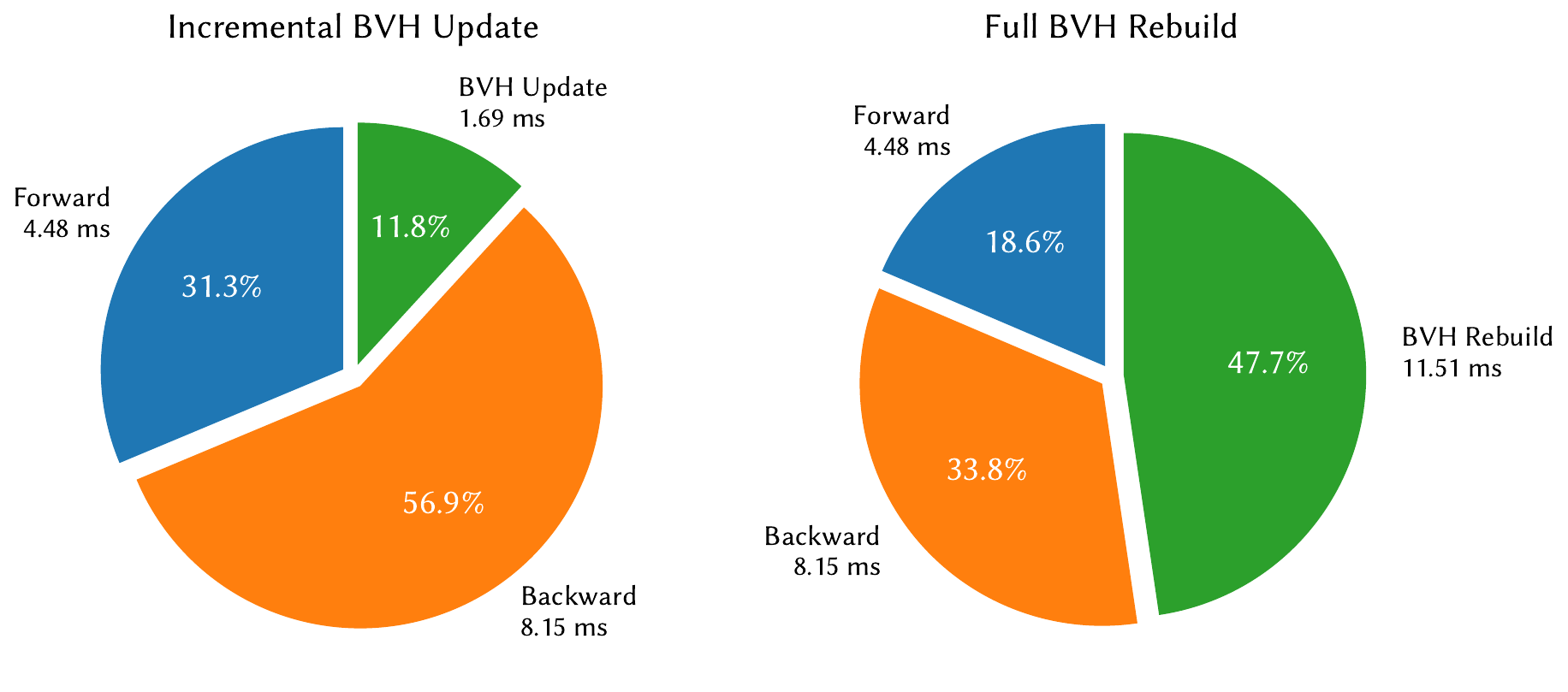}
    \caption{\changed{Per-view training step breakdown, showing the relative importance of BVH updates compared to the forward and backward pass.}}
    \label{fig:breakdown}
\end{figure}

\changed{We measured the speed of BVH updates, forward pass, and backward pass, averaged across all test views and further averaged across all scenes.\footnote{Note that these measurements were obtained by measuring the forward, backward, and update separately; the effective runtime might differ slightly.} Results in~\cref{fig:breakdown} show that incremental updates only take up $\approx\!12\%$ of the combined BVH update + forward pass + backward pass time during optimization. Full rebuilds are much slower, but very rare. We only made a minimal effort to optimize this code, and it is possible the updates could be made more efficient.}

\subsection{\changed{Memory Use at Higher Resolution}}
\changed{As discussed in the main text, working with large preallocated per-pixel linked lists is efficient, but consumes a lot of memory. Rendering very high image resolutions requires a different approach. One simple and effective strategy is tile-based rendering: the image is divided into a few large tiles, and tiles are rendered one-by-one. This keeps linked list memory use constant and remains efficient when tiles are large.}

\begin{figure}[h!]
    \centering
    \includegraphics[width=0.49\textwidth]{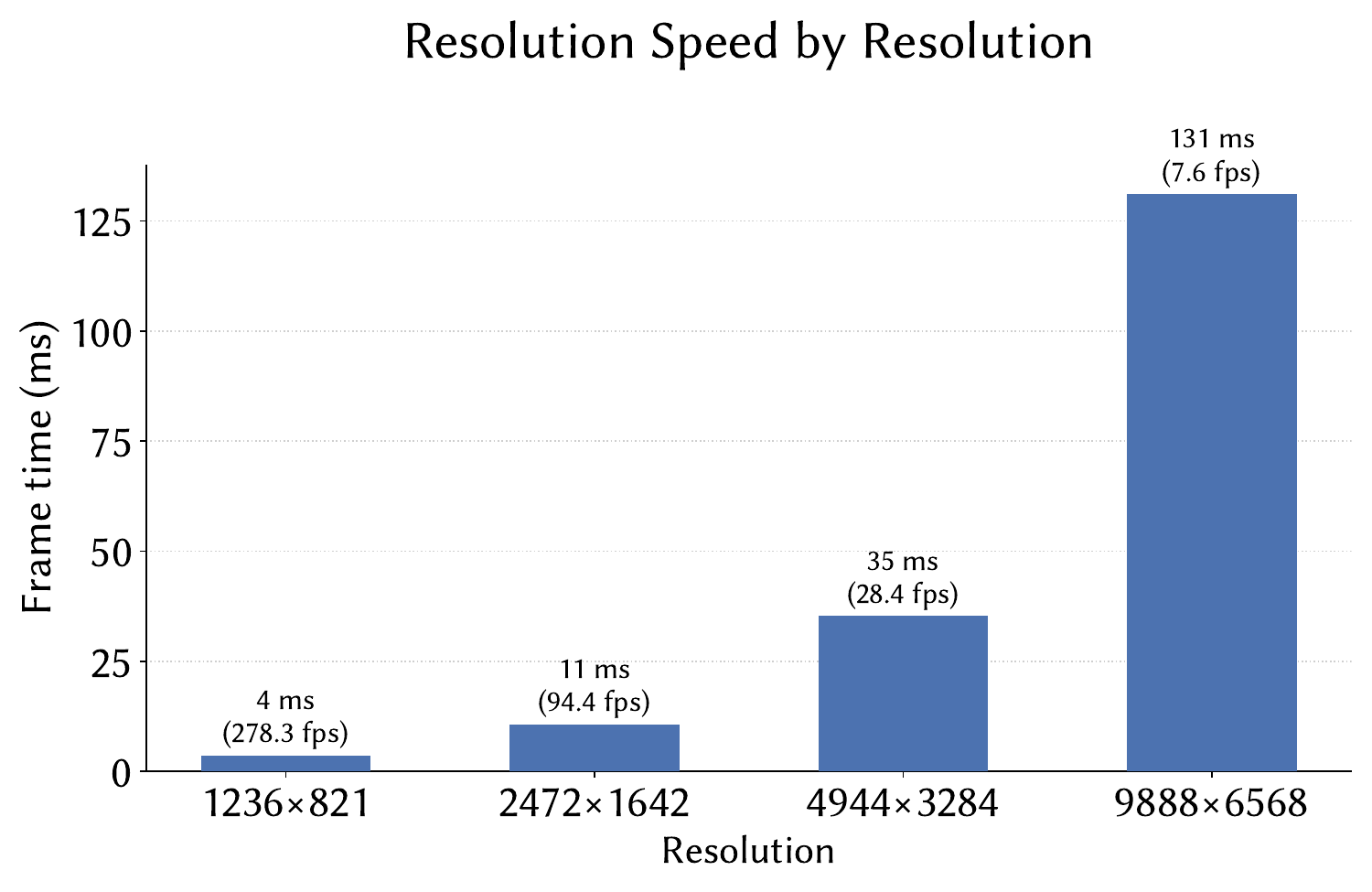}
    \includegraphics[width=0.49\textwidth]{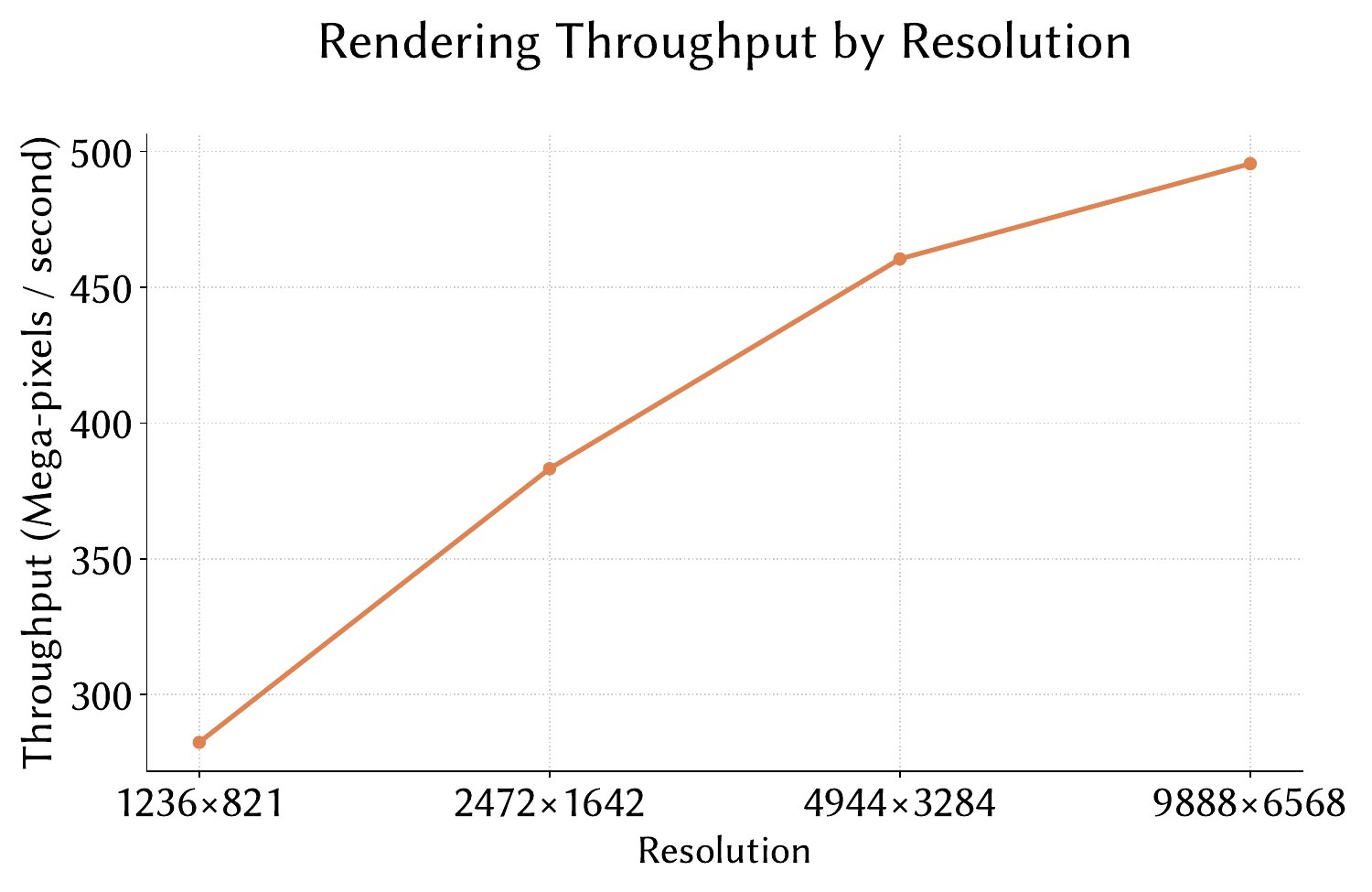}
    \caption{\changed{Rendering speed and efficiency (throughput in megapixels per second) when rendering the Mip-NeRF360 Bicycle scene at increasing resolutions with tile-based rendering.}}
    \label{fig:tiling}
\end{figure}

\changed{We implemented a prototype of this approach and rendered the Mip-NeRF360 Bicycle scene at increasing resolutions, from $1236\!\times\!821$ to $9888\!\times\!6568$, showing that high resolutions can be rendered with our method without running out of memory. \Cref{fig:tiling} further shows that this approach remains efficient; rendering throughput actually improves at higher resolutions (ostensibly due to increased memory coherence and/or better GPU utilization).}

\changed{Extending this approach to high-resolution training could be done either by training on image tiles rather than full images, or by processing the forward and backward passes tile-by-tile.}

\subsection{Explaining the Convergence Failure with DHT Disabled}\label{sec:dht-convergence}
When training at high transmittance thresholds without DHT, Gaussians expand pathologically, often crashing due to running out of memory; \changed{this is shown in \cref{fig:no-dht-crash}} \deleted{[Figure repeated from the main text was removed.]}. DHT fixes this issue by providing a more accurate gradient for the heading Gaussians, by estimating the color of what comes behind them. Without it, these Gaussians continually expand to try to cover a seemingly black background which falsely appears because of early termination. Yet expanding them further increases this problem, since bigger Gaussians intersect more pixels, which tends to make rays terminate even earlier.

\section{Detailed Ablations}\label{sec:detailed-ablations}
This section contains detailed ablations exploring the design space of ray tracing under DI, and serves to validate our design choices and final hyperparameter selection. \deleted{Most of these ablations were presented in the main text; some text and results are repeated.} This detailed view presents full hyperparameter sweeps and all quantitative measures for every parameter \changed{presented in the main text}.

\begin{table}[h!]
\centering
\small
\caption{3DGRT BVH update times in millisecond at iteration 1 for both initialization types, comparing OBBs vs icosahedron bounds.}
\label{tbl:bvh-update-time}
\begin{tabular}{@{}lcccc@{}}
\toprule
Method & BVH Update (ms)$_\downarrow$ &  \#Gaussians$_\downarrow$   \\
\midrule
	\textsc{3DGRT}\textsubscript{SI}  & \textbf{0.9} & \textbf{0.11M}   \\
	\textsc{3DGRT}\textsubscript{SI+Icosahedrons}  & 4.5 & 0.11M   \\
\midrule
	\textsc{3DGRT}\textsubscript{DI}  & \textbf{11.4} & \textbf{3.97M} \\
	\textsc{3DGRT}\textsubscript{DI+Icosahedrons}  & 125.7 & 3.97M \\
\bottomrule
\end{tabular}
\end{table}

\begin{table}[h!]
\centering
\small
\caption{Performance of 3DGRT under dense initialization with oriented bounding boxes vs. icosahedrons for bounding Gaussians.}
\label{tbl:obb-vs-icosahedron}
\begin{tabular}{@{}lc|cc@{}}
\toprule
Variant & Bound Type & Opt Time$_\downarrow$ & FPS$_\uparrow$ \\
\midrule
\textsc{3DGRT}\textsubscript{DI} & OBB & \textbf{00:40:05} & 131 \\
\textsc{3DGRT}\textsubscript{DI+Icosahedrons} & Icosahedron &  01:22:12 & \textbf{145} \\
\bottomrule
\end{tabular}
\end{table}

\paragraph{\textbf{Restricting Gaussian Support is Faster and Worse}} 
We first review the hyperparameters settings $\alpha_{\mathrm{min}}$ and $n_{\mathrm{expo}}$ which 3DGRT introduced to restrict Gaussian support. Results for $\alpha_{\mathrm{min}}$ are presented in \cref{tbl:alpha-min-sweep}: as expected, higher values result in a marked increase in FPS counterbalanced by substantial decreased quality. Setting $\alpha_{\mathrm{min}}=0.001$, a very low value, increases quality minimally over the suggested value $0.01$ which we retain (PSNR $26.47 \rightarrow 26.52$, LPIPS $0.236 \rightarrow 0.234$) while also dropping FPS considerably ($248 \rightarrow 156)$; corresponding cells are highlighted in \colorbox{yellow!60}{yellow}. \deleted{Interestingly, higher values result in increased pruning reflected by a lower final Gaussian count, which we did not expect. It may be possible to utilize higher values effectively if this effect is counteracted.}

Results for $n_{\mathrm{expo}}$ are shown in \cref{tbl:expo-sweep}. Again, selecting higher values improved speed but decreased quality noticeably, and so we retained the proposed values $n_{\mathrm{expo}}=2$. Matching 3DGS with $n_{\mathrm{expo}}=1$ clearly improves quality (PSNR $26.47 \rightarrow 26.59$, LPIPS $0.236 \rightarrow 0.233$), but also decreased FPS considerably ($248 \rightarrow 175)$; corresponding cells again highlighted in \colorbox{yellow!60}{yellow}.

\begin{table}[h!]
\centering
\small
\caption{Quality and performance of our method under different settings of the opacity threshold $\alpha_{\mathrm{min}}$.}
\label{tbl:alpha-min-sweep}
\begin{tabular}{@{}lc|ccccccc@{}}
\toprule
Variant & $\alpha_{\mathrm{min}}$ & PSNR$_\uparrow$ & SSIM$_\uparrow$ & LPIPS$_\downarrow$ & Opt. Time$_\downarrow$ & FPS$_\uparrow$ & Init \#G$_\downarrow$ & Final \#G$_\downarrow$ \\
\midrule
$\mathrm{GRay}_{\alpha_{\mathrm{min}}=0.001}$  & 0.001 & \colorbox{yellow!60}{26.52} & 0.820 & \colorbox{yellow!60}{0.234} & 08{:}09 & \colorbox{yellow!60}{156} & 3.27M & 1.70M \\
$\mathrm{GRay}$                               & 0.01  & \colorbox{yellow!60}{26.47} & 0.819 & \colorbox{yellow!60}{0.236} & 05{:}40 & \colorbox{yellow!60}{248} & 3.27M & 1.52M \\
% $\mathrm{GRay}_{\alpha_{\mathrm{min}}=0.0125}$& 0.0125& 26.44 & 0.818 & 0.238 & 05{:}23 & 265 & 3.27M & 1.47M \\
$\mathrm{GRay}_{\alpha_{\mathrm{min}}=0.02}$  & 0.02  & 26.35 & 0.816 & 0.242 & 04{:}43 & 313 & 3.27M & 1.32M \\
% $\mathrm{GRay}_{\alpha_{\mathrm{min}}=0.025}$ & 0.025 & 26.28 & 0.814 & 0.246 & 04{:}24 & 343 & 3.27M & 1.23M \\
$\mathrm{GRay}_{\alpha_{\mathrm{min}}=0.04}$  & 0.04  & 26.14 & 0.809 & 0.259 & 03{:}45 & 429 & 3.27M & 1.00M \\
$\mathrm{GRay}_{\alpha_{\mathrm{min}}=0.05}$  & 0.05  & 26.03 & 0.805 & 0.270 & 03{:}27 & 481 & 3.27M & 0.87M \\
\bottomrule
\end{tabular}
\end{table}

\begin{table}[h!]
\centering
\small
\caption{Quality and performance of our method under different settings of the exponent parameter $n_{\mathrm{expo}}$.}
\label{tbl:expo-sweep}
\begin{tabular}{@{}lc|ccccccc@{}}
\toprule
Variant & $n_{\mathrm{expo}}$ & PSNR$_\uparrow$ & SSIM$_\uparrow$ & LPIPS$_\downarrow$ & Opt. Time$_\downarrow$ & FPS$_\uparrow$ & Init \#G$_\downarrow$ & Final \#G$_\downarrow$ \\
\midrule
$\mathrm{GRay}_{n_{\mathrm{expo}}=0.5}$ & 0.5 & 26.56 & 0.817 & 0.243 & 10{:}33 & 95  & 3.27M & 1.50M \\
$\mathrm{GRay}_{n_{\mathrm{expo}}=1}$   & 1   & \colorbox{yellow!60}{26.59} & 0.821 & \colorbox{yellow!60}{0.233} & 07{:}05 & \colorbox{yellow!60}{175} & 3.27M & 1.58M \\
$\mathrm{GRay}$                         & 2   & \colorbox{yellow!60}{26.47} & 0.819 & \colorbox{yellow!60}{0.236} & 05{:}40 & \colorbox{yellow!60}{247} & 3.27M & 1.52M \\
$\mathrm{GRay}_{n_{\mathrm{expo}}=3}$   & 3   & 26.32 & 0.815 & 0.242 & 05{:}10 & 282 & 3.27M & 1.44M \\
$\mathrm{GRay}_{n_{\mathrm{expo}}=4}$   & 4   & 26.28 & 0.812 & 0.247 & 04{:}55 & 302 & 3.27M & 1.37M \\
\bottomrule
\end{tabular}
\end{table}

\paragraph{\textbf{BVH Update Times Become Critical Under DI}}\label{sec:bvh-updates} \deleted{While large Gaussians counts might reduce the ray tracing workload, they necessarily result in slower BVH updates since incremental updates scale linearly with the Gaussian count. While icosahedron bounds can be effective at run-time for static scenes, we found that with DI the cost of updating them at each training iteration became substantial since doing so requires transforming 12 vertices for each icosahedron.} In \Cref{tbl:bvh-update-time}  we compare the BVH incremental update times at initialization for both icosahedron bounds and emulated OBBs in 3DGRT. We reported averages over all test views and over all scenes, run $100$ times and averaged. While update times remain negligible at ${<}5\mathrm{ms}$ with SI, they grow to over $125\mathrm{ms}$ under DI becoming a substantial part of the training cost. Correspondingly, results in \cref{tbl:obb-vs-icosahedron} conclude that training with OBBs is two times faster than icosahedrons under DI. As such, we only considered OBBs for our ray tracer, even though frame rates with icosahedrons appear slightly higher.
\begin{table}[h!]
\centering
\small
\caption{Quality and performance of our method under different settings of the bin size $\psi_{bin}$ used for initial Gaussian binning.}
\label{tbl:init-binning-sweep}
\begin{tabular}{@{}lc|ccccccc@{}}
\toprule
Variant & $\psi_{\mathrm{bin}}$ & PSNR$_\uparrow$ & SSIM$_\uparrow$ & LPIPS$_\downarrow$ & Opt. Time$_\downarrow$ & FPS$_\uparrow$ & Init \#G$_\downarrow$ & Final \#G$_\downarrow$ \\
\midrule
$\mathrm{GRay}_{\psi_{\mathrm{bin}}=0.01}$  & 0.01 & 26.47 & 0.819 & 0.235 & 06{:}04 & 235 & 3.94M & 1.63M \\
$\mathrm{GRay}_{\psi_{\mathrm{bin}}=0.02}$  & 0.02 & 26.48 & 0.819 & 0.235 & 06{:}01 & 238 & 3.78M & 1.61M \\
$\mathrm{GRay}_{\psi_{\mathrm{bin}}=0.03}$  & 0.03 & 26.44 & 0.819 & 0.235 & 05{:}51 & 242 & 3.53M & 1.58M \\
$\mathrm{GRay}$                            & 0.04 & \colorbox{yellow!60}{26.45} & 0.818 & \colorbox{yellow!60}{0.237} & 05{:}40 & 247 & 
\colorbox{yellow!60}{3.27M} & 1.52M \\
$\mathrm{GRay}_{\psi_{\mathrm{bin}}=0.05}$  & 0.05 & 26.41 & 0.817 & 0.239 & 05{:}28 & 255 & 3.03M & 1.46M \\
$\mathrm{GRay}_{\psi_{\mathrm{bin}}=0.06}$  & 0.06 & 26.42 & 0.816 & 0.241 & 05{:}16 & 263 & 2.82M & 1.40M \\
$\mathrm{GRay}_{\psi_{\mathrm{bin}}=0.07}$  & 0.07 & 26.35 & 0.815 & 0.245 & 05{:}04 & 271 & 2.64M & 1.34M \\
\midrule
$\mathrm{GRay}_{\mathrm{NoBinning}}$        & --   & \colorbox{yellow!60}{26.49} & 0.819 & \colorbox{yellow!60}{0.234} & 06{:}23 & 234 & \colorbox{yellow!60}{3.97M} & 1.63M \\
\bottomrule
\end{tabular}
\end{table}

\paragraph{\textbf{Many DI Gaussians Are Useless}} \label{sec:many-gaussians-useless} \deleted{DI densely covers surfaces with Gaussians, but can result in highly redundant Gaussians when the selected reference views are similar. Accordingly, our experiments show that a large number of Gaussians can be removed with only minimal impact on quality.} \Cref{tbl:init-binning-sweep} shows the value of our binning-based merging scheme: our final setting $\psi_{\textrm{bin}}=0.04$ removes over $15\%$ of the initial Gaussians at limited cost to PSNR ($26.49 \rightarrow 26.45$) and LPIPS ($0.234 \rightarrow 0.237$); corresponding cells are highlighted in \colorbox{yellow!60}{yellow}.

\begin{table}[h!]
\centering
\small
\caption{Quality and performance of our method under different settings of the pruning threshold $\gamma_{prune}$ used for weight-based pruning.}
\label{tbl:pruning-sweep}
\begin{tabular}{@{}lc|ccccccc@{}}
\toprule
Variant & $\gamma_{\mathrm{prune}}$ & PSNR$_\uparrow$ & SSIM$_\uparrow$ & LPIPS$_\downarrow$ & Opt. Time$_\downarrow$ & FPS$_\uparrow$ & Init \#G$_\downarrow$ & Final \#G$_\downarrow$ \\
\midrule
$\mathrm{GRay}_{\gamma_{\mathrm{prune}}=10^{-8}}$  & $10^{-8}$  & 26.42 & 0.814 & 0.234 & 07{:}00 & 202 & 3.27M & 2.61M \\
$\mathrm{GRay}_{\gamma_{\mathrm{prune}}=5\times10^{-8}}$ & $5{\times}10^{-8}$ & 26.47 & 0.817 & 0.234 & 06{:}09 & 228 & 3.27M & 1.90M \\
$\mathrm{GRay}_{\gamma_{\mathrm{prune}}=7.5\times10^{-8}}$ & $7.5{\times}10^{-8}$ & 26.47 & 0.818 & 0.235 & 05{:}52 & 239 & 3.27M & 1.69M \\
$\mathrm{GRay}$                            &  $10^{-7}$ & \colorbox{yellow!60}{26.47} & 0.819 & \colorbox{yellow!60}{0.236} & 05{:}40 & 248 & 3.27M & \colorbox{yellow!60}{1.52M} \\
$\mathrm{GRay}_{\gamma_{\mathrm{prune}}=2.5\times10^{-7}}$ & $2.5{\times}10^{-7}$ & 26.39 & 0.817 & 0.249 & 04{:}46 & 291 & 3.27M & 0.97M \\
$\mathrm{GRay}_{\gamma_{\mathrm{prune}}=5\times10^{-7}}$ & $5{\times}10^{-7}$ & 26.22 & 0.809 & 0.273 & 04{:}04 & 340 & 3.27M & 0.58M \\
$\mathrm{GRay}_{\gamma_{\mathrm{prune}}=10^{-6}}$  & $10^{-6}$  & 25.75 & 0.783 & 0.321 & 03{:}26 & 414 & 3.27M & 0.28M \\
\midrule
$\mathrm{GRay}_{\mathrm{NoPruning}}$ & -- & \colorbox{yellow!60}{26.42} & 0.813 & \colorbox{yellow!60}{0.234} & 07{:}26 & 197 & 3.27M & \colorbox{yellow!60}{3.27M} \\
\bottomrule
\end{tabular}
\end{table}

\paragraph{\textbf{Aggressive Weight-Based Pruning Maintains Good Quality.}}
\deleted{High Gaussian counts can also be mitigated through more aggressive pruning.} \Cref{tbl:pruning-sweep} compares different values of our weight-based pruning threshold $\gamma_{\textrm{prune}}$. The value $\gamma_{\textrm{prune}}=10^{-7}$ reduces the Gaussian count to less than half ($3.27\text{M} \rightarrow 1.52\text{M}$) while \emph{improving} PSNR, and only worsening LPIPS by small amounts ($0.234 \rightarrow 0.236)$;  corresponding cells are again highlighted in \colorbox{yellow!60}{yellow}. This results in substantial improvement to FPS ($197 \rightarrow 248$) and to optimization times ($7{:}26 \rightarrow 5{:}40$). Our weight-based pruning significantly outperforms opacity-based pruning, which we show in \cref{additional-results}.

\begin{table}[h!]
\centering
\small
\caption{Quality and performance of our method under different settings of the scale decay parameter $\eta_{\mathrm{decay}}$. }
\label{tbl:scale-decay-sweep}
\begin{adjustbox}{width=\textwidth}
\begin{tabular}{@{}lc|ccccccc@{}}
\toprule
Variant & $\eta_{\mathrm{decay}}$ & PSNR$_\uparrow$ & SSIM$_\uparrow$ & LPIPS$_\downarrow$ & Opt. Time$_\downarrow$ & FPS$_\uparrow$ & Final \#G$_\downarrow$ & Final Avg Scale \\
\midrule
\textsc{GRay}\textsubscript{$\eta_{\mathrm{decay}}{=}0.999800$} & 0.999800 & 26.35 & 0.816 & 0.242 & 05{:}11 & 306 & 1.38\,M & 0.0148 \\
\textsc{GRay}\textsubscript{$\eta_{\mathrm{decay}}{=}0.999825$} & 0.999825 & 26.39 & 0.817 & 0.240 & 05{:}20 & 286 & 1.43\,M & 0.0153 \\
\textsc{GRay}\textsubscript{$\eta_{\mathrm{decay}}{=}0.999850$} & 0.999850 & 26.41 & 0.818 & 0.238 & 05{:}29 & 267 & 1.47\,M & 0.0159 \\
\textsc{GRay} & 0.999875 & \colorbox{cyan!30}{26.47} & 0.819 & \colorbox{cyan!30}{0.236} & 05{:}40 & \colorbox{yellow!60}{248} & \colorbox{yellow!60}{1.52\,M} & \colorbox{yellow!60}{0.0166} \\
\textsc{GRay}\textsubscript{$\eta_{\mathrm{decay}}{=}0.999900$}  & 0.999900 & 26.52 & 0.819 & 0.235 & 05{:}51 & 229 & 1.57\,M & 0.0173 \\
\textsc{GRay}\textsubscript{$\eta_{\mathrm{decay}}{=}0.999925$} & 0.999925 & 26.50 & 0.820 & 0.234 & 06{:}04 & 209 & 1.61\,M & 0.0180 \\
\textsc{GRay}\textsubscript{$\eta_{\mathrm{decay}}{=}0.999950$} & 0.999950 & 26.53 & 0.821 & 0.232 & 06{:}14 & 192 & 1.65\,M & 0.0190 \\
\midrule
\textsc{GRay}$_{\mathrm{NoDecay}}$ & -- & \colorbox{cyan!30}{26.60} & 0.822 & \colorbox{cyan!30}{0.231} & 06{:}43 & \colorbox{yellow!60}{157} & \colorbox{yellow!60}{1.73\,M} & \colorbox{yellow!60}{0.0211} \\
\bottomrule
\end{tabular}
\end{adjustbox}
\end{table}

\paragraph{\textbf{Scale Decay Efficiently Trades Quality for Speed}} \label{sec:smaller-initial-gaussians} \changed{Results for the scale decay parameter $\eta_{\mathrm{decay}}$ are presented in \cref{tbl:scale-decay-sweep}.} Increasing decay improved FPS consistently and by large amounts ($\textsc{GRay}_{\mathrm{NoDecay}}@157 \rightarrow \textsc{GRay}@248$), which is explained by both reduced final Gaussian average scales ($0.0211 \rightarrow 0.0166$) and an increase in pruning ($1.73\textrm{M} \rightarrow 1.52\textrm{M}$ final Gaussian count); corresponding cells highlighted in \colorbox{yellow!60}{yellow}. However, decay also worsens quality slightly \changed{(cells highlighted in \colorbox{cyan!30}{blue})}. \deleted{Since this reduction is fairly minor (cells highlighted in \colorbox{cyan!30}{blue}), we conclude that scale decay is a highly effective lever for balancing quality and speed.}  

\paragraph{\textbf{Fewer Iterations can Still Converge Fully}}\label{sec:training-in-fewer-iterations}  \deleted{One of EDGS's main claims is that DI converges more rapidly than SI~\cite{edgs}. Yet, training at reduced iteration counts can lead to reduced quality.} \Cref{tbl:reduced-iter-count} shows that \changed{it is possible} to train at $15\mathrm{K}$ iterations with slightly better LPIPS score than $30\mathrm{K}$ iterations ($0.243 \rightarrow 0.236$) and only a slight decrease in PSNR ($26.51 \rightarrow 26.47$), leading to halved training times ($10{:}24 \rightarrow 05{:}40$) almost for free; corresponding cells highlighted in \colorbox{yellow!60}{yellow}. We achieve this with the additional learning rate schedules described in \cref{implementation-details}, which allow rapid initial growth that compensates for the reduced iteration count. \deleted{Notice how variants without these schedules reach significantly worse LPIPS but better FPS; this is likely because the Gaussians do not have enough iterations to scale up sufficiently to minimize loss, resulting in small Gaussians that render fast without modeling the scene well. }
\begin{table}[h!]
\centering
\small
\caption{Quality and performance of our method at reduced iteration counts, both with and without modified learning rate schedules.}
\label{tbl:reduced-iter-count}
\begin{adjustbox}{width=\textwidth}
\begin{tabular}{@{}lcc|ccccccc@{}}
\toprule
Variant & Iterations & Extra LR Schedules & PSNR$_\uparrow$ & SSIM$_\uparrow$ & LPIPS$_\downarrow$ & Opt. Time$_\downarrow$ & FPS$_\uparrow$ & Final Avg Scale $\downarrow$ \\
\midrule
\textsc{GRay}\textsubscript{30K+NoSchedules} & 30000 & \ding{55} & 25.97 & 0.809 & 0.251 & 07{:}17 & 396 & 0.0140 \\
\textsc{GRay}\textsubscript{30K}             & 30000 & \ding{51} & \colorbox{yellow!60}{26.51} & 0.817 & \colorbox{yellow!60}{0.243} & \colorbox{yellow!60}{10{:}24} & 280 & 0.0186 \\
\textsc{GRay}\textsubscript{15K+NoSchedules} & 15000 & \ding{55} & 25.71 & 0.806 & 0.251 & 04{:}06 & 356 & 0.0118 \\
\textsc{GRay}                               & 15000 & \ding{51} & \colorbox{yellow!60}{26.47} & 0.819 & \colorbox{yellow!60}{0.236} & \colorbox{yellow!60}{05{:}40} & 248 & 0.0166 \\
\textsc{GRay}\textsubscript{7.5K+NoSchedules} & 7500 & \ding{55} & 25.21 & 0.799 & 0.261 & 02{:}12 & 329 & 0.0111 \\
\textsc{GRay}\textsubscript{7.5K}             & 7500 & \ding{51} & 26.23 & 0.817 & 0.238 & 02{:}58 & 229 & 0.0153 \\
\bottomrule
\end{tabular}
\end{adjustbox}
\end{table}

\paragraph{\textbf{DHT Stabilizes Early Ray Termination During Training}} \label{sec:early-ray-termination} 

We ran our method with different levels of early ray termination with and without DHT; results are presented in \cref{tbl:dht}. \deleted{Our method trains stably at high transmittance thresholds ($\tau_{\mathrm{min}}{=}0.03$) while skipping the accumulation of $40\%$ of hit Gaussians.} \changed{Our method} reaches slightly better LPIPS than when early termination is disabled ($\tau_{\mathrm{min}} = 0.0$); corresponding cells are highlighted in \colorbox{yellow!60}{yellow}. Since these Gaussians are further skipped during the backward pass, optimization time drops significantly ($6{:}33 \rightarrow 5{:}40$). Finally, running our method with DHT disabled shows rapid deterioration of quality metrics when truncation levels increase. For instance, \textsc{GRay}$_{\mathrm{NoDHT}+\tau_{\mathrm{min}}{=}0.03}$ reaches catastrophically low PSNR (24.54) and LPIPS (0.284); corresponding cells are highlighted in \colorbox{cyan!30}{blue}. Certain runs at high thresholds even fail to converge due to rapid Gaussian expansions; see \cref{additional-analysis}.

\begin{table}[h!]
\centering
\small
\caption{Quality and performance of our method at different early ray termination transmittance thresholds $\tau_{\mathrm{min}}$ with and without DHT. Empty rows indicate failed training runs in at least one of the scenes. }
\label{tbl:dht}
\begin{tabular}{@{}lc c|cccccc@{}}
\toprule
Variant & $\tau_{\mathrm{min}}$ & DHT & PSNR$_\uparrow$ & SSIM$_\uparrow$ & LPIPS$_\downarrow$ & Opt Time$_\downarrow$ & FPS$_\uparrow$ & \% Skipped \\
\midrule
\textsc{GRay}$_{\tau_{\mathrm{min}}{=}0.001}$ & 0.001 & \ding{51} & 26.46 & 0.819 & 0.238 & 06{:}07 & 257 & 17.09 \\
\textsc{GRay}$_{\tau_{\mathrm{min}}{=}0.01}$  & 0.01  & \ding{51} & 26.49 & 0.819 & 0.236 & 05{:}51 & 255 & 29.38 \\
\textsc{GRay}$_{\tau_{\mathrm{min}}{=}0.02}$  & 0.02  & \ding{51} & 26.45 & 0.818 & 0.236 & 05{:}45 & 251 & 35.85 \\
\textsc{GRay} & 0.03 & \ding{51} & 26.47 & 0.819 & \colorbox{yellow!60}{0.236} & \colorbox{yellow!60}{05{:}40} & 248 & \colorbox{yellow!60}{40.28} \\
\textsc{GRay}$_{\tau_{\mathrm{min}}{=}0.04}$  & 0.04  & \ding{51} & 26.42 & 0.818 & 0.237 & 05{:}36 & 245 & 43.74 \\
\textsc{GRay}$_{\tau_{\mathrm{min}}{=}0.05}$  & 0.05  & \ding{51} & 26.37 & 0.817 & 0.239 & 05{:}33 & 243 & 46.50 \\
\textsc{GRay}$_{\tau_{\mathrm{min}}{=}0.1}$   & 0.1   & \ding{51} & 26.05 & 0.808 & 0.255 & 05{:}15 & 242 & 55.45 \\
\midrule
\textsc{GRay}$_{\mathrm{NoDHT}+\tau_{\mathrm{min}}{=}0.001}$ & 0.001 & \ding{55} & 26.48 & 0.819 & 0.237 & 06{:}00 & 269 & 17.08 \\
\textsc{GRay}$_{\mathrm{NoDHT}+\tau_{\mathrm{min}}{=}0.01}$  & 0.01  & \ding{55} & 26.36 & 0.818 & 0.239 & 05{:}44 & 249 & 35.47 \\
\textsc{GRay}$_{\mathrm{NoDHT}+\tau_{\mathrm{min}}{=}0.02}$  & 0.02  & \ding{55} & 25.60 & 0.809 & 0.257 & 05{:}38 & 222 & 47.85 \\
\textsc{GRay}$_{\mathrm{NoDHT}+\tau_{\mathrm{min}}{=}0.03}$  & 0.03  & \ding{55} & \colorbox{cyan!30}{24.54} & 0.793 & \colorbox{cyan!30}{0.284} & 05{:}34 & 210 & 54.53 \\
\textsc{GRay}$_{\mathrm{NoDHT}+\tau_{\mathrm{min}}{=}0.04}$  & 0.04  & \ding{55} & - & - & - & - & - & - \\
\textsc{GRay}$_{\mathrm{NoDHT}+\tau_{\mathrm{min}}{=}0.05}$  & 0.05  & \ding{55} & 22.30 & 0.752 & 0.340 & 05{:}29 & 195 & 63.10 \\
\textsc{GRay}$_{\mathrm{NoDHT}+\tau_{\mathrm{min}}{=}0.1}$   & 0.1   & \ding{55} & - & - & - & - & - & - \\
\midrule
\textsc{GRay}$_{\mathrm{NoEarlyTerm.}}$ & 0.0 & - & 26.46 & 0.819 & \colorbox{yellow!60}{0.241} & \colorbox{yellow!60}{06{:}33} & 248 & \colorbox{yellow!60}{0.00} \\
\bottomrule
\end{tabular}
\end{table}

\deleted{
\paragraph{\textbf{PPLLs Still Outmatch Repeated Traversals Under DI}}\label{sec:pplls-outmatch} Finally, we validated that PPLLs are still worth using since DI reduces the required number of BVH traversals when using 3DGRT's Gaussian collection strategy. \Cref{tbl:ppll-vs-repeated-traversals} shows that PPLLs are still much faster even in the DI regime, increasing frame rates by about $40\%$ ($188 \rightarrow 248$). Note that we disabled DHT for repeated traversals since the list of unprocessed Gaussians isn't readily available; implementing DHT for it would require an additional traversal, further decreasing speed. Also note that our method uses PPLLs for the backward pass as well; although we did not measure it, the performance difference during training is likely larger than at inference. Hence beyond DI and our modified training regimen we attribute most of the remaining performance difference between GRay and 3DGRT to PPLLs.
}

\begin{table}[h!]
\deleted{\centering
\small
\caption{\deleted{Comparing the frame rate of our method on pretrained scenes with multiple BVH traversals vs. per-pixel linked lists for Gaussian collection.}}
\label{tbl:ppll-vs-repeated-traversals}
\begin{tabular}{@{}l c|cccccc@{}}
\toprule
Variant & {Collection Method} & FPS$_\uparrow$ \\
\midrule
\textsc{GRay}\textsubscript{MultipleTraversal} & {Multiple BVH Traversals} & 188 \\
\textsc{GRay}                         & {Per-Pixel Linked Lists}                & \textbf{248} \\
\bottomrule
\end{tabular}}
\end{table}

\section{Additional Results}\label{additional-results}

This section contains additional results that were not presented in the main text.

\subsection{Balancing Quality And Speed}\label{sec:can-gray-match-3dgs}

Through the accumulation of techniques that increase speed with a minimal cost to quality, we were able to obtain ray-tracing optimization times comparable to 3DGS while maintaining better LPIPS. But how fast can we go, and how slow are we without these techniques? This subsection compares both extremes of the quality/speed tradeoff. Due to sometimes high memory requirements, these experiments were run at half resolution.

\begin{table}[h!]
\caption{Quality and performance of our ray tracer at different configuration presets (low to high quality). \textbf{Experiments run at half resolution.}  }
\label{tbl:quality-speed-presets}
\centering
\begin{adjustbox}{width=\linewidth}
\begin{tabular}{@{}lcc|cccccccc@{}}
\toprule
Method & Resolution & \#Iterations & PSNR$_\uparrow$ & SSIM$_\uparrow$ & LPIPS$_\downarrow$ & Init Time$_\downarrow$ & Opt Time$_\downarrow$ & FPS$_\uparrow$ & Init \#G$_\downarrow$ & Final \#G$_\downarrow$ \\
\midrule
$\mathrm{GRay}_{\mathrm{LQ}}$
& 1/2 & 7500  & 26.23 & 0.839 & 0.191 & 01{:}58 & 00{:}46 & 1051 & 3.27M & 0.48M \\

$\mathrm{GRay}$
& 1/2 & 15000 & 27.06 & 0.852 & 0.160 & 01{:}58 & 03{:}15 & 519 & 3.27M & 1.47M \\

$\mathrm{GRay}_{\mathrm{HQ}}$
& 1/2 & 30000 & 27.41 & 0.847 & 0.154 & 01{:}58 & 36{:}11 & 55 & 3.97M & 3.79M \\

\midrule 

$\mathrm{3DGS}$  
    & 1/2 & 30000 & 28.02 & 0.875 & 0.166
    & 00{:}00 & 04:22 & 399
    & 0.11M & 1.88M \\

$\mathrm{3DGS}_{\mathrm{DI}}$  
    & 1/2 & 30000 & 27.83 & 0.866 & 0.146
    & 00{:}00 & 09{:}13 & 260
    & 0.11M & 3.23M \\

\bottomrule
\end{tabular}
\end{adjustbox}
\end{table}

\paragraph{\textbf{Low-Quality Ray Tracing}}\label{sec:low-quality-rt} 
We further speed up our method by increasing the alpha threshold to $\alpha_{\mathrm{min}} = 0.02$ and the generalized exponent to $n_{\mathrm{expo}}=3$, increasing the pruning threshold to $\gamma_{\mathrm{prune}} = 5{\times}10^{-7}$, increasing the scale decay to $\eta_{\mathrm{decay}}=0.9998$, and enable the SH stepping laziness at $K_{\mathrm{lazy}} = 12$ (described in \cref{sec:cuda-implementation}). Finally, we further halve training iterations to $7500$. We dub this version $\mathrm{GRay}_{\mathrm{LQ}}$. Results in \cref{tbl:quality-speed-presets} show that$\mathrm{GRay}_{\mathrm{LQ}}$ can train in under a minute and more than double 3DGS's frame rate, at the cost of significantly reduced quality.
 
\paragraph{\textbf{High-Quality Ray Tracing}}\label{sec:high-quality-rt}  
To maximize quality, we lower the alpha threshold to $\alpha_{\mathrm{min}} = 0.001$, the generalized exponent to $n_{\mathrm{expo}}=1$ (matching 3DGS), and set $\tau_{\min} = 0.001$ (matching 3DGRT), disable initialization binning, and lower the pruning threshold to $\gamma_{\mathrm{prune}} = 10^{-9}$ (which nearly disables it), and finally disable scale decay entirely. We dub this version $\mathrm{GRay}_{\mathrm{HQ}}$. Results in \cref{tbl:quality-speed-presets} show that$\mathrm{GRay}_{\mathrm{HQ}}$ can further improve quality over $\mathrm{GRay}$ and approach $3DGS_{\mathrm{DI}}$ without reaching it. In principle, it should be possible to obtain near-identical quality; we leave resolving this as future work.

\begin{table}[h!]
\centering
\small
\caption{Quality and performance of our method with opacity-based pruning, under different settings of the pruning threshold $\gamma^{\mathrm{prune}}_{\mathrm{opacity}}$ compared to our weight-based final solution.}
\label{tbl:opacity-pruning-sweep}
\begin{tabular}{@{}lcccccccc@{}}
\toprule
Variant & $\gamma^{\mathrm{prune}}_{\mathrm{opacity}}$ & PSNR$_\uparrow$ & SSIM$_\uparrow$ & LPIPS$_\downarrow$ & Opt Time$_\downarrow$ & FPS$_\uparrow$ & Init \#G$_\downarrow$ & Final \#G$_\downarrow$ \\
\midrule
$\mathrm{GRay}_{\gamma^{\mathrm{prune}}_{\mathrm{opacity}=0.400}}$ & 0.400 & 21.78 & 0.599 & 0.545 & 02{:}34 & 838 & 3.27M & 0.01M \\
$\mathrm{GRay}_{\gamma^{\mathrm{prune}}_{\mathrm{opacity}=0.200}}$ & 0.200 & 25.65 & 0.782 & 0.329 & 03{:}07 & 449 & 3.27M & 0.32M \\
$\mathrm{GRay}_{\gamma^{\mathrm{prune}}_{\mathrm{opacity}=0.150}}$ & 0.150 & 26.24 & 0.811 & 0.272 & 04{:}07 & 337 & 3.27M & 0.85M \\
$\mathrm{GRay}_{\gamma^{\mathrm{prune}}_{\mathrm{opacity}=0.125}}$ & 0.125 & 26.34 & 0.816 & 0.254 & 04{:}47 & 291 & 3.27M & 1.24M \\
$\mathrm{GRay}_{\gamma^{\mathrm{prune}}_{\mathrm{opacity}=0.100}}$ & 0.100 & 26.40 & 0.818 & 0.242 & 05{:}32 & 254 & 3.27M & 1.69M \\
$\mathrm{GRay}_{\gamma^{\mathrm{prune}}_{\mathrm{opacity}=0.075}}$ & 0.075 & 26.45 & 0.818 & \colorbox{yellow!60}{0.235} & \colorbox{yellow!60}{06{:}16} & \colorbox{yellow!60}{225} & 3.27M & 2.18M \\
$\mathrm{GRay}_{\gamma^{\mathrm{prune}}_{\mathrm{opacity}=0.050}}$ & 0.050 & 26.44 & 0.816 & 0.233 & 06{:}52 & 206 & 3.27M & 2.65M \\
$\mathrm{GRay}_{\gamma^{\mathrm{prune}}_{\mathrm{opacity}=0.025}}$ & 0.025 & 26.42 & 0.813 & 0.233 & 07{:}16 & 194 & 3.27M & 3.06M \\
\midrule
$\mathrm{GRay}$ & -- & 26.47 & 0.819 & \colorbox{yellow!60}{0.236} & \colorbox{yellow!60}{05{:}40} & \colorbox{yellow!60}{248} & 3.27M & 1.52M \\
\bottomrule
\end{tabular}
\end{table}

\subsection{Comparison to Opacity-Based Pruning}

We compared our weight-based pruning to a standard opacity-based baseline, where $\gamma^{\mathrm{prune}}_{\mathrm{opacity}}$ denotes the opacity pruning threshold. Note that in our method higher opacity pruning thresholds are required than with 3DGS to get effective pruning, ostensibly because we do not use opacity resets. As such we ran a sweep over many different settings; results are presented in \cref{tbl:opacity-pruning-sweep}. An LPIPS score comparable to our final solution is attained at  $\gamma^{\mathrm{prune}}_{\mathrm{opacity}=0.075}$, but this results in significantly slower training ($05{:}40 \rightarrow 06{:}16$) and lower final FPS ($248 \rightarrow 225$); corresponding cells are highlighted in \colorbox{yellow!60}{yellow}. Pruning more aggressively with $\gamma^{\mathrm{prune}}_{\mathrm{opacity}=0.100}$ results in similar training times, but a noticeable drop in quality. In short, weight-based pruning appears superior to opacity-based pruning for our method.

\subsection{Variability}
To verify the stability and reproducibility of our evaluation pipeline, we ran our method 3 additional times with identical hyperparameters. Results reported in \cref{tbl:gray-repro} indicate minimal variability.

\begin{table}[h!]
\centering
\small
\caption{Three additional independent runs of our method.}
\label{tbl:gray-repro}
\begin{tabular}{@{}lccccc@{}}
\toprule
Method & PSNR$_\uparrow$ & SSIM$_\uparrow$ & LPIPS$_\downarrow$ & Opt Time$_\downarrow$ & FPS$_\uparrow$ \\
\midrule
	\textsc{GRay} (extra run 1) & 26.45 & 0.818 & 0.237 & 05{:}40 & 247 \\
	\textsc{GRay} (extra run 2) & 26.48 & 0.819 & 0.236 & 05{:}41 & 247 \\
	\textsc{GRay} (extra run 3) & 26.45 & 0.818 & 0.236 & 05{:}38 & 247 \\
\bottomrule
\end{tabular}
\end{table}

\section{Implementation Details}\label{implementation-details}

This section clarifies implementation and hyperparameters.

\subsection{\changed{CUDA Implementation Differences}}\label{sec:cuda-implementation} 
\changed{As compared to EPBRR, we do not fuse the forward and backward pass to allow for the SSIM loss and easy addition of other custom losses. We also use a CUDA-side Adam optimizer that launches a separate thread for every spherical harmonic coefficient instead of just launching one per Gaussian.}

\changed{We use the fused SSIM loss proposed in Taming 3DGS~\cite{taming-3dgs} as do most existing works. In experiments in \cref{sec:low-quality-rt}, we also use the lazy stepping for non-DC Spherical Harmonic (SH) bands they propose, although we leave it disabled by default; this technique consists of only taking optimization steps every $K_{\mathrm{lazy}}$ iterations (Taming 3DGS uses $K_{\mathrm{lazy}} = 12$).}

\subsection{Learning Rate Schedules}\label{sec:lr-schedules}
For the Gaussian mean parameters, 3DGS employs an exponential (log-linear) learning-rate schedule of the form
$$
\exp\big((1-t)\log(\lambda_{\mathrm{init}}) + t\log(\lambda_{\mathrm{final}})\big),
$$
where t increases linearly from $0$ at the start of training to $1$ at iteration $30000$ and where $\lambda_{\mathrm{init}}$ is the learning rate's initial value and $\lambda_{\mathrm{final}}$ its final value. We first adjusted this schedule so that $t=1$ is reached at the final training iteration regardless of its value; for example, at iteration $15000$ when training for $15000$ iterations. We further introduce analogous schedules for the Gaussian scale ($0.02 \rightarrow 0.005$), rotation ($0.004 \rightarrow 0.001$), and SH DC coefficients ($0.04 \rightarrow 0.0025$); in each of these cases, the final learning rates were left unchanged, while the initial rates were increased.

\subsection{Additional Hyperparameter Details}\label{sec:additional-hyperparams}
\changed{We opted for an initial opacity of 0.1, matching 3DGS rather than the configuration proposed by EDGS (0.5). We used 0.5 for other baselines under dense initialization.} Our method does not prune large Gaussians based on their size, and does not feature opacity resets. We briefly experimented with adding world-space Gaussian pruning to our method, but this resulted in lower quality metrics. As such, when running dense initialization for 3DGS, we also opted to exclude this form of pruning, which also matches the configuration from 3DGRT. Note however that EDGS further excludes opacity resets, and includes additional hyperparameter changes: an opacity decay (scaling all Gaussian opacities down by $0.99$ every $10$ iterations) and a modified positional learning rate (clamping its value to at most the value obtained at step $8000$). We did not incorporate these changes in any other method. We briefly experimented with disabling opacity resets for 3DGS$_{\text{DI}}$ but this did not seem to affect quality much. The exact configuration used for all methods is clarified in \cref{tbl:method-features}. 

\begin{table}[h!]
\centering
\caption{Clarification of additional hyperparameter settings across all methods.}
\label{tbl:method-features}
% \begin{adjustbox}{width=\linewidth}
\begin{tabular}{@{}lcccccc@{}}
\toprule
 & Ours & 3DGS$_{\text{SI}}$ & 3DGS$_{\text{DI}}$ & EDGS & 3DGRT$_{\text{SI}}$ & 3DGRT$_{\text{DI}}$ \\
\midrule
Viewport-size Gaussian pruning        & \ding{55} & \ding{51} & \ding{55} & \ding{55} & \ding{55} & \ding{55} \\
World-size Gaussian pruning           & \ding{55} & \ding{51} & \ding{55} & \ding{55} & \ding{55} & \ding{55} \\
Opacity resets                         & \ding{55} & \ding{51} & \ding{51} & \ding{55} & \ding{51} & \ding{51} \\
EDGS's modified LR + opacity decay       & \ding{55} & \ding{55} & \ding{55} & \ding{51} & \ding{55} & \ding{55} \\
\bottomrule
\end{tabular}
% \end{adjustbox}
\end{table}

As a sanity check, we also ran 3DGS with dense initialization and EDGS's additional changes. Results are in \cref{tbl:match-edgs}; $\cdot_{\mathrm{MatchEDGS}}$ indicates such a run with these changes and opacity resets disabled such as to match EDGS closely. When they are included, our port of 3DGS with dense initialization ($\mathrm{3DGS}_{\mathrm{DI}+\mathrm{MatchEDGS}}$) obtains quality metrics very close to EDGS.

\begin{table}[h!]
\caption{Reconstruction quality and performance of 3DGS with and without matching EDGS's additional modifications.}
\label{tbl:match-edgs}
\centering
\begin{adjustbox}{width=\linewidth}
\begin{tabular}{@{}lccccccccc@{}}
\toprule
Method & PSNR$_\uparrow$ & SSIM$_\uparrow$ & LPIPS$_\downarrow$ & Init Time$_\downarrow$ & Opt Time$_\downarrow$ & FPS$_\uparrow$ & Init \#G$_\downarrow$ & Final \#G$_\downarrow$ & \#Iterations \\
\midrule
$\mathrm{3DGS}_{\mathrm{SI}}$          & 27.10 & 0.831 & 0.262 & 00{:}00 & 06{:}18 & 253 & 0.11M  & 2.25M  & 30000 \\
$\mathrm{3DGS}_{\mathrm{DI}}$          & 26.97 & 0.827 & 0.226 &  01{:}58 & 10{:}09 & 241 & 3.97M  & 3.28M  & 30000 \\
$\mathrm{3DGS}_{\mathrm{SI}+\mathrm{MatchEDGS}}$ & 27.40 & 0.830 & 0.276 & 00{:}00 & 05{:}03 & 299 & 0.11M  & 1.41M  & 30000 \\
$\mathrm{3DGS}_{\mathrm{DI}+\mathrm{MatchEDGS}}$ & 27.45 & 0.849 & 0.211 & 01{:}58 & 07{:}47 & 311 & 3.97M & 1.72M  & 30000 \\
$\mathrm{EDGS}_{\mathrm{RegularAdam}}$  & 27.55 & 0.850 & 0.204 & 00{:}00 & 30{:}38 & 173 & 3.98M & 2.26M & 30000 \\
\bottomrule
\end{tabular}
\end{adjustbox}
\end{table}

\begin{table}[h]
\centering
\caption{Near plane distance ($z_{\mathrm{near}}$) used for each scene.}
\label{tbl:znear-values}
\begin{tabular}{lc}
\toprule
Scene & $z_{\mathrm{near}}$ \\
\midrule
\textsc{bicycle}    & 2.00 \\
\textsc{bonsai}     & 3.00 \\
\textsc{counter}    & 1.00 \\
\textsc{drjohnson}  & 2.00 \\
\textsc{flowers}    & 1.00 \\
\textsc{garden}     & 2.00 \\
\textsc{kitchen}    & 2.00 \\
\textsc{playroom}   & 5.00 \\
\textsc{room}       & 1.00 \\
\textsc{stump}      & 2.50 \\
\textsc{train}      & 0.65 \\
\textsc{treehill}   & 3.25 \\
\textsc{truck}      & 1.45 \\
\bottomrule
\end{tabular}
\end{table}

\subsection{Floater Masking with Near Clipping Planes}
Dense initialization is floater-prone, and to alleviate this, we used near clipping planes to reduce their visual impact in rendered videos. We applied the same clipping planes to all baselines and our method. The near clipping planes $z_{\mathrm{near}}$ values we used are listed in \cref{tbl:znear-values}.  To implement these for both ray tracers, we simply started tracing further along the initial ray, while for splatting, we culled based if the centroid was closer than $z_{\mathrm{near}}$ which can result in some negligible differences. 

Please bear in mind that we did not use clipping planes when evaluating quality: all metrics throughout the paper were measured without them. Measuring PSNR with clipping planes offers negligible improvements ($26.91 \rightarrow 26.93$ for \textsc{TreeHill} at $z_{\mathrm{near}}=1.0$; $21.89 \rightarrow 21.94$ for \textsc{Garden} at $z_{\mathrm{near}}=2.0$).

\section{Detailed Results}
This section contains detailed results for all individual scenes for our method in \cref{tbl:per-scene}, for 3DGRT with SI in \cref{tbl:per-scene-3dgrt-si}, with DI in \cref{tbl:per-scene-3dgrt-di}, for 3DGS with SI in \cref{tbl:per-scene-3dgs-si}, and with DI in \cref{tbl:per-scene-3dgs-di}.

\vspace{1em}
\begin{table}[h!]
\centering
\small
\caption{Per-scene results of \textsc{GRay}.}
\label{tbl:per-scene}
\begin{adjustbox}{width=0.8\linewidth}
\begin{tabular}{@{}lccccccc@{}}
\toprule
Scene & PSNR$_\uparrow$ & SSIM$_\uparrow$ & LPIPS$_\downarrow$ & Opt Time$_\downarrow$ & FPS$_\uparrow$ & Start \#G$_\downarrow$ & Final \#G$_\downarrow$ \\
\midrule
\textsc{Bicycle}   & 24.38 & 0.730 & 0.233 & 05{:}38 & 263 & 3.60M & 2.22M \\
\textsc{Bonsai}    & 31.68 & 0.942 & 0.202 & 08{:}00 & 126 & 3.93M & 1.83M \\
\textsc{Counter}   & 28.76 & 0.910 & 0.213 & 08{:}15 & 108 & 3.22M & 1.30M \\
\textsc{Drjohnson} & 28.65 & 0.886 & 0.321 & 04{:}52 & 345 & 3.87M & 1.33M \\
\textsc{Flowers}   & 20.01 & 0.536 & 0.357 & 05{:}24 & 247 & 2.96M & 1.69M \\
\textsc{Garden}    & 26.91 & 0.858 & 0.118 & 05{:}47 & 241 & 3.21M & 2.05M \\
\textsc{Kitchen}   & 30.59 & 0.926 & 0.141 & 07{:}29 & 151 & 2.04M & 1.22M \\
\textsc{Playroom}  & 29.87 & 0.898 & 0.279 & 04{:}31 & 360 & 3.65M & 1.24M \\
\textsc{Room}      & 30.26 & 0.921 & 0.246 & 06{:}17 & 214 & 4.22M & 1.28M \\
\textsc{Stump}     & 26.35 & 0.771 & 0.234 & 05{:}12 & 262 & 2.37M & 1.78M \\
\textsc{Train}     & 21.04 & 0.798 & 0.238 & 03{:}27 & 321 & 3.40M & 0.93M \\
\textsc{Treehill}  & 21.89 & 0.619 & 0.333 & 05{:}00 & 250 & 2.42M & 1.62M \\
\textsc{Truck}     & 23.77 & 0.850 & 0.157 & 03{:}46 & 331 & 3.60M & 1.27M \\
\bottomrule
\end{tabular}
\end{adjustbox}
\end{table}

\vspace{1em}
\begin{table}[h!]
\centering
\small
\caption{Per-scene results of \textsc{3DGRT}\textsubscript{SI}.}
\label{tbl:per-scene-3dgrt-si}
\begin{adjustbox}{width=0.8\linewidth}
\begin{tabular}{@{}lccccccc@{}}
\toprule
Scene & PSNR$_\uparrow$ & SSIM$_\uparrow$ & LPIPS$_\downarrow$ & Opt Time$_\downarrow$ & FPS$_\uparrow$ & Start \#G$_\downarrow$ & Final \#G$_\downarrow$ \\
\midrule
\textsc{Bicycle}   & 24.61 & 0.741 & 0.260 & 56{:}30 & 63 & 0.05M & 5.62M \\
\textsc{Bonsai}    & 31.62 & 0.940 & 0.239 & 61{:}02 & 72 & 0.21M & 1.47M \\
\textsc{Counter}   & 28.74 & 0.907 & 0.244 & 64{:}43 & 53 & 0.16M & 1.26M \\
\textsc{Drjohnson} & 29.24 & 0.900 & 0.317 & 66{:}21 & 46 & 0.08M & 2.05M \\
\textsc{Flowers}   & 21.50 & 0.613 & 0.340 & 49{:}36 & 78 & 0.04M & 4.36M \\
\textsc{Garden}    & 26.68 & 0.849 & 0.140 & 51{:}33 & 64 & 0.14M & 4.06M \\
\textsc{Kitchen}   & 30.71 & 0.924 & 0.155 & 72{:}43 & 47 & 0.24M & 1.72M \\
\textsc{Playroom}  & 30.25 & 0.904 & 0.332 & 54{:}24 & 59 & 0.04M & 1.22M \\
\textsc{Room}      & 30.55 & 0.916 & 0.281 & 60{:}36 & 63 & 0.11M & 1.25M \\
\textsc{Stump}     & 26.22 & 0.767 & 0.257 & 53{:}20 & 65 & 0.03M & 6.10M \\
\textsc{Train}     & 21.21 & 0.812 & 0.240 & 31{:}31 & 114 & 0.18M & 2.34M \\
\textsc{Treehill}  & 22.38 & 0.625 & 0.377 & 58{:}15 & 56 & 0.05M & 7.35M \\
\textsc{Truck}     & 24.27 & 0.867 & 0.176 & 34{:}37 & 98 & 0.14M & 3.27M \\
\bottomrule
\end{tabular}
\end{adjustbox}
\end{table}

\vspace{1em}
\begin{table}[h!]
\centering
\small
\caption{Per-scene results of \textsc{3DGRT}\textsubscript{DI}.}
\label{tbl:per-scene-3dgrt-di}
\begin{adjustbox}{width=0.8\linewidth}
\begin{tabular}{@{}lccccccc@{}}
\toprule
Scene & PSNR$_\uparrow$ & SSIM$_\uparrow$ & LPIPS$_\downarrow$ & Opt Time$_\downarrow$ & FPS$_\uparrow$ & Start \#G$_\downarrow$ & Final \#G$_\downarrow$ \\
\midrule
\textsc{Bicycle}   & 24.72 & 0.762 & 0.208 & 31{:}57 & 140 & 4.22M & 2.79M \\
\textsc{Bonsai}    & 31.46 & 0.940 & 0.195 & 56{:}27 & 77 & 4.50M & 3.18M \\
\textsc{Counter}   & 28.40 & 0.907 & 0.204 & 59{:}09 & 68 & 4.49M & 2.95M \\
\textsc{Drjohnson} & 28.47 & 0.880 & 0.310 & 35{:}56 & 122 & 4.09M & 2.68M \\
\textsc{Flowers}   & 20.79 & 0.605 & 0.316 & 31{:}39 & 171 & 3.68M & 2.16M \\
\textsc{Garden}    & 26.79 & 0.856 & 0.119 & 40{:}05 & 110 & 4.02M & 2.93M \\
\textsc{Kitchen}   & 30.37 & 0.925 & 0.139 & 69{:}22 & 58 & 4.50M & 3.41M \\
\textsc{Playroom}  & 29.60 & 0.882 & 0.287 & 36{:}44 & 113 & 3.70M & 2.82M \\
\textsc{Room}      & 30.22 & 0.917 & 0.240 & 56{:}27 & 74 & 4.49M & 3.30M \\
\textsc{Stump}     & 26.35 & 0.766 & 0.236 & 26{:}35 & 164 & 2.70M & 2.14M \\
\textsc{Train}     & 20.80 & 0.801 & 0.242 & 20{:}29 & 281 & 3.84M & 1.19M \\
\textsc{Treehill}  & 21.79 & 0.611 & 0.332 & 28{:}39 & 158 & 3.01M & 2.23M \\
\textsc{Truck}     & 24.28 & 0.866 & 0.141 & 27{:}31 & 172 & 4.38M & 2.50M \\
\bottomrule
\end{tabular}
\end{adjustbox}
\end{table}

\vspace{1em}
\begin{table}[h!]
\centering
\small
\caption{Per-scene results of \textsc{3DGS}\textsubscript{SI}.}
\label{tbl:per-scene-3dgs-si}
\begin{adjustbox}{width=0.8\linewidth}
\begin{tabular}{@{}lccccccc@{}}
\toprule
Scene & PSNR$_\uparrow$ & SSIM$_\uparrow$ & LPIPS$_\downarrow$ & Opt Time$_\downarrow$ & FPS$_\uparrow$ & Start \#G$_\downarrow$ & Final \#G$_\downarrow$ \\
\midrule
\textsc{Bicycle}   & 25.10 & 0.752 & 0.263 & 08{:}17 & 171 & 0.05M & 4.00M \\
\textsc{Bonsai}    & 32.07 & 0.940 & 0.249 & 05{:}02 & 311 & 0.21M & 1.17M \\
\textsc{Counter}   & 29.06 & 0.910 & 0.253 & 05{:}56 & 237 & 0.16M & 1.05M \\
\textsc{Drjohnson} & 29.30 & 0.905 & 0.312 & 06{:}26 & 202 & 0.08M & 2.92M \\
\textsc{Flowers}   & 21.43 & 0.597 & 0.375 & 05{:}52 & 319 & 0.04M & 2.51M \\
\textsc{Garden}    & 27.23 & 0.861 & 0.131 & 08{:}36 & 203 & 0.14M & 3.38M \\
\textsc{Kitchen}   & 30.58 & 0.925 & 0.155 & 08{:}55 & 177 & 0.24M & 1.72M \\
\textsc{Playroom}  & 29.76 & 0.907 & 0.316 & 04{:}40 & 309 & 0.04M & 1.71M \\
\textsc{Room}      & 31.40 & 0.920 & 0.287 & 05{:}14 & 248 & 0.11M & 1.16M \\
\textsc{Stump}     & 26.55 & 0.766 & 0.262 & 06{:}30 & 269 & 0.03M & 3.96M \\
\textsc{Train}     & 21.96 & 0.815 & 0.241 & 04{:}58 & 282 & 0.18M & 1.08M \\
\textsc{Treehill}  & 22.54 & 0.630 & 0.384 & 05{:}58 & 267 & 0.05M & 2.67M \\
\textsc{Truck}     & 25.29 & 0.881 & 0.176 & 05{:}36 & 294 & 0.14M & 1.98M \\
\bottomrule
\end{tabular}
\end{adjustbox}
\end{table}

\vspace{1em}
\begin{table}[h!]
\centering
\small
\caption{Per-scene results of \textsc{3DGS}\textsubscript{DI}.}
\label{tbl:per-scene-3dgs-di}
\begin{adjustbox}{width=0.8\linewidth}
\begin{tabular}{@{}lccccccc@{}}
\toprule
Scene & PSNR$_\uparrow$ & SSIM$_\uparrow$ & LPIPS$_\downarrow$ & Opt Time$_\downarrow$ & FPS$_\uparrow$ & Start \#G$_\downarrow$ & Final \#G$_\downarrow$ \\
\midrule
\textsc{Bicycle}   & 24.91 & 0.760 & 0.203 & 07{:}53 & 261 & 4.22M & 3.06M \\
\textsc{Bonsai}    & 32.07 & 0.941 & 0.206 & 11{:}09 & 195 & 4.50M & 3.51M \\
\textsc{Counter}   & 28.97 & 0.912 & 0.205 & 15{:}19 & 140 & 4.49M & 3.73M \\
\textsc{Drjohnson} & 28.64 & 0.879 & 0.322 & 06{:}45 & 342 & 4.09M & 4.07M \\
\textsc{Flowers}   & 20.71 & 0.595 & 0.317 & 08{:}29 & 276 & 3.68M & 2.49M \\
\textsc{Garden}    & 27.34 & 0.866 & 0.111 & 11{:}55 & 195 & 4.02M & 3.45M \\
\textsc{Kitchen}   & 31.33 & 0.929 & 0.137 & 16{:}44 & 131 & 4.50M & 3.84M \\
\textsc{Playroom}  & 29.79 & 0.897 & 0.289 & 06{:}27 & 386 & 3.70M & 3.67M \\
\textsc{Room}      & 31.29 & 0.927 & 0.240 & 10{:}57 & 192 & 4.49M & 4.01M \\
\textsc{Stump}     & 26.39 & 0.761 & 0.241 & 06{:}13 & 309 & 2.70M & 2.31M \\
\textsc{Train}     & 21.77 & 0.808 & 0.217 & 09{:}38 & 250 & 3.84M & 2.55M \\
\textsc{Treehill}  & 21.87 & 0.598 & 0.324 & 07{:}29 & 278 & 3.01M & 2.52M \\
\textsc{Truck}     & 25.46 & 0.879 & 0.125 & 12{:}58 & 176 & 4.38M & 3.48M \\
\bottomrule
\end{tabular}
\end{adjustbox}
\end{table}

% \subsection{Additional Ablations}
% \todo{Fill in and explain these}
% \begin{table}[h!]
% \centering
% \small
% \caption{Quality and performance \todo{finish caption}.}
% \label{tbl:reduced-iter-count}
% \begin{tabular}{@{}l|cccccc@{}}
% \toprule
% Method & PSNR & SSIM & LPIPS & Opt Time & FPS \\
% \midrule
% \textsc{GRay} & - & - & - & - & - \\
% \textsc{GRay}\textsubscript{Prune@500} & - & - & - & - & - \\
% \textsc{GRay}\textsubscript{DefaultSHLR} & - & - & - & - & - \\
% \bottomrule
% \end{tabular}
% \end{table}

\end{document}